\documentclass[letterpaper,notitlepage,10pt]{article}
\pdfoutput=1
\usepackage{booktabs,ulem}

\usepackage{natbib}
\usepackage{epsfig}
\usepackage{fullpage}
\usepackage{graphicx}
\usepackage{color}
\usepackage{subfigure}
\usepackage{rotating}
\usepackage{geometry}
\usepackage{setspace}
\usepackage{enumerate}
\usepackage{alltt,boites} 
\usepackage{hyperref}
\usepackage{rotating}
\usepackage{verbatim}

\usepackage{amsmath}
\usepackage{amssymb}
\usepackage{amsthm}

\usepackage[usenames,dvipsnames,table]{xcolor}

\RequirePackage{amsfonts,mathrsfs,algorithm,algorithmic,lscape,longtable}

\RequirePackage{soul}
\usepackage{tabularx}
\newcolumntype{C}[2]{>{\hsize=#1\hsize\\centering\arraybackslash}X}%

\title{A Hierarchical Framework for State-Space Matrix Inference and Clustering\footnote{This work was supported by  National Institutes of Health Grants (HG006716 and HG007019) to S.K.}}

\author{Chandler Zuo$^{1,2}$\footnote{E-mail: zuo@stat.wisc.edu}, Kailei
  Chen$^{1,2}$, Kyle J. Hewitt$^3$, Emery H. Bresnick$^3$, and S\"und\"uz Kele\c{s}$^{1,2}$\footnote{E-mail: keles@stat.wisc.edu}}

\date{}

\begin{document}

\maketitle

\noindent
$^1$ Department of Statistics, University of Wisconsin, Madison, WI, U.S.A.

\noindent
$^2$ Department of Biostatistics and Medical Informatics, University of Wisconsin, Madison, WI, U.S.A.

\noindent
$^3$ Department of Cell and Regenerative Biology, University of Wisconsin, Madison, WI, U.S.A.

\begin{abstract}
In recent years, a large number of genomic and epigenomic studies have been focusing on the integrative analysis of multiple experimental datasets measured over a large number of observational units. The objectives of such studies include not only inferring a hidden state of activity for each unit over individual experiments, but also detecting highly associated clusters of units based on their inferred states. Although there are a number of methods tailored for specific datasets,  there is currently no state-of-the-art modeling framework for this general class of problems. In this paper, we develop the MBASIC (\textbf{M}atrix \textbf{B}ased \textbf{A}nalysis for \textbf{S}tate-space \textbf{I}nference and \textbf{C}lustering) framework. MBASIC consists of two parts: state-space mapping and state-space clustering. In state-space mapping, it maps observations onto a finite state-space, representing the activation states of units across conditions. In state-space clustering, MBASIC incorporates a finite mixture model to cluster the units based on their inferred state-space profiles across all conditions. Both the state-space mapping and clustering can be simultaneously estimated through an Expectation-Maximization algorithm. MBASIC flexibly adapts to a large number of parametric distributions for the observed data, as well as the heterogeneity in replicate experiments. It allows for imposing structural assumptions on each cluster, and enables model selection using information criterion. In our data-driven  simulation studies, MBASIC showed significant accuracy in recovering both the underlying state-space variables and clustering structures. We applied MBASIC to two genome research problems using large numbers of datasets from the ENCODE project. The first application grouped genes based on transcription factor occupancy profiles of their promoter regions in two different cell types. The second application focused on identifying groups of loci that are similar to a GATA2 binding site that is functional at its endogenous locus by  utilizing transcription factor occupancy data and illustrated applicability of MBASIC in a wide variety of problems. 
In both studies, MBASIC showed  higher levels of raw data fidelity than analyzing these data with a two-step approach  using ENCODE results on transcription factor occupancy data.
\end{abstract}

\section{Introduction}

The flow of genetic information through DNA
  transcription and RNA translation is a highly regulated process. The underlying mechanisms of regulation by both genomic and epigenomic factors are central targets in large numbers of genomic and epigenomic
  studies. This paper is motivated by a number of such studies
  that aim to elucidate genomic regulatory mechanisms across multiple
  biological conditions. Experiments that investigate such processes produce a
  plethora of data types. For example, relevant to DNA transcription is  transcription factor occupancy
   data  that quantify which  regions of DNA are
  occupied by DNA binding proteins that
  can enhance or reduce gene expression. Histone modification data map
  covalent post-translational modifications to histone proteins, core proteins that make up the nucleosome structure of DNA.
 Such modifications impact DNA transcription by altering chromatin structure.  
  
 Computational and statistical analysis of these data often involve
 identifying genomic loci that show significant signal, i.e.,
 enrichment, compared to background noise in the experimental
 measurements, with the operating principle that multiple loci might exhibit similar signal profile over different biological conditions.

Improvements in the next-generation sequencing technology further  accelerated rapid generation of these types of data. In return, the vast availability of such data has revolutionized the scope of genome regulation studies. Previous analyses  had been restricted to detecting regions of genome that were associated with one particular factor in one single organism. Many recent studies focus on detecting more complex functional patterns that integrate data from multiple organisms under multiple conditions. Namely,  the associations between DNA elements and how they change across biological and/or experimental conditions 
have been the focus of many integrative modeling approaches. Examples of these studies include:

\begin{description}
\item [Differential binding analysis among multiple ChIP-seq data.] 
One of the key mechanisms of gene expression  regulation is through differential
  activities of transcription factors and epigenetic modifications. Currently, chromatin
immunoprecipitation followed by high throughput sequencing (ChIP-seq)
is  the state-of-the-art method for  genome-wide profiling of protein-DNA interactions. Such two key interactions are  DNA 
occupancy  by transcription factors and histone modifications. 
Most transcription factors, i.e., DNA binding proteins, recognize DNA in a sequence specific manner and 
  promote or represses gene expression. Similarly, histone modifications  can induce diverse biological consequences such as transcriptional  activation/deactivation.
The study of gene regulation often involves comparing transcription factor occupancy and histone modifications across multiple biological conditions. Such conditions can be different treatment levels, time points of measurements, or different dosage levels \citep{DBChIP, DEseq, histonePCA, cormotif}. 

\item [Transcription factor regulatory network analysis.] The combinatorial nature of transcription factor regulation  underlies the large diversity observed in eukaryotic gene control. This largely motivates  construction of
  regulatory networks that model gene expression
  as a combinatorial function of regulatory interactions between DNA and different
  transcription factors. The large-scale data from the ENCODE project \citep{ENCODE} now enable joint analyses of over one hundred human transcription factors across multiple cell types. Such analyses are posed to reveal a great amount of information about co-association patterns between different TFs, hierarchical network organizations, and systems-level integration of complex cellular signals \citep{Neph12, Gernstein12, Gernstein11,jMosaics}. While the large number of TFs makes it computationally formidable to exhaust all possible combinatorial associations for such analyses, it is important to detect the most significant combinatorial patterns that preserve global regulatory dynamics.
\item [Comparative functional genomic studies across different species.] Functional genomics analysis compares gene expressions or TF occupancy profiles between multiple species. The main task is to identify divergent and conserved functional modules that are central to evolutionary relationships (e.g., \cite{bourque2010}, \cite{odom2010}). Existing methods, that build on hidden Markov models \citep{arboretum} or biclustering \citep{bicluster}, implicitly assume that the functional modules should at least have similar signal profiles (i.e., expression, occupancy) among some subsets of the species under consideration. For these analyses, it is also important to identify functional modules that are fully divergent across species. These regions play an equally important role in understanding connectivity among species over the evolutionary history.
\end{description}

Although the types of data for these different studies vary, the underlying statistical principles are largely shared. Therefore,  we propose a unified framework for the analysis of such data by formalizing the shared aspects. We formulate the underlying statistical problem as follows. Suppose a dataset  $\{Y_{ik}\}$ is collected over a set of observational units (e.g., loci in genomic experiments) $i=1,2,\cdots,I$ under conditions $k=1,2,\cdots,K$. Inferring the association patterns within a single experiment involves mapping the corresponding set of observations $\{Y_{ik}:i=1,2,\cdots,I\}$ to a finite discrete state-space, $\mathscr S = \{1,2,\cdots,S\}$. This space contains different levels of association (e.g., enrichment/non-enrichment indicating the status of occupancy in ChIP-seq experiments, expressed/not expressed in RNA-seq gene expression experiments). This falls under the classical finite-mixture modeling framework, where a latent state variable $\theta_{ik}\in \mathscr S$ is inferred for each observational unit $Y_{ik}$. A higher level of modeling on the matrix $\Theta=(\theta_{ik})_{ 1\leq i \leq I, 1\leq k \leq K}$ is required for  integrating the association patterns under different conditions. We call this matrix the \textit{state-space matrix} since it describes the latent states of individual observations.

We propose the following framework to model the state-space matrix $\Theta$. We assume that  rows of $\Theta$ can be partitioned into $J+1$ subsets: $\{1,\cdots,I\}=\mathscr C_0\cup \mathscr C_1\cup\cdots \cup \mathscr C_J$. Rows of $\Theta$ within partition $\mathscr C_j$, $j\geq 1$, are generated by the same distribution parametrized by $w_{j\cdot} =(w_{jk})_{1 \leq k \leq K }$:
\[
\theta_{ik}\sim g(\cdot|w_{jk}), \quad i\in \mathscr C_j,
\]
while the rows of $\mathscr C_0$, which denotes the group of "singleton" units, i.e., units that do not cluster in any of the  $J$ groups, are generated by row specific distributions. The goal of this model is thus to estimate a partitioning that best characterizes the row associations of state-space matrix $\Theta$.

We refer to the proposed framework as \textbf{M}atrix \textbf{B}ased \textbf{A}nalysis for \textbf{S}tate-space \textbf{I}nference and \textbf{C}lustering (\textbf{MBASIC}). MBASIC is related to classical factor analysis which considers the problem of projecting one dimension (either row or column) of large noisy matrices into low-dimensional spaces. MBASIC has two distinguished features compared to the existing literature in these areas. First, MBASIC deals with matrices with discrete entries, while most existing methods are designed for matrices on continuous scales. Second, MBASIC estimates the low-dimensional projection by grouping the rows of the original matrix in contrast to the Principle Component Analysis (PCA) approaches which form linear combinations of the rows (e.g., \cite{histonePCA}, \cite{binMatrix}). This is motivated by the following arguments:

\begin{enumerate}
\item In MBASIC, each factor estimate $w_{j\cdot}$ characterizes the commonality of a group of rows and is easily interpretable in practice. Such interpretability can further be enhanced by imposing structural restrictions on the $w_{j\cdot}$ vector for practical purposes. Examples of such constraints are  described in Section \ref{sec:struct};
\item PCA for high dimensional matrices are often accompanied by regularization techniques, which are computationally prohibitive for many epigenetic datasets. In contrast, clustering the matrix rows  can be implemented very efficiently and  in a straightforward manner.
\end{enumerate}

 The hierarchical structure of MBASIC is similar to two other recently proposed statistical models: iASeq \citep{iaseq} and Cormotif \citep{cormotif}. Both these models incorporate a state-space clustering structure similar to MBASIC. MBASIC extends these models in several critically essential directions. First, MBASIC is developed for general purposes and can be easily implemented for a wide range of parametric distributions, while Cormotif and iASeq operate with specific distributions targeting the problems of differential expression and allele-specific binding. Second, neither of these models include a group of singletons with idiosyncratic state-space profiles. When we are agnostic about the ``true'' clustering structure in applications, separating the singletons can reduce their influence on the estimation of clustering parameters. Third, both iASeq and Cormotif separate estimation for the distributional parameters from the clustering structure, while MBASIC jointly fits all model parameters. A limiting assumption of MBASIC compared to these models is that MBASIC does not allow the distributional parameters within the same state to be heterogeneous. However, a pre-processing step that accounts for the the heterogeneity can overcome such a limitation. We evaluate and discuss all of these features with extensive simulation studies in this paper.

This paper is organized as follows. We start with a formal description of MBASIC in Section \ref{sec:model}, followed by model estimation and selection methods in Section \ref{sec:estimation}. We also investigate general features of MBASIC compared to iASeq and  Cormotif with extensive simulations in this section. Section \ref{sec:realdata} presents results from several real data examples. Mathematical details of the algorithm are included in Appendix \ref{sec:em}.

\section{The Hierarchical Mixture Model Framework\label{sec:model}}

Consider a dataset with observations from $I$ different \textit{observational units} under $K$ different \textit{conditions}. For each condition $k\in\{1,2,\cdots,K\}$, there are $n_k$ replicate experiments, indexed by $l=1,2,\cdots,n_k$. We use $Y_{ikl}$ to denote the observation for the $l$-th replicate of  unit $i$ under condition $k$. For each condition $k$ at unit $i$, there exists a hidden state variable $\theta_{ik}\in \mathscr S=\{1,2,\cdots,S\}$. The MBASIC model consists of the following components:

\begin{enumerate}
\item \textbf{State-space Mapping}:
\begin{equation}\label{eq:mixdist}
Y_{ikl} | \theta_{ik}=s {\buildrel\rm ind.\over\sim} f_s(\cdot|\mu_{kls},\sigma_{kls},\gamma_{ikls}). 
\end{equation}

\item \textbf{State-space Clustering}: $\theta_{ik}$'s are independently sampled from $\mathscr S$ with the sampling probability:
\begin{equation}\label{eq:cluster}
P(\theta_{ik}=s)=\zeta p_{is} + (1-\zeta)\sum_{j=1}^J\pi_jw_{jks}.
\end{equation}
\end{enumerate}

In (\ref{eq:mixdist}), $\mu_{kls}$ and $\sigma_{kls}$ are the parameters related to the mean and dispersion for the $s$-th state for replicate $l$ under condition $k$, and $\gamma_{ikls}$ is the covariate encoding known information for unit $i$. In (\ref{eq:cluster}), $p_{is}$, $\zeta$, $\pi_j$, and $w_{jks}$ are additional non-negative parameters subject to restrictions: 
\[
\quad 0\leq \zeta \leq 1; \quad \sum_{j=1}^J\pi_j = 1; \quad \sum_{s=1}^Sw_{jks}=1, \forall j, k; \quad \sum_{s=1}^Sp_{is}=1, \forall i.
\]

We further discuss these parameters in Section \ref{sec:clustermodel}.

\subsection{State-space Mapping}

Equation \ref{eq:mixdist} partitions observational units $i=1,\cdots,I$ into $S$ subsets according to their hidden states. Within the same replicate, data from the same hidden state follow the same distribution $f_s(\cdot|\mu_{kls},\sigma_{kls},\gamma_{ikls})$. MBASIC assumes that the hidden states $\theta_{ik}$'s are independent of the replicate index $l$, which means all replicates under the same condition have the same set of hidden states. However, distributional parameters for a given state can be different among replicates. Such a setting allows for the flexibility of modeling the heterogeneity in replicate experiments.

The density function $f$ can be from an arbitrary  parametric distribution. We consider three fundamental families of distributions commonly used for genomic data analysis:

\begin{itemize}
\item \textit{Log-normal Distribution.} $LN(\mu_{kls}\gamma_{ikls},\sigma_{kls})$ with a density function: 
\begin{equation}\label{eq:lognormal}
f_s(y|\mu_{kls},\sigma_{kls},\gamma_{ikls})=\frac{1}{\sqrt{2\pi}\sigma_{kls}}\exp\left \{-\frac{(\log(y+1)-\mu_{kls}\gamma_{ikls})^2}{2\sigma_{kls}^2} \right \}.
\end{equation}
\item \textit{Negative Binomial Distribution.}  $NB(\mu_{kls}\gamma_{ikls},\sigma_{kls})$ with a density function:
\begin{equation}\label{eq:negbin}
f_s(y|\mu_{kls},\sigma_{kls},\gamma_{ikls})=\frac{\Gamma(y+\sigma_{kls})}{\Gamma(\sigma_{kls})\Gamma(y)}\frac{(\mu_{kls}\gamma_{ikls})^y\sigma_{kls}^{\sigma_{kls}}}{(\mu_{kls}\gamma_{ikls}+\sigma_{kls})^{y+\sigma_{kls}}}.
\end{equation}
\item \textit{Binomial Distribution.} $Binom(\gamma_{ikls},\mu_{kls})$ with a density function:
\begin{equation}\label{eq:binom}
f_s(y|\mu_{kls},\gamma_{ikls})={\gamma_{ikls}\choose y}\mu_{kls}^y(1-\mu_{kls})^{\gamma_{ikls}-y}.
\end{equation}

\end{itemize}

In these three examples, $\gamma_{ikls}$ represents the known heterogeneity across loci whereas $\mu_{kls}$ and $\sigma_{kls}$ are unknown parameters. For example, when using Eqn. (\ref{eq:lognormal}) or (\ref{eq:negbin}) in a ChIP-seq analysis with $S=2$ states, we can estimate $\gamma_{ikl1}$ using data from the control samples so that 
the ChIP sample read counts in the background state scale with the control sample data (e.g., as in \cite{cssp}), and assume $\gamma_{ikl2}=1$ for the enriched states. Eqn. (\ref{eq:binom}) can be used to analyze allele-specific binding data, where $\gamma_{ikls}$ is the total read counts from both paternal and maternal alleles and is constant across $s$.  Application with the binomial distribution also requires that $\mu_{kls}\sum_{i=1}^I\gamma_{ikls}$, $\forall k, l$, is strictly increasing in $s$ for model identification.

The MBASIC can be easily extended to other classes of parametric distributions and estimation for these distributions follows the same Expectation-Maximization skeleton. While Section \ref{sec:estimation} relies on these three distributions to describe the model and the estimation algorithms, the second real data example in Section \ref{sec:realdata}  utilizes a more complex parametrization, which demonstrates the wide applicability of the MBASIC framework. Furthermore, we consider the following degenerate distribution:
\begin{equation}\label{eq:degen}
f_s(y|\mu_{kls},\sigma_{kls},\gamma_{ikls})=I(y=s),
\end{equation}
where I(.) denotes the indicator function. This degenerate form corresponds to the situation where the states, $\theta_{ik}$'s, are directly observed rather than inferred from $Y_{ikl}$'s. We utilize this parametrization for comparing MBASIC to alternative two-step analysis approaches in Section~\ref{sec:simulation}.  Parameter estimation for this case follows a slightly modified procedure from the non-degenerate cases, which is described in Section \ref{sec:estimation}.

\subsection{State-space Clustering\label{sec:clustermodel}}

Equation (\ref{eq:cluster}) models the distribution of $\theta_{ik}$ as a mixture of multiple distributions. To illustrate this model we introduce additional variables. The goal is to identify $J$ clusters from the set of observation units $1\leq i \leq I$. Let $b_i = I(\mbox{unit $i$ does not belong to any cluster})$ and $z_{ij}= I(\mbox{unit $i$ belongs to cluster $j$})$. The $b_i$ variables entertain the possibility that some observations are "singletons", i.e., they do not cluster with any other observational units. With these additional variables, the distribution in Equation (\ref{eq:cluster}) can be hierarchically decomposed  as follows:

\begin{itemize}
\item $b_i{\buildrel\rm i.i.d. \over\sim} Bernoulli(\zeta)$;
\item $(z_{i1},~z_{i2},\cdots,~z_{iJ}){\buildrel\rm i.i.d. \over\sim} MultiNom(1,(\pi_1,~\pi_2,\cdots,~\pi_J))$;
\item Conditional on $b_i$ and $z_{ij}$, $\theta_{ik}$'s are independent samples from $\mathscr S$, with sampling probabilities $P(\theta_{ik}=s|b_i=1)=p_{is}$, $P(\theta_{ik}=s|b_i=0,z_{ij}=1)=w_{jks}$.
\end{itemize}

In this set up, although the singleton state-space probabilities
$p_{is}$ are assumed to be constant across conditions, i.e.,
$P(\theta_{ik}=s) = p_{is}$, $\forall k$, this assumption is mildly
restrictive since it accommodates $(P(\theta_{ik}=1, \cdots,
P(\theta_{ik}=S))$ to follow an arbitrary prior distribution (e.g.,
$(P(\theta_{ik}=1, \cdots, P(\theta_{ik}=S)) ~\sim Dirichlet(\alpha,
\cdots, \alpha), \forall k$) as long as it leads to the same marginal
distribution for $\theta_{ik}$ for all $k$.

It is worth noting that this hierarchical structure essentially seeks a low-rank representation for the matrix $\Theta=(\theta_{ik})_{1\leq i \leq I, 1\leq k \leq K}$. To illustrate this, we introduce additional matrices $\Theta_s=(I(\theta_{ik}=s))_{1\leq i\leq I, 1\leq k\leq K}$, $W_s=(w_{jks})_{1\leq j \leq J, 1\leq k \leq K }$, $Z=(z_{ij})_{1\leq i \leq I, 1 \leq j \leq J}$ and vectors $p_s=(p_{is})_{1\leq i \leq I}$, $B=(b_i)_{1\leq i \leq I}$. Then, the conditional expectation of $\Theta_s$ is:

\begin{equation}
E(\Theta_s|Z,B)=(ZW_s)\circ((1-B)1_K^T)+(p_s\circ B)1_K^T,
\end{equation}

where ``$\circ$'' denotes the Hadamard product. We note that $E(\Theta_s|Z,B)$ is a matrix of rank $J+1$, which is usually much smaller than the dimension of the matrix $\Theta_s$. Similar models for low-rank representation of discrete matrices were considered in \cite{binMatrix}, and turned out to be challenging both theoretically and computationally. The row-clustering structure for the matrices $E(\Theta_s|Z,B)$ in MBASIC is more restrictive than the general low-rank structure. Such additional restrictions not only reduce the difficulty in parameter estimation but also enable the flexibility in many useful ways. For example, while \cite{binMatrix} can only estimate one matrix at a time and thus is only applicable  when $S=2$, MBASIC can be applied to arbitrary values of $S$.

\section{Model Estimation and Selection\label{sec:estimation}}

\subsection{Likelihood Functions}

In the MBASIC model, the  likelihood function for both the observed random variables $Y_{ikl}$'s and the unobserved $\theta_{ik}$'s, $z_{ij}$'s, $b_i$'s, i.e., full data likelihood,  is given by:

\begin{equation}\label{eq:fulllik}
\begin{aligned}
l&(\mu,\sigma, \pi,p,\zeta,w;y,\theta,z,b) = \\
& \prod_{i=1}^I\zeta^{b_i}(1-\zeta)^{1-b_i} 
 \cdot \prod_{i=1}^I\prod_{k=1}^K\prod_{s=1}^Sp_{is}^{I(\theta_{ik}=s)b_i}
 \cdot \prod_{i=1}^I\prod_{j=1}^J \pi_j^{z_{ij}} \\
\cdot & \prod_{i=1}^I\prod_{k=1}^K\prod_{s=1}^S\left [\prod_{l=1}^{n_k}f_s(y_{ikl}|\mu_{kls},\sigma_{kls},\gamma_{ikls}) \right ]^{I(\theta_{ik}=s)} 
 \cdot \prod_{i=1}^I\prod_{j=1}^J\prod_{k=1}^K\prod_{s=1}^Sw_{jks}^{I(\theta_{ik}=s)(1-b_i)z_{ij}}.\\
\end{aligned}
\end{equation}

For non-degenerate distributions, we can show that the marginal likelihood is:

\begin{equation}\label{eq:marginlik}
\begin{aligned}
  l(\mu,\sigma, \pi,p,\zeta,w;y)=&\prod_{i=1}^I \left \{\zeta\prod_{k=1}^K \left [\sum_{s=1}^Sp_{is}\prod_{l=1}^{n_k}f_s(y_{ikl}|\mu_{kls},\sigma_{kls},\gamma_{ikls}) \right ] \right. \\
+ & \left. (1-\zeta)\sum_{j=1}^J\pi_j\prod_{k=1}^K \left [\sum_{s=1}^Sw_{jks}\prod_{l=1}^{n_k}f_s(y_{ikl}|\mu_{kls},\sigma_{kls},\gamma_{ikls}) \right ] \right \}.
\end{aligned}
\end{equation}

Equation (\ref{eq:marginlik}) is easily interpretable. Conditional on $b_i$ and $z_{ij}$, the joint distribution for each $Y_{ikl},~1\leq l \leq n_k$ is a mixture of $S$ components, where the weight on the s-th component is either $p_{is}$ (when $b_i = 1$) or $w_{jks}$ (when $b_i=0$ and $z_{ij}=1$). This yields the expressions in the square brackets. Integrating out $b_i$ and $z_{ij}$, the joint distribution for $Y_{ikl}$ of fixed $i$ is a mixture of $J+1$ components, with probability $\zeta$ of being a singleton and probability $(1-\zeta)\pi_j$ of belonging to cluster $j$. 

For the degenerate case, by substituting (\ref{eq:degen}) into (\ref{eq:marginlik}), it can be shown that the marginal likelihood is:

\begin{equation}\label{eq:thetalik}
l(\mu,\sigma, \pi,p,\zeta,w;\theta)=\prod_{i=1}^I \left \{\zeta\prod_{k=1}^K\prod_{s=1}^Sp_{is}^{I(\theta_{ik}=s)}+(1-\zeta)\sum_{j=1}^J\pi_j\prod_{k=1}^K\prod_{s=1}^Sw_{jks}^{I(\theta_{ik}=s)} \right \}.
\end{equation}

\subsection{An Expectation and Maximization (E-M) Algorithm}

The hierarchical structure of MBASIC naturally fits in the Expectation-Maximization algorithm \cite{em}, which maximizes the marginal likelihood (equations (\ref{eq:marginlik}) or (\ref{eq:thetalik})) by iteratively maximizing the complete data log-likelihood function.  We let $\phi$ to denote a vector including all unknown parameters $\mu$, $\sigma$, $\pi$, $p$, $\zeta$, $w$, and $\hat \phi ^{(t)}$ to denote the parameter estimates at the $t$-th iteration. The complete data log-likelihood function is:

\begin{equation}\label{eq:Q}
\begin{aligned}
Q(\phi |\hat \phi^{(t-1)}) = & \sum_{i=1}^I\sum_{k=1}^K\sum_{s=1}^S \left [\sum_{l=1}^{n_k}\log f_s(y_{ikl}|\mu_{kls},\sigma_{kls},\gamma_{ikls}) \right ]E[I(\theta_{ik}=s)|\hat \phi^{(t-1)}] \\
 + & \sum_{i=1}^I\sum_{k=1}^K\sum_{s=1}^S\log p_{is} E[I(\theta_{ik}=s)b_i|\hat \phi^{(t-1)}] +  \sum_{i=1}^I\sum_{j=1}^J \log \pi_jE[z_{ij}(1-b_i)|\hat \phi^{(t-1)}] \\
 + & \sum_{i=1}^I\{ \log\zeta E[b_i|\hat \phi^{(t-1)}] + \log(1-\zeta)(1-E[b_i|\hat \phi^{(t-1)}])\}\\
 + & \sum_{i=1}^I\sum_{k=1}^K\sum_{j=1}^J\sum_{s=1}^SE[I(\theta_{ik}=s)z_{ij}(1-b_i)|\hat \phi^{(t-1)}]\log w_{jks}.
\end{aligned}
\end{equation}

The E-M algorithm for MBASIC is outlined by Algorithm \ref{alg:em}. E-step updates are  listed in equations (\ref{eq:estep:b})-(\ref{eq:estep:thetab}) and their derivations are provided in Appendix \ref{sec:appalg}.

\begin{algorithm}                      
\caption{Expectation-Maximization (EM)}          
\label{alg:em}                           
\begin{algorithmic}                    
  \FOR{$t=1,~2,\cdots$ until convergence}
  \STATE \textit{Expectation Step}: Compute the conditional expectations  $E[I(\theta_{ik}=s)|\hat \phi^{(t-1)}]$, $E[b_i|\hat \phi^{(t-1)}]$, $E[I(\theta_{ik}=s)b_i|\hat \phi^{(t-1)}]$, $E[z_{ij}(1-b_i)|\hat \phi^{(t-1)}]$, $E[I(\theta_{ik}=s)z_{ij}(1-b_i)|\hat \phi^{(t-1)}]$;
   \STATE \textit{Maximization Step}: Update estimates for parameters $\mu_{kls}$, $\sigma_{kls}$, $\zeta$, $\pi_j$, $w_{jks}$, $p_{is}$ as maximizers for (\ref{eq:Q}) .
   \ENDFOR
\end{algorithmic}
\end{algorithm}

\begin{equation}\label{eq:estep:b}
\begin{aligned}
E(&b_i|\hat\phi^{(t-1)})=\\
&\frac{\hat\zeta^{(t-1)} \prod_{k=1}^K \left (\sum_{s=1}^S\hat f_{iks}^{(t-1)}\hat p_{is}^{(t-1)}\right )}{(1-\hat\zeta^{(t-1)})\sum_{j=1}^J\hat\pi_j^{(t-1)}\prod_{k=1}^K\left (\sum_{s=1}^S\hat f_{iks}^{(t-1)}\hat w_{jks}^{(t-1)} \right )+\hat\zeta^{(t-1)} \prod_{k=1}^K \left (\sum_{s=1}^S\hat f_{iks}^{(t-1)}\hat p_{is}^{(t-1)} \right )},\\
\end{aligned}
\end{equation}

\begin{equation}\label{eq:estep:zb}
E(z_{ij}(1-b_i)|\hat\phi^{(t-1)})=\frac{\hat\pi_j^{(t-1)}\prod_{k=1}^K \left (\sum_{s=1}^S\hat f_{iks}^{(t-1)}\hat w_{jks}^{(t-1)} \right )}{\sum_{j=1}^J\hat \pi^{(t-1)}_j\prod_{k=1}^K \left (\sum_{s=1}^S\hat f_{iks}^{(t-1)}\hat w_{jks}^{(t-1)} \right )}[1-E(b_i|\hat\phi^{(t-1)})],
\end{equation}

\begin{equation}\label{eq:estep:thetazb}
\begin{aligned}
&E(I(\theta_{ik}=s)z_{ij}(1-b_i)|\hat\phi^{(t-1)})=\\
&[1-E(b_i|\hat\phi^{(t-1)})]
\frac{\hat \pi_j^{(t-1)}\prod_{k=1}^K \left (\sum_{s=1}^S\hat f_{iks}^{(t-1)}\hat w_{jks}^{(t-1)} \right )}{\sum_{j=1}^J\hat \pi_j^{(t-1)}\prod_{k=1}^K \left (\sum_{s=1}^S\hat f_{iks}^{(t-1)}\hat w_{jks}^{(t-1)} \right )} \cdot \frac{\hat f_{iks}^{(t-1)}\hat w_{jks}^{(t-1)}}{\sum_{s=1}^S\hat f_{iks}^{(t-1)}\hat w_{jks}^{(t-1)}},\\
\end{aligned}
\end{equation}

\begin{equation}\label{eq:estep:thetab}
E(I(\theta_{ik}=s)b_i|\hat\phi^{(t-1)})=E(b_i|\hat\phi^{(t-1)})\cdot \frac{\hat f_{iks}^{(t-1)}\hat p_{is}^{(t-1)}}{\sum_{s=1}^S\hat f_{iks}^{(t-1)}\hat p_{is}^{(t-1)}},
\end{equation}

where $\hat f_{iks}^{(t-1)}=\prod_{l=1}^{n_k}f(y_{ikl}|\hat\mu_{kls}^{(t-1)},\hat\sigma_{kls}^{(t-1)},\gamma_{ikls})$. Given these results from the E-step, updates of $\zeta$, $\pi_j$, $w_{jks}$, $p_{is}$ in the M-step are straight forward as in equations (\ref{eq:updatezeta}), (\ref{eq:updatepi}), (\ref{eq:updatep}), and (\ref{eq:updatew}). 

\begin{eqnarray}
\hat \zeta^{(t)}&=&\frac{\sum_{i=1}^IE[b_i|\hat \phi^{(t-1)}]}{I}, \label{eq:updatezeta}\\
\hat \pi_j^{(t)}&=&\frac{\sum_{i=1}^IE[z_{ij}(1-b_i)|\hat \phi^{(t-1)}] }{\sum_{i=1}^I\sum_{j=1}^JE[z_{ij}(1-b_i)|\hat \phi^{(t-1)}] }, \label{eq:updatepi}\\
\hat p_{is}^{(t)}&=&\frac{\sum_{k=1}^KE[I(\theta_{ik}=s)b_i|\hat \phi^{(t-1)}]}{\sum_{s=1}^S\sum_{k=1}^KE[I(\theta_{ik}=s)b_i|\hat \phi^{(t-1)}]}, \label{eq:updatep}\\
\hat w_{jks}^{(t)}&=&\frac{\sum_{i=1}^IE[I(\theta_{ik}=s)z_{ij}(1-b_i)|\hat \phi^{(t-1)}]}{\sum_{s=1}^S\sum_{i=1}^IE[I(\theta_{ik}=s)z_{ij}(1-b_i)|\hat \phi^{(t-1)}]}. \label{eq:updatew}
\end{eqnarray}

Updates for $\mu_{kls}$ and $\sigma_{kls}$ have to be treated according to the specific distributions. For the log-normal distributions (\ref{eq:lognormal}), we have:

\begin{eqnarray}
\quad \hat \mu_{kls}^{(t)}&=&\frac{\sum_{i=1}^I\log (y_{ikl}+1)P[\theta_{ik}=s|\hat \phi^{(t-1)}]}{\sum_{i=1}^I\gamma_{ikls}P[\theta_{ik}=s|\hat \phi^{(t-1)}]}, \label{eq:updatemu}\\
\quad \hat \sigma_{kls}^{(t)2}&=&\frac{\sum_{i=1}^IP[\theta_{ik}=s|\hat \phi^{(t-1)}] [\log (y_{ikl}+1)-\hat \mu_{kls}^{(t)}\gamma_{ikls}]^2}{\sum_{i=1}^IP[\theta_{ik}=s|\hat \phi^{(t-1)}]}. \label{eq:updatesigma}
\end{eqnarray}

For the binomial distributions (\ref{eq:binom}), we have:

\begin{equation}
\quad \hat \mu_{kls}^{(t)}=\frac{\sum_{i=1}^Iy_{ikl}P[\theta_{ik}=s|\hat \phi^{(t-1)}]}{\sum_{i=1}^I\gamma_{ikls}P[\theta_{ik}=s|\hat \phi^{(t-1)}]}. 
\end{equation}

Closed form maximizers of $\mu$ and $\sigma$ do not exist for the negative binomial distribution (\ref{eq:negbin}). We adopt the method of moment estimates as in \cite{Kuan11, cssp}, where the updated values $\hat \mu_{kls}^{(t)}$ and $\hat \sigma_{kls}^{(t)}$ are the solutions of the following equations: 

\begin{eqnarray*}
\hat \mu_{kls}^{(t)}\sum_{i=1}^I\gamma_{ikls}P[\theta_{ik}=s|\hat \phi^{(t-1)}]&=&\sum_{i=1}^Iy_{ikl}P[\theta_{ik}=s|\hat \phi^{(t-1)}], \\
\sum_{i=1}^I \left [\hat \mu_{kls}^{(t)2}\gamma_{ikls}^2  \left ( 1+\frac{1}{\hat \sigma_{kls}^{(t)}} \right )+\hat \mu_{kls}^{(t)}\gamma_{ikls} \right ]P[\theta_{ik}=s|\hat \phi^{(t-1)}] &=&\sum_{i=1}^Iy_{ikl}^2P[\theta_{ik}=s|\hat \phi^{(t-1)}  ].
\end{eqnarray*}

For the degenerate distributions as in (\ref{eq:degen}), $\theta_{ik}$'s are directly observed. Therefore, the E-M algorithm for this case  requires slight modifications: we skip the estimation for $E[I(\theta_{ik}=s)|\hat \phi^{(t-1)}]$ in the E-step and for $\mu$, $\sigma$ in the M-step.

\subsection{Estimating Structured Clusters}\label{sec:struct}

In integrative functional genomics studies, the set of experimental conditions usually consists of interactions of multiple experimental factors; hence, it is often important to identify clusters,  states of which are homogeneous across the levels of one or more factors. For example, in a typical transcription factor network analysis, experimental conditions include the combination of different cell types and TFs. It is often desirable to separate loci groups
  whose states are homogeneous within each cell type from those with
  cell type specific states for the purpose of cell type
  comparison. Depending on the cell types involved, such comparison
  can yield insights on cell development, pathology and/or cell-specific functions.
  We refer to clusters with homogeneous states within each cell type
  as \textit{TF-homogeneous}. Another example is encountered in  comparative functional genomics studies across different species, where experimental conditions range across both species and TFs. Clusters of loci, states of which are homogeneous across species conditional on each TF, constitute  conserved functional modules. The \textit{TF-homogeneous} clusters in this context represent the marginal effect of the species factor, and play a central role in understanding the evolutionary relationships.

To estimate a cluster with homogeneity for a particular experimental factor, MBASIC allows structural constraints on its state-space parameters. Recall that the  parameters of cluster $j$ are represented by $w_{j.s}=(w_{j1s}, w_{j2s}, \cdots, w_{jKs})$. Marginalizing the effect of this factor, the $K$ experimental conditions can be partitioned into $M$ sets, $\{1,2,\cdots,K\} = T_1\cup T_2\cup \cdots \cup T_M$, where conditions within each set differ only in the levels of this factor. The  parameters of this cluster satisfy the following constraints:

\begin{equation}\label{eq:constraint}
w_{jk_1s}=w_{jk_2s},\quad \mbox{if}  \quad \exists~m  \quad s.t. \quad k_1, k_2 \in T_m.
\end{equation}

\begin{figure}[hbt]
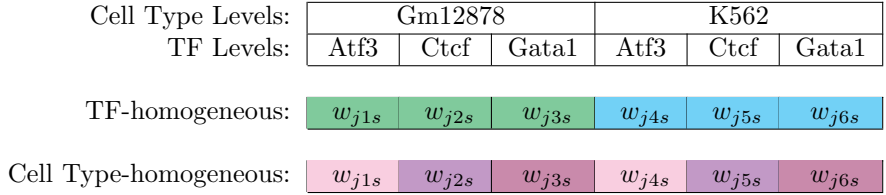

\centering
\begin{tabular}{r|c|c|c|c|c|c|}
\cline{2-7}
\multicolumn{1}{r|}{Cell Type Levels:} & \multicolumn{3}{c|}{Gm12878} & \multicolumn{3}{c|}{K562} \\
\cline{2-7}
TF Levels: & Atf3 & Ctcf & Gata1 & Atf3 & Ctcf & \multicolumn{1}{c|}{Gata1} \\
\cline{2-7}
\multicolumn{1}{c}{   }&\multicolumn{1}{c}{   }&\multicolumn{1}{c}{   }&\multicolumn{1}{c}{   }&\multicolumn{1}{c}{   }&\multicolumn{1}{c}{   }&\multicolumn{1}{c}{   }\\
\cline{2-7}
TF-homogeneous: & \cellcolor{Green!50} $w_{j1s}$ & \cellcolor{Green!50} $w_{j2s}$ & \cellcolor{Green!50} $w_{j3s}$ & \cellcolor{ProcessBlue!50} $w_{j4s}$ & \cellcolor{ProcessBlue!50} $w_{j5s} $& \cellcolor{ProcessBlue!50} $w_{j6s}$ \\
\cline{2-7}
\multicolumn{1}{c}{   }&\multicolumn{1}{c}{   }&\multicolumn{1}{c}{   }&\multicolumn{1}{c}{   }&\multicolumn{1}{c}{   }&\multicolumn{1}{c}{   }&\multicolumn{1}{c}{   }\\
\cline{2-7}
Cell Type-homogeneous: & \cellcolor{Lavender!50} $w_{j1s}$ & \cellcolor{Purple!50} $w_{j2s}$ &\cellcolor{RedViolet!50} $w_{j3s}$& \cellcolor{Lavender!50} $w_{j4s}$ & \cellcolor{Purple!50} $w_{j5s}$ &\cellcolor{RedViolet!50} $w_{j6s}$\\
\cline{2-7}
\end{tabular}
\caption{A graphical description for a parametrization with structural constraints.  Interactions of 2 cell types and 3 TFs result in six experimental conditions. Parameters with homogeneous values are shaded by the same color.}\label{fig:structure}
\end{figure}

A pictorial depiction with six experimental conditions due to full interaction between 2 cell types and 3 TFs is depicted in Figure \ref{fig:structure}. Estimating structured clustering models follows the previous E-M algorithm with a modification in Equation (\ref{eq:updatew}). A constrained maximizer  for $w_{jks}$ subject to constraint  (\ref{eq:constraint})  is computed as:
\[
\hat w_{jks}^{(t)}=\frac{\sum_{k':k'\in T_m}\sum_{i=1}^IE[I(\theta_{ik'}=s)z_{ij}(1-b_i)|\hat \phi^{(t-1)}]}{\#\{T_m\}\sum_{s=1}^S\sum_{i=1}^IE[I(\theta_{ik}=s)z_{ij}(1-b_i)|\hat \phi^{(t-1)}]},~~k\in T_m.
\]

MBASIC requires that  such structural constraints must be specified a priori and remain fixed during model fitting. MBASIC incorporates a model selection procedure to compare models with different hypothesized structural constraints and numbers of clusters. 
We next describe the details of this model selection procedure.

\subsection{Model Selection}\label{sec:modelselect}

The MBASIC framework so far assumes that the total number of clusters $J$ is known a priori. In practice, models with varying values of $J$ need to be fitted independently and compared with each other according to some information criterion to determine the best value of $J$. Since the E-M algorithm aims to maximize the data likelihood function, AIC and BIC criteria can be utilized with MBASIC. The degrees of freedom for a model with $J$ clusters is $df=F_1S\sum_{k=1}^Kn_k+(S-1)I+J+F_2$, where $F_1=2$ for distributions (\ref{eq:lognormal}) and (\ref{eq:negbin}), $F_1=1$ for (\ref{eq:binom}), and $F_2$ is the total number of free variables among $w_{jks}$'s. If there are no structured clusters, we have $F_2=JK(S-1)$.

When there is no prior information available, both the total number of clusters and the number of clusters following structural constraints have to be determined. This results in a prohibitively large number of candidate models, and computing the information criterion for each of them is not practical. In such cases we incorporate the following two-phase strategy to limit the number of candidate models:

\begin{enumerate}
\item Evaluate models with varying total number of clusters without any structural constraints. Select $J_{opt}$ according to the minimal AIC or BIC value.
\item Evaluate models with the fixed number of $J_{opt}$ clusters while varying the number of clusters following each structural constraint. Select the number of clusters following each structural constraint based on the minimal AIC or BIC value.
\end{enumerate}

We acknowledge that the above two-step strategy is only a practical compromise to restrict the space of candidate models and does not guarantee finding the best model that globally minimizes the information criterion. However, we have conducted extensive simulation studies which illustrated that the proposed two-phase strategy  performs well in a wide variety of settings.

\subsection{Simulation Studies}\label{sec:simulation}

\begin{table}
\centering
\caption{Design of the simulation studies. 
$S$: size of the state-space; $\zeta$: Proportion of singletons; $I$: number of units; $J$: number of clusters;  $K$: number of experimental conditions. 
}\label{tbl:simdesign}
\begin{tabular}{ccccccp{1.3cm}}
\toprule
Study & Distribution & $S$ & $\zeta$ & $I$ & $(J, K)$ & Model Selection \\
\midrule
1 & LN, NB, Bin & 2, 3, 4 & 0, 0.1, 0.4 & 4000 & (20, 30) & No \\
2 & LN, NB, Bin & 2 & 0.1, 0.4 & 4000 & (20, 30) & Yes \\
3 & iASeq & 3 & 0, 0.1, 0.4 & 4000 & (10, 20), (20, 30) & Yes \\
4 & Cormotif & 2 & 0 & 10,000 & (4, 4), (5, 8), (5, 10) & Yes \\
5 & Cormotif & 2 & 0, 0.1, 0.4 & 4000 & (10, 20) & Yes \\
6 & LN & 2 & 0, 0.33 & 4120, 4600, 6120 & (8, 30) & Yes\\
\bottomrule
\end{tabular}
\end{table}

We conducted 6 model-based simulation studies to investigate the performance of MBASIC in various settings as summarized in Table \ref{tbl:simdesign}. Each simulation study has multiple settings that vary the distributional assumptions, size of the state-space ($S$), proportion of singletons ($\zeta$), number of units ($I$), number of clusters ($J$), and number of conditions ($K$).  
We provide the details of these simulation studies in Appendix \ref{sec:appsimulation} and highlight the overall conclusions in this section. 

Data in Simulation Studies 1-2 were simulated according to MBASIC's distributional assumptions. In Simulation Study 1, we emphasized two most important features of MBASIC: the joint estimation procedure of all model parameters and the inclusion of a singleton cluster. We derived six alternative algorithms (Table \ref{tbl:s1summary}) to benchmark MBASIC's performance in various settings. Three of the algorithms (SE-HC, SE-MC, PE-MC) use two-stage procedures for model estimation, decoupling either the estimation of the state-space variables or the distributional parameters from the mixture modeling of clustering analysis. The other three algorithms are created as variations on these by excluding the singleton feature (SE-MC0, PE-MC0, MBASIC0). These benchmark algorithms are in spirit analogous to procedures in many applied genomic data analyses where the association between observational units are estimated separately from the estimation of individual data set specific parameters (e.g., \cite{Gernstein12}, \cite{iaseq}, \cite{cormotif}). 

Figures \ref{fig:sim_ln}-\ref{fig:sim_bin} summarize the performance comparisons in Simulation Study 1. We observed that MBASIC's joint estimation feature improved the inference for both the clustering structure and the individual states. In the presence of many singletons, the inclusion of their idiosyncratic state-space profiles was essential for robust clustering.
In Simulation Study 2, we evaluated the effect of using BIC to select the number of clusters as well as the structural constraints within each cluster. Tables \ref{tbl:unstruct} and \ref{tbl:struct} indicate that MBASIC was always able to select models with similar structures with the simulated truth.

In Simulation Studies 3 to 5, we  simulated data according to the models proposed by iASeq \citep{iaseq} and Cormotif \citep{cormotif}. These models allow heterogeneous distributional parameters within the same state, a potential advantage over MBASIC in specific data analysis  such as differential expression or allele-specific binding. Comparison to these two models is intended to enable investigation of whether MBASIC is robust against such within-state heterogeneity. In Simulation Study 3, we showed that MBASIC with the binomial distribution could directly handle data generated under the iASeq framework and achieve competitive performance (Figure \ref{fig:iaseq}). In Simulation Study 4, we inherited the simulation settings from \cite{cormotif}, where distributions from different states were weakly separable, but the individual states were completely deterministic from the clustering. We explored more dynamic settings in Simulation Study 5, where we had easier separation between different states, but randomness among the states within the same cluster. We showed that a pre-processing step homogenizing the within-state units followed by MBASIC leads to comparable performance to Cormotif in Simulation Study 4 (Figure \ref{fig:cormotif1}), and much better performance in Simulation Study 5 (Figure \ref{fig:cormotif2}).

 \cite{cormotif} discusses an interesting point that when the clustering model does not accommodate singletons, small clusters tend to be merged together to form spurious clusters, estimated state-space patterns of which are the averages among several true clusters. In order to investigate whether such a phenomena exists for MBASIC, we conducted Simulation Study 6, where we simulated data with two large clusters and six small clusters, and compared the performance of MBASIC and MBASIC0 to highlight the effect of including a singleton cluster. We found that compared to MBASIC0, MBASIC was significantly less aggressive in merging small clusters. Overall, it captured large clusters and  allocated the small cluster units as singletons (Figures \ref{fig:s6sim1} and \ref{fig:s6sim2}, Tables \ref{tbl:confusionmat1}, \ref{tbl:confusionmat2} and \ref{tbl:s6metric}). This study highlighted the utility of a singleton cluster as  a potential remedy for  merging of small clusters.

Combining  results from all of our simulation studies, we conclude that MBASIC is a powerful model for both state-space estimation and clustering structure recovery. Its adaptability to singletons, effectiveness in model selection, and robustness against within-state heterogeneity strongly support its applicability for real data sets.

\section{Applications of MBASIC to Genome Research Problems}\label{sec:realdata}

\subsection{Transcription Factor Enrichment Network}

Regulation of gene expression relies heavily on the context-specific combinatorial activities of TFs. Gene clustering analysis based on TF occupancy data, i.e., ChIP-seq, aims to identify combinatorial patterns of TF occupancy and group genes based on such patterns.
The ENCODE consortium \citep{ENCODE} has generated TF ChIP-seq datasets for over 100 TFs across multiple cell types, and has motivated several integrative studies for learning regulation patterns \citep{Gernstein12, Weng12}. 
In this study, we applied MBASIC to the analysis of such data. Specifically, we focused on the TF enrichment patterns at the promoter regions, i.e., -5000 bps and +1000 bps the transcription start site, of the 10290 genes that had significant expression, as measured by RNA-seq, in either the Gm12878 or the K562 cells. The input data to MBASIC were the mapped numbers of reads at these promoter regions from the uniformly processed ChIP-seq data by \cite{Gernstein12}. We chose the cell types  Gm12878 and K562 because they had the largest numbers of  TF ChIP-seq experiments. The final dataset utilized  included ChIP-seq data for $I=10290$ observational units over 30 TFs corresponding to $K=60$ experimental conditions (cell type $\times$ TF)
with a total of  166 replicate experiments.


We fitted MBASIC with $S=2$ states and used log normal distributions as in Equation (\ref{eq:lognormal}). $s=1$ corresponded to the unenriched state, and we let $\gamma_{ikl1}=\log(1+x_{ik})$, where $x_{ik}$ is the count from the matching control experiment at unit $i$. $s=2$ corresponded to the enrichment state, and we let $\gamma_{ikl2}=1$ for all loci. 

\begin{figure}[htbp]
\centering
\begin{tabular}{ll}
(a) & (b) \\
\includegraphics[width=0.45\textwidth,page=1]{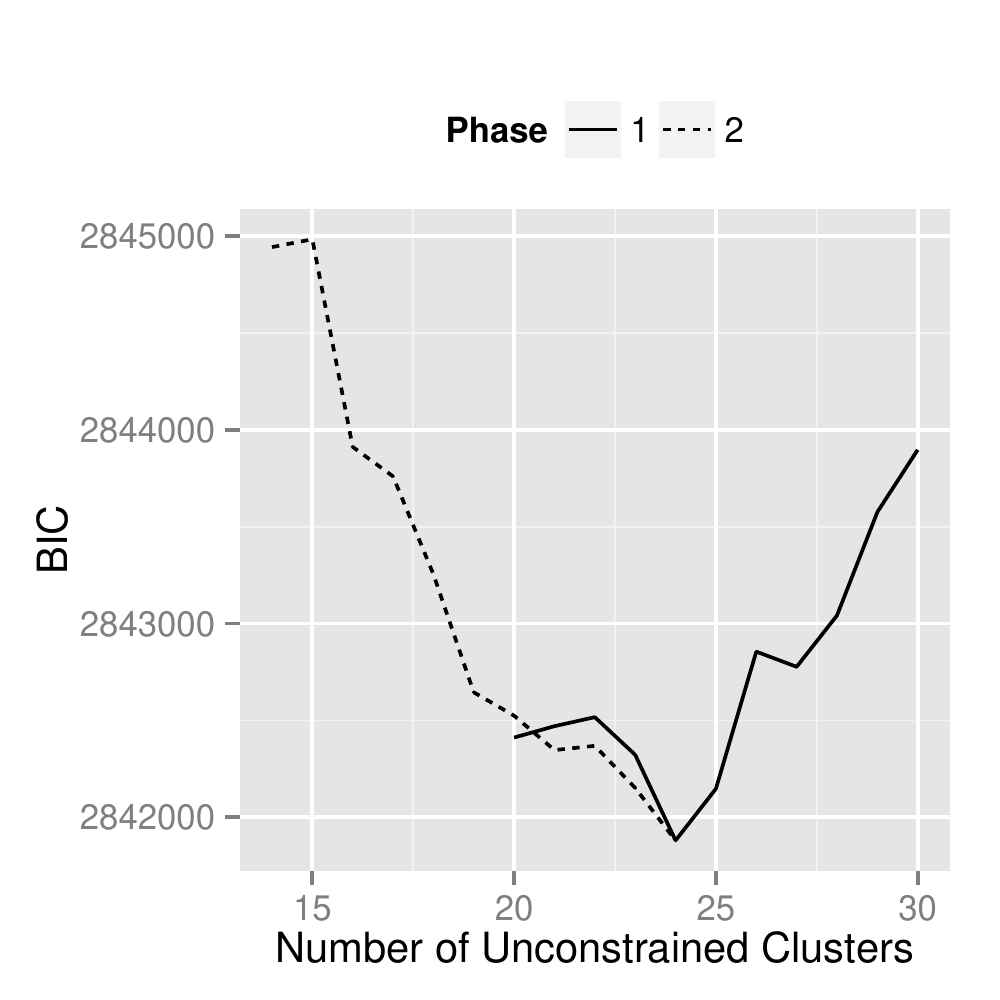} & 
\includegraphics[width=0.45\textwidth,page=2]{ModelSelection.pdf} \\
\end{tabular}
\caption{(a) BIC  and (b) log-likelihood values for models with different structures. All the clusters are unstructured in the Phase 1 models and the  x-axis denotes the total number of clusters. The total number of clusters is 24 for Phase 2 models and the x-axis denotes the number of unconstrained clusters. The remaining clusters have TF-homogeneity.}\label{fig:modelselection}
\end{figure}

\begin{figure}[htbp]
\centering
\begin{tabular}{cc}
(a) & (b) \\
\includegraphics[width=0.45\textwidth,height = 0.8\textheight,page=1]{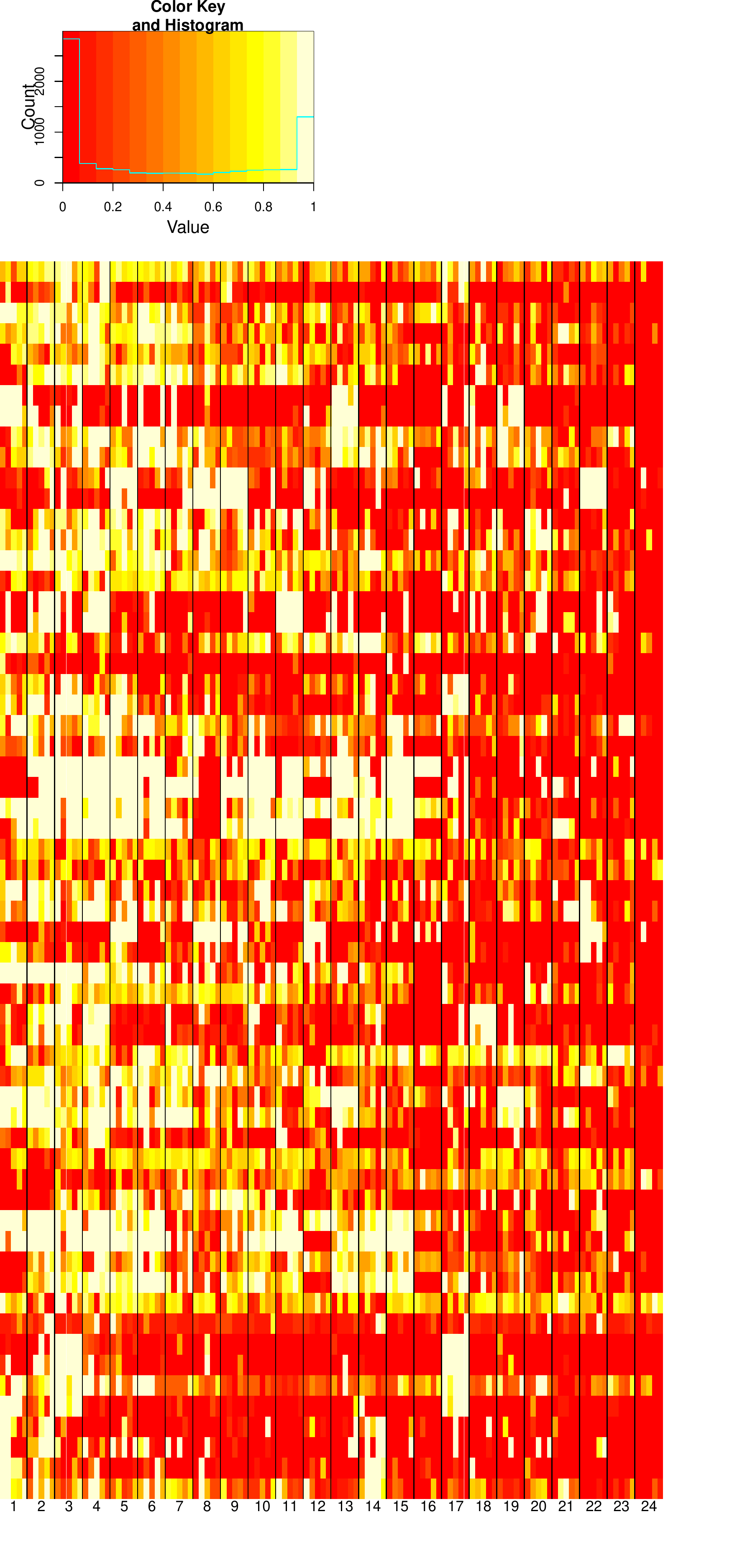}&
\includegraphics[width=0.52\textwidth,height = 0.8\textheight,page=2]{W_sample.pdf}\\
\end{tabular}
\caption{(a) Normalized data for each cell-TF combination at five sub-sampled loci within each cluster. (b) Estimated enrichment probability at each cell-TF combination for each cluster.}\label{fig:W_sample}
\end{figure}

We followed the two-phase procedure using BIC from Section \ref{sec:modelselect} to select both the number of clusters and the structure of each cluster. In Phase 1, we selected the number of clusters as 24. In Phase 2, we considered two types of structural constraints for each cluster, referred to by \textit{TF-homogeneity} and \textit{cell type-homogeneity} and defined as $w_{jk_1s}=w_{jk_2s}$ if $k_1$ and $k_2$ corresponded to the same TF or cell type. We found that imposing cell type-homogeneity to any cluster would cause that cluster to be degenerate (i.e., no unit was assigned to that cluster). Therefore, we  chose the final model among those with TF-homogeneity structures. The BIC and log likelihood values for different models fitted in both phases are shown in Figure \ref{fig:modelselection}. The final model  had 24 unconstrained clusters, consisting  of $1-\zeta=89.8\%$ of the 10290 loci.  The ranges of the estimated distribution parameters among replicates within the same cell type-TF combination is shown in Figure \ref{fig:par_gene}. We notice that these parameters can be substantially different among replicated experiments. This provides further support for our replicate specific parametrization.

To compare the normalized data and the predicted enrichment probability for each cluster, we computed the normalized signals \footnote{{The normalized signal for unit $i$ and condition $k$ is:\[
\tilde \theta_{ik}=\frac{\prod_{l=1}^{n_k}f_s(y_{ikl}|\hat \mu_{kl2},\hat\sigma_{kl2},\gamma_{ikl1})}{\prod_{l=1}^{n_k}f_s(y_{ikl}|\hat\mu_{kl2},\hat\sigma_{kl2}) + \prod_{l=1}^{n_k}f_s(y_{ikl}|\hat \mu_{kl1},\hat \sigma_{kl1},\gamma_{ikl2})}
\]}}
and compared them to the estimated cluster parameters. Figure \ref{fig:W_sample} depicts such normalized signals from five randomly selected loci within each predicted cluster (Figure~\ref{fig:W_sample}(a)), as well as the predicted enrichment probabilities at the corresponding condition and cluster ($w_{jk2}$'s) (Figure~\ref{fig:W_sample}(b)). We observe that the estimated enrichment probabilities at the cluster level  capture the commonality among loci within each cluster. In addition, each loci cluster exhibits distinct combinatorial patterns of activity across all cell type-TF combination. The  cell type-TF combination enriched within each cluster is listed in Table \ref{tbl:profile}. 

\begin{figure}
\centering
\begin{tabular}{ll}
(a) & (b) \\
\includegraphics[width=0.45\textwidth,page=1]{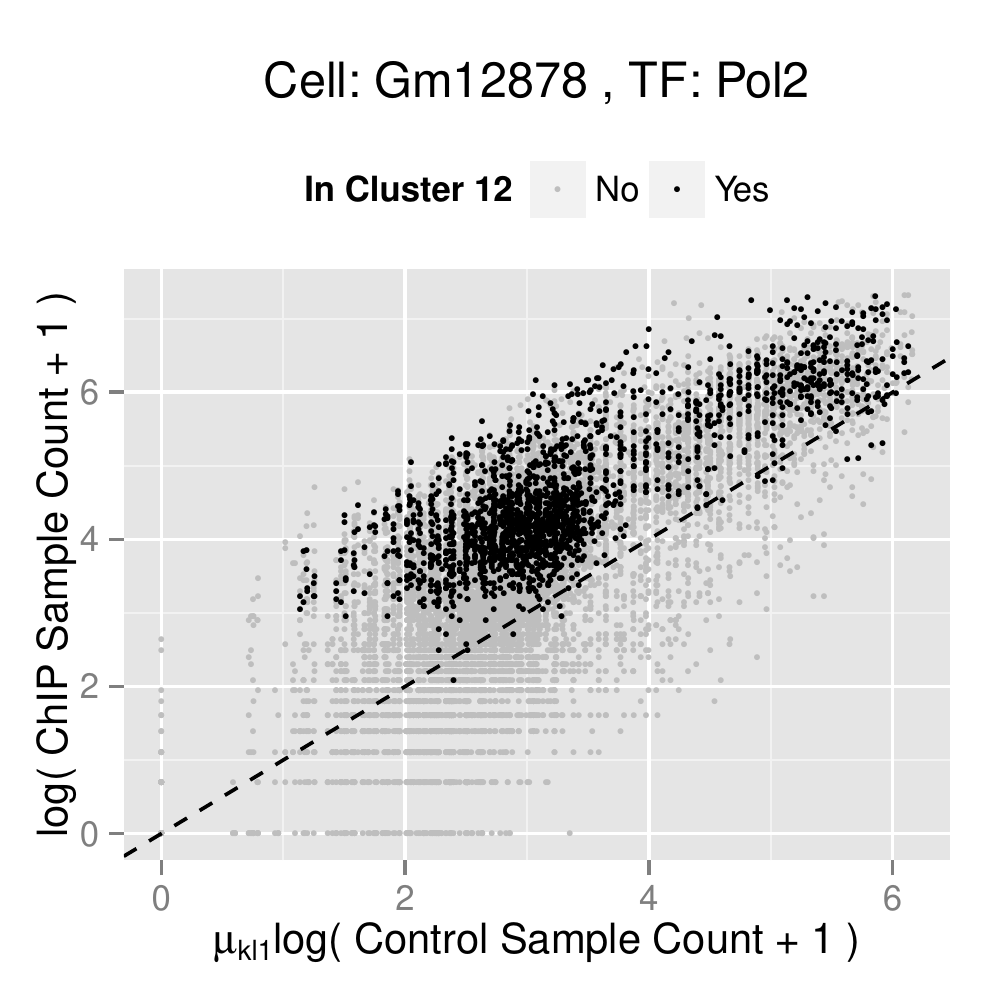}&
\includegraphics[width=0.45\textwidth,page=2]{data_compare_Gm12878_K562_Pol2_c12.pdf}\\
\end{tabular}
\caption{(a, b) Plots of the transformed Pol2 ChIP sample read counts against the transformed control sample read counts for all units in (a) Gm12878 and (b) K562 cells. Data from unenriched units are expected  to reside around  the 45 degree dashed line.}\label{fig:pol2}
\end{figure}

Our clustering results are consistent with the existing literature on the TF enrichment networks. For example, cooperating TFs  tend to be enriched at the same loci. This pattern can be observed in Figure \ref{fig:W_sample}(b) between  Bcl3 and Bclaf1. 
Pol2 and Pol24h8 represent  Pol2 experiments  with different antibodies. As expected, we observe enrichment at the same loci for these two different version of Pol2 experiments.
Moreover, pairs of TFs that have similar binding motifs have similar enrichment probabilities over the clusters. For example, \cite{Weng12} discovered the UA1 motif as common to both Chd2 and Ets1 and the USF motif for Max, Usf1, and Usf2. Interactions between Taf1 and Tbp have also been studied by \cite{Taf1}. Similar enrichment probabilities of these TFs across clusters can be observed in Figure \ref{fig:W_sample}(b). In addition to these observations that are consistent with the literature, our results illustrate how the genome-wide TF association patterns can be attributed to specific clusters. We explored the loci clusters with distinct patterns between cell types (e.g., Pol2 in Cluster 12, Figure \ref{fig:pol2}), TFs from the same families (e.g., Bcl3 v.s. Bclaf1 in Cluster 3, Figure \ref{fig:bcl}), and TFs with similar genome-wide enrichment (e.g., Max v.s. Usf1 in Cluster 2, Figure \ref{fig:maxusf1}) using raw data.  
We further evaluated each cluster of genes  for their  KEGG pathway enrichment \citep{enrichtest}, and identified 8 KEGG pathways that are significantly enriched in individual clusters (Table \ref{tbl:kegg}). Three of our clusters (Clusters 7, 9, and 19) have more than half  of their genes  in one single pathway. Since KEGG pathways curate the known knowledge of molecular interaction systems, these clusters may be driven by unknown  biological processes that warrant further investigation.

\begin{table}[hbt]
\centering
\caption{Significantly enriched KEGG pathways across the 24 clusters.}\label{tbl:kegg}
\begin{tabular}{cccccc}
\hline
KEGG.name & \# Genes & Z Score & Cluster & Cluster\\
& Overlapped & & & Size \\
\hline
Protein processing in endoplasmic reticulum & 156 & 5.652 & 7 & 391\\
Fatty acid elongation in mitochondria  & 7 & 7.518 & 8 & 133\\
B cell receptor signaling pathway  & 74 & 6.016 & 9 & 146\\
Lysine biosynthesis  & 3 & 6.53 & 9 & 146\\
D-Glutamine and D-glutamate metabolism  & 3 & 5.548 & 12 & 184\\
Vitamin B6 metabolism  & 4 & 5.28 & 14 & 156\\
Non-homologous end-joining  & 12 & 7.539 & 17 & 213\\
Lysosome &  116 & 5.402 & 19 & 187\\
\hline
\end{tabular}
\end{table}



\begin{figure}[htbp]
\centering
\begin{tabular}{cc}
(a) & (b) \\
\multicolumn{2}{c}{ \includegraphics[width=\textwidth,page=1]{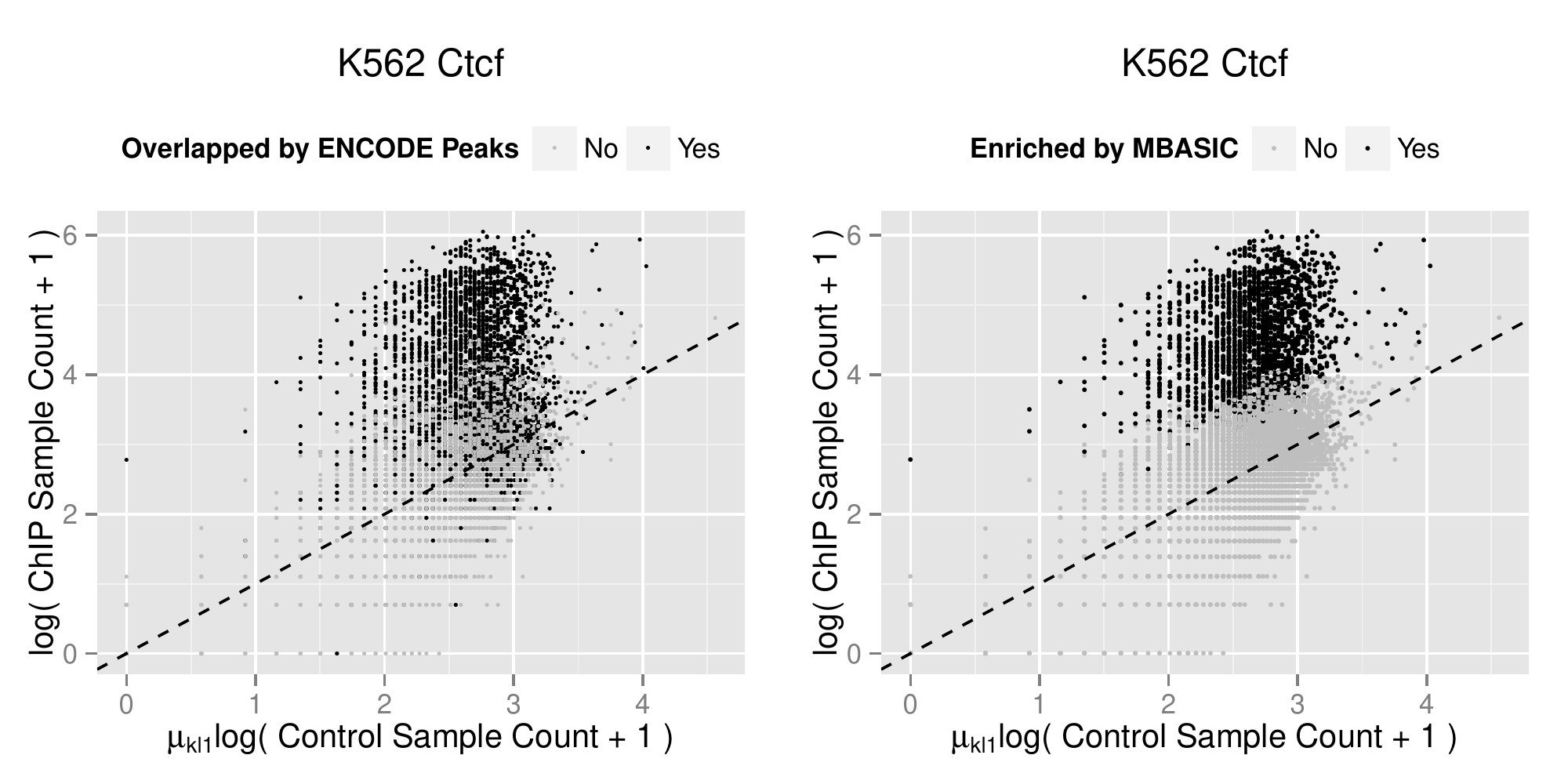} }\\
\end{tabular}
\caption{ (a, b) Transformed ChIP versus control sample read counts from a Gm12878-Ctcf dataset. 
Enrichment states are annotated by (a) ENCODE peak profiles  and (b) MBASIC estimation. In MBASIC, an observational unit is estimated to be enriched if its enrichment probability satisfies $P(\theta_{ik}=2|Y)>0.5$. }\label{fig:TF_gene}
\end{figure}

MBASIC infers the clustering structure based on its own estimates of the state-space profiles. The ENCODE consortium provides the estimated enrichment regions (i.e., \textit{peaks}) for each experiment in this study. 
Then, a natural question is whether MBASIC reveals more information compared to clustering of genes based on ENCODE-estimated binary enrichment profiles of TFs. To address this, we created a binary vector for each gene by overlapping its promoter with the ENCODE peaks. Then, we applied the state-of-the-art MClust model \citep{mclust} to cluster the 10290 promoter regions based on these peak profiles. MClust selected 90 clusters based on BIC. Figure \ref{fig:W_mclust} displays cluster-level estimated enrichment probabilities of TFs across the conditions considered. Compared to Figure \ref{fig:W_sample}, we can see that many of the MClust clusters have very similar enrichment profiles. For example, Clusters 51, 7, 8, 32, 54 contained almost no enrichment for any TFs, but are classified as distinct clusters. The association between units across these clusters are thus non-trivial to interpret. In addition, we found that for some conditions, the enrichment states predicted by MBASIC are quite different than those from the ENCODE peak profiles (e.g., Figure \ref{fig:TF_gene}). This is because the ENCODE peaks are identified by whole genome-wide analysis and may not reflect the differences between the ChIP and control samples at the local promoter regions. MBASIC attains larger raw data fidelity 
 by directly modeling the counts at each unit rather than inheriting results from existing analyses.

\subsection{Genome-wide Identification of +9.5-like Composite Elements}

\cite{Johnsonetal12} and \cite{Gaoetal2013}  described the requirement of 
the intronic +9.5 site, an Ebox-GATA composite element located at  chr6: 88143884-88157023 in the mouse genome (genome version mm9),  to establish the hematopoietic stem/progenitor cell (HSC) compartment in the fetal liver and for hematopoietic stem cell genesis in the aorta-gonad-mesonephros (AGM), respectively.
Furthermore, \cite{Johnsonetal12} and \cite{Hsu15032013}  showed that heterozygous +9.5 mutations cause a human immunodeficiency associated with myelodysplastic syndrome (MDS) and acute myeloid leukemia (AML). Because the +9.5 site is the only known \textit{cis}-element  deletion of which  
depletes fetal liver HSCs and is lethal at E13-14 of embryogenesis,
 identifying additional loci  that have similar functionality is extremely important  for
establishing mechanisms that enable GATA factor-bound regions with nonredundant activity 
 and have the potential to reveal
novel targets for therapeutic modulation of hematopoiesis. In this application, we identified 4803 genomic regions with the Ebox-GATA motif (CATCTG-N[7-9]-AGATAA where N[7-9] denotes a variable size spacer of  7 and 9 nucleotides) in the human genome (genome version hg19).  We considered a 150 bps window anchored at each of the 4803 composite elements as the observational unit. 
To analyze the TF occupancy activities at these units and identify a group of composite elements with occupancy profiles similar to that of the +9.5 composite element, we downloaded all ChIP-seq data for the Huvec and K562 cells from \cite{Gernstein12}. In total, the data set contained 224 replicates spanning $K=84$ experimental conditions and 77 TFs.

We used negative binomial distributions  with $S=2$ states, where $s=1$ denoted  the unenriched (unoccupied) state,   in the MBASIC framework.  We chose $\gamma_{ikl1}=1+x_{ik}$, where $x_{ik}$ is the count for unit $i$ from the matching control experiment for condition $k$, to incorporate data from the accompanying control experiments of the ChIP samples. For $s=2$, we utilized the following mixture distribution to account for the heavy tails observed in the raw data:

\[
Y_{ikl}-3 | \theta_{ik}=2 {\buildrel\rm ind.\over\sim} \nu_{ikl}NB(\mu_{kl2},\sigma_{kl2})+(1-\nu_{ikl})NB(\mu_{kl3},\sigma_{kl3}),
\]
\[
\nu_{ik} {\buildrel\rm i.i.d.\over\sim} Bernoulli( v_{kl} ).
\]

Here, the constant 3 represents the minimum count threshold for enrichment estimation. The use of mixture distributions to capture heavy tailed count data was previously considered by \cite{cssp}. We note that an alternative approach to capture heavy tailed counts would be to fit a model using $S=3$ states, with $s=2,3$ representing two distinct enrichment components. Such an approach would differ from the proposed approach in a subtle yet important way. In this alternative approach, allocation of each unit to different enrichment components would affect the clustering estimation, while in our approach, clustering is only determined by the enrichment status of the individual unit regardless of which enrichment component it follows. The E-M algorithm for this setting requires a slight modification as discussed in Section \ref{sec:multicomp}.

\begin{figure}
\centering
\includegraphics[width=0.8\textwidth,height=\textheight,page=1]{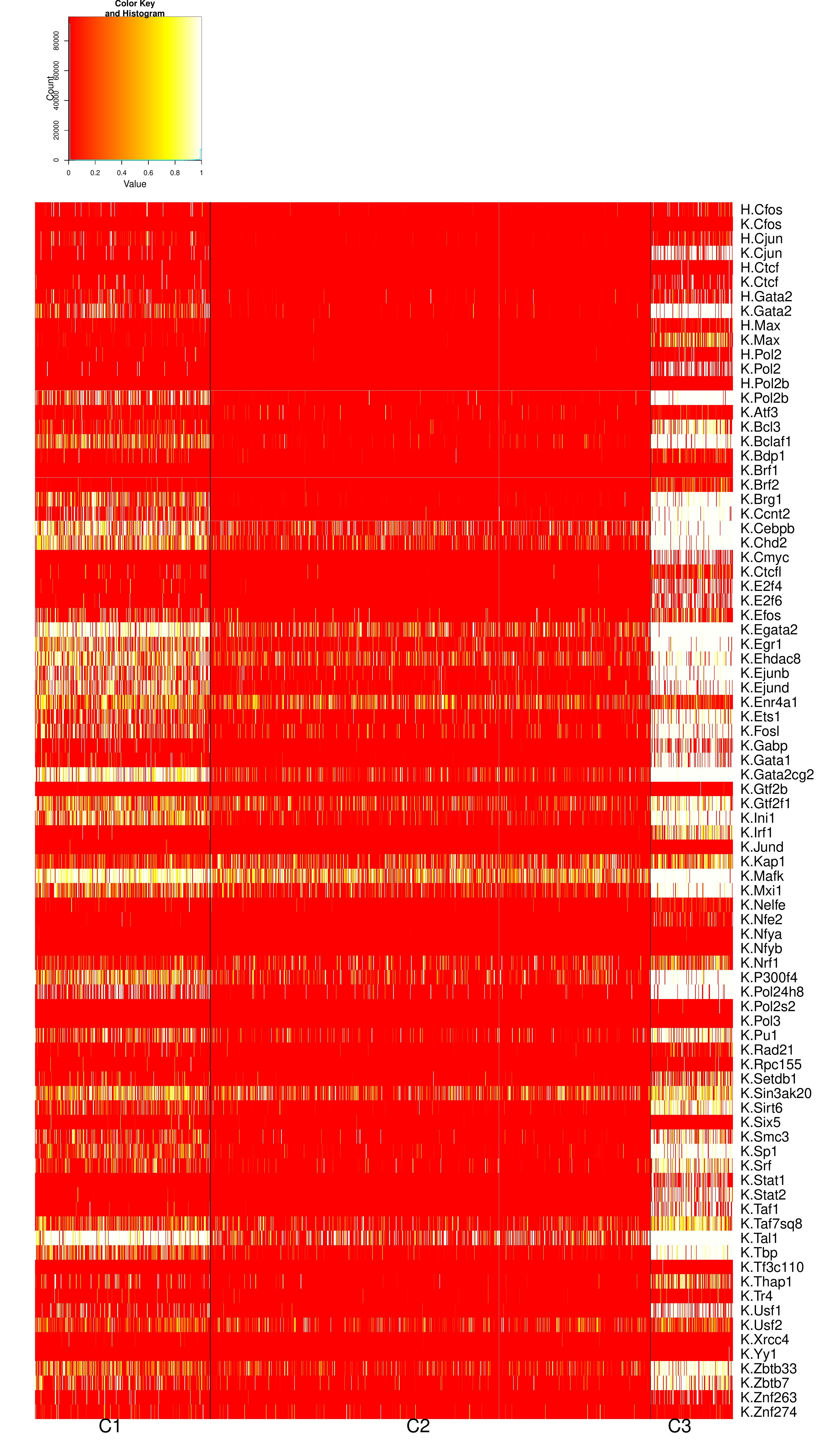}
\caption{Posterior enrichment probability (i.e,. $P(\theta_{ik}=2|Y)$) for all units in the three clusters. The right most column of the C3 cluster corresponds to the +9.5 element.}\label{fig:bind_cluster}
\end{figure}

Following the two phase model selection procedure using BIC, we selected the model with 3 clusters, 2 of which were cell type-homogeneous.  The ranges of the estimated distribution parameters among replicates within the same condition are displayed in Figures \ref{fig:par_ebox_mu} and \ref{fig:par_ebox_sigma}. The three clusters (denoted by C1, C2, and C3) included 332, 837, 157 composite elements, respectively, and the remaining 3477 composite elements were identified as singletons. A heatmap for the enrichment probability of each unit under each cell type-TF combination across the three clusters is shown in Figure \ref{fig:bind_cluster}. The +9.5 element is a member of cluster C3 which consists of a total of 157 +9.5-like composite elements. 
A detailed genomic annotation of these  elements are provided in Table \ref{tbl:pa_result}. Notably, 46\%  of the C3 elements reside in intronic regions and 42\% of these are within first intron. Only 15\% of the cluster are located up to 10Kb upstream of  transcription start sites. 

A detailed analysis of Figure \ref{fig:bind_cluster} reveals that cluster C3 is driven by several transcription factors with known associations to GATA2. First,
we note that a large fraction of the C3 loci are bound by BRG1. The chromatin remodeler BRG1 is involved in GATA1-mediated chromatin looping \citep{Kimetal2009NAR, Kimetal2009} 
and co-localizes with GATA1 at some chromatin sites \citep{Hu01102011}. 
BRG1 has broad functions in many cell types; however, conditional knockouts of BRG1 reveal its importance in specific cell and tissue contexts \citep{JCP:JCP24414}. Another factor that clearly stands out as having a GATA2-like profile in cluster C3 is ETS1. Our prior work identified  the propensity of occupied GATA motifs to reside near Ets motifs \citep{Linneman11} and \cite{Crispino2012} has highlighted GATA2-ETS co-localization.



\begin{figure}[htbp]
\centering
\includegraphics[width=0.45\textwidth,page=2]{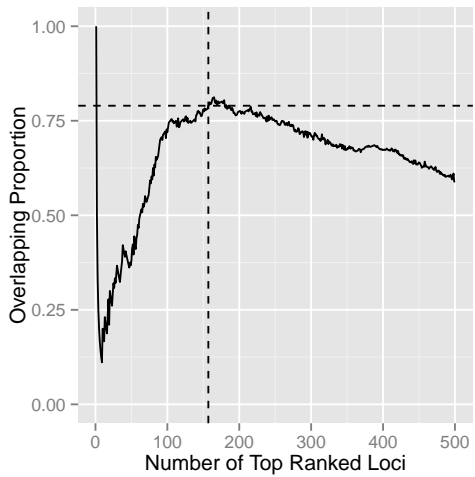}
\caption{Proportion of overlap between the top ranked +9.5-like composite elements identified by MBASIC and ENCODE peak profiles. The overlap proportion is calculated by considering the  same number of top ranked units (x-axis)  in both the ENCODE-based and MBASIC-based similarities to the +9.5 site. The dashed lines mark that 78.3\% of the C3 units are ranked in the top 157 based on the ENCODE peak profiles.  }\label{fig:rank_overlap}
\end{figure}

We next performed an alternative naive analysis by utilizing the list of peaks provided by the ENCODE project. As in the case of the Transcription Factor Enrichment Network example of Section 4.1, these peaks, provided by the ENCODE consortium, were identified by analyzing each dataset individually with ENCODE's uniform ChIP-seq processing pipeline. Figure \ref{fig:encode_ebox} displays the ENCODE peak profiles for our cell type-TF conditions. 
For each of the 4803 composite elements, we constructed a \textit{peak profile}, which is a binary vector indicating whether the element  overlaps with the ENCODE peaks for each cell type-TF combination.
We then computed the peak profile based similarity between the +9.5 site and each the of the composite elements using the \texttt{R} function \texttt{dist.binary} with the "Jaccard index" option. For comparison, we computed \textit{pseudo-binary similarities} between each element  and the +9.5 site using the MBASIC estimated enrichment probabilities across all conditions\footnote{The pseudo-binary similarity between two units $i_1$ and $i_2$ is calculated as $s(i_1,i_2)=\frac{\sum_kP\{\theta_{i_1k}=1|Y\}P\{\theta_{i_2k}=1|Y\}}{\sum_k P\{\theta_{i_1k}=1|Y\}+P\{\theta_{i_2k}=1|Y\}-P\{\theta_{i_1k}=1|Y\}P\{\theta_{i_2k}=1|Y\}}$.}.
We then ranked the composite elements based on both ENCODE  and MBASIC estimated similarities. 
Figure \ref{fig:rank_overlap} provides a comparison of the two lists as a function of top ranking composite elements. 
Overall, we observe that the rankings based on MBASIC estimation are consistent with the rankings based on the ENCODE peak profiles. 

\begin{figure}[htbp]
\centering
\begin{tabular}{cc}
\multicolumn{2}{c}{(a)}\\
\multicolumn{2}{c}{
\includegraphics[width=\textwidth,page=3]{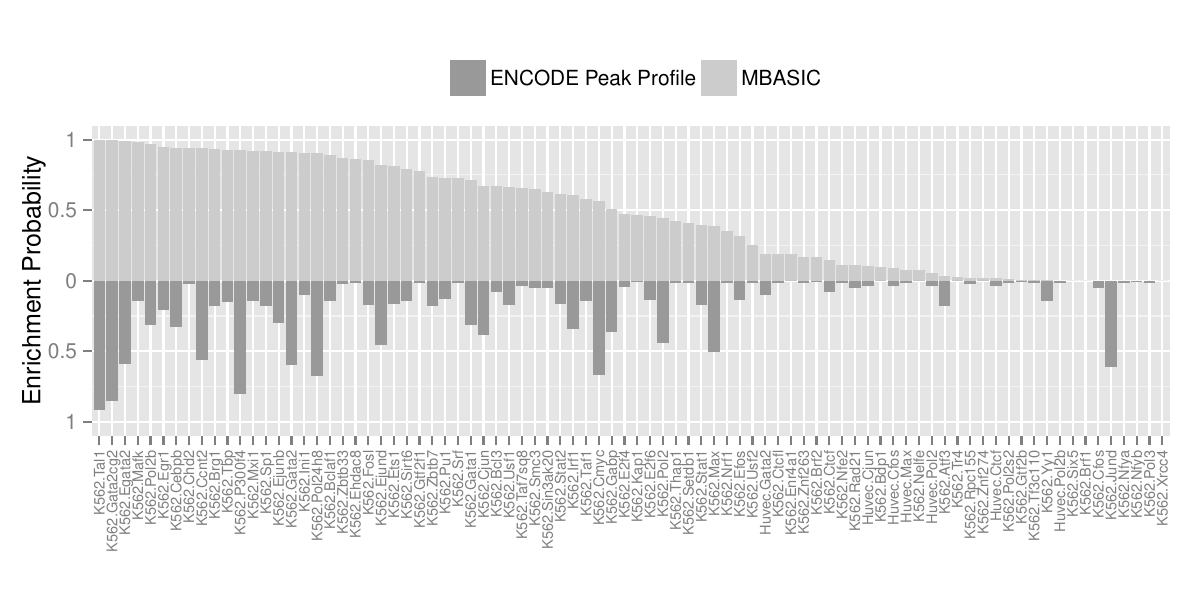}
}\\
(b) & (c) \\
\multicolumn{2}{c}{\includegraphics[width=\textwidth,page=5]{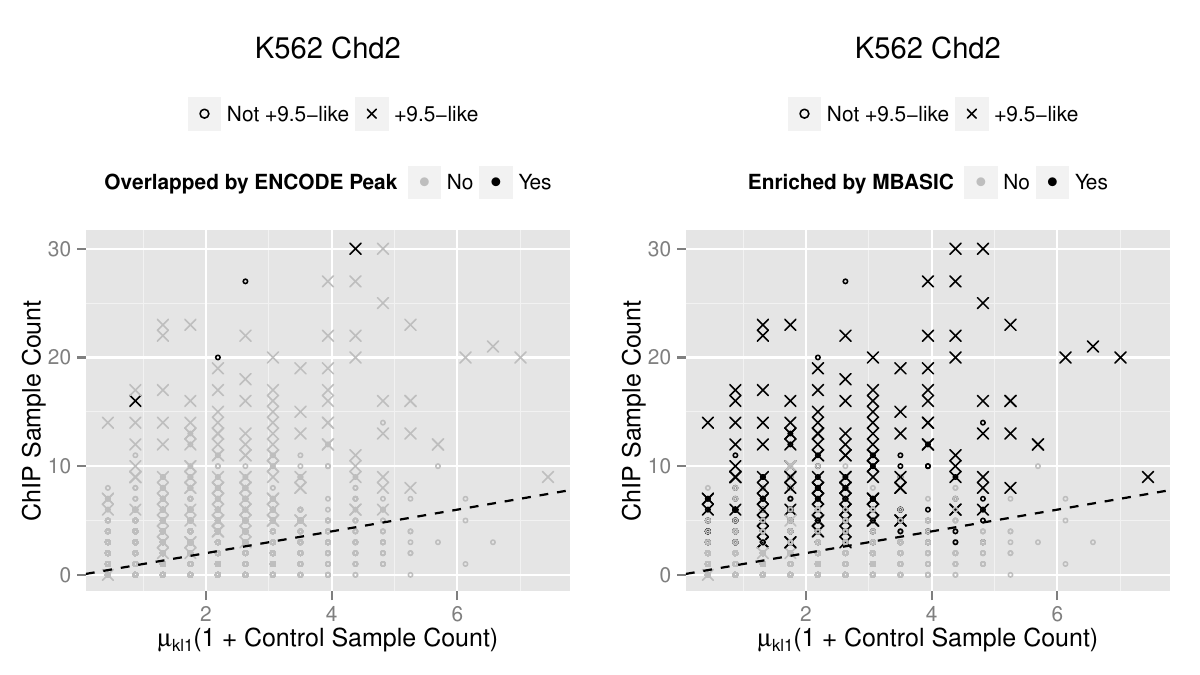}}
\end{tabular}
\caption{(a) Top half: Enrichment probabilities for the C3 units across all experimental conditions estimated by MBASIC. Bottom half: Proportion of C3 units that are overlapped by the ENCODE peaks for each condition. (b, c) ChIP sample read counts against normalized control sample read counts for one replicate of K562-Chd2 dataset. Enrichment status are annotated by (a) the ENCODE peak profiles and (c) MBASIC prediction.}\label{fig:peak_compare}
\end{figure}

Although the rankings of the composite elements with respect to their +9.5 similarity using both the ENCODE peak profiles and MBASIC estimation were quite similar, the two approaches resulted in different enrichment estimation at the individual TF-cell combination level. Figure \ref{fig:peak_compare}(a)  compares the estimated cluster-level enrichment probabilities of each cell type-TF combination for cluster C3  against their average ENCODE peak profiles and highlights the difference between the two procedures. To further investigate these differences, we plotted the raw data for individual replicates and compared the composite elements that were estimated to be enriched by the two methods. An example using data from K562-Chd2 is displayed in Figures \ref{fig:peak_compare}(b) and (c). 
Although many elements have significantly higher counts in the ChIP sample  compared to the control sample, they are not identified as occupied by Chd2 in K562 according to ENCODE peak annotation. Another example using a replicate from K562-Yy1 is shown in Figure \ref{fig:yy1}, where several elements with zero ChIP count are overlapped by ENCODE peaks. These results indicate that MBASIC provides a grouping of the Ebox-GATA composite elements that is more consistent with the raw data compared to grouping based on ENCODE peak annotation.


\section{Conclusions and Discussion}

Clustering analysis based on an underlying state-space is a common problem for many genomic and epigenomic studies where multiple data sets over many observational units are integrated. In this paper, we developed a unified statistical framework, called MBASIC, for addressing these class of problems. MBASIC simultaneously projects the observations onto a hidden state-space and infers clustered units in this space. The hierarchical structure of MBASIC enables the information of the state-space clusters to be fed back into the projection of the raw data, thus reinforces the accuracy of predicting the state-space states of individual units.
The MBASIC framework offers flexibility in a number of aspects of experimental design, such as different numbers of replicates under individual experimental conditions and missing values. Additionally, it is applicable to many parametric distributions. Our computational studies highlighted good operating characteristics of MBASIC and the two genomic applications illustrated how large numbers of ChIP-seq datasets can be integrated for addressing specific problems. In both of the applications, MBASIC algorithm converged within 20 minutes for a fixed model on a 64 bit machine
with Intel Xeon 3.0GHz processor and 64GB of RAM. For model selection,  we  utilized \texttt{R}  package \texttt{snow} to implement  the 2-phase procedure with parallel fitting of different candidate models  using a  8-core  64 bit, 64GB  RAM machine
with 8 Intel Xeon 3.0GHz processors. These runs were completed under 2 hours.  
The computational efficiency of our model depends on the simple, closed-form updates in our E-M algorithm. Such a mathematical form is due, at least in part, to our modeling assumption that the rows of our state-space matrix is clustered. We have argued that this assumption, as compared to the PCA-type model structures, offers easier interpretation and is well suited for many genomic applications. 
MBASIC is available as \texttt{R} package \texttt{mbasic} at https://github.com/chandlerzuo/mbasic.
\newpage

\appendix

\renewcommand\thefigure{\thesection.\arabic{figure}}
\setcounter{figure}{0}
\renewcommand\thetable{\thesection.\arabic{table}}
\setcounter{table}{0}

\section{Details of the Expectation-Maximization (EM) Algorithms}\label{sec:appalg}

\subsection{Derivation for the E-step}\label{sec:em}

We derive the expressions for the E-step updates of our algorithm in Eqns. (\ref{eq:estep:b}), (\ref{eq:estep:zb}), (\ref{eq:estep:thetazb}), (\ref{eq:estep:thetab}) as well as the marginal likelihood in Eqns. (\ref{eq:marginlik}) and (\ref{eq:thetalik}).

In what follows, we let $\theta_{iks}$  denote $1\{\theta_{ik}=s\}$. The joint density of $(z,b,\theta,Y)$ is given by:
\begin{equation}\label{eq:full}
f(z, b,\theta, y) =  \prod_{i=1}^I\zeta^{b_i}(1-\zeta)^{1-b_i}\cdot \prod_{i=1}^I\prod_{j=1}^J\pi_j^{z_{ij}}\cdot \prod_{i=1}^I\prod_{k=1}^K\prod_{s=1}^S\left [ f_{iks}(\prod_{j=1}^Jw_{jks}^{z_{ij}})^{1-b_{i}}p_{is}^{b_i} \right ]^{\theta_{iks}},
\end{equation}
where  $f_{iks}=\prod_{l=1}^{n_k}f(y_{ikl}|\mu_{kls},\sigma_{kls},\gamma_{ikls})$. 
The following elementary equality is used repeatedly throughout the rest of the derivations in this section. 

\[
\sum_{\sum_j a_{ij}=1, a_{ij}\in\{0,1\}}\prod_i\prod_j b_{ij}^{a_{ij}} = \prod_i \left (\sum_jb_{ij} \right ).
\]

The joint density of $(z, b, Y)$ can be calculated from Eqn. (\ref{eq:full}):
\begin{equation}\label{eq:Zb}
\begin{aligned}
f(z,b,y) = &\sum_{\sum_s\theta_{iks} = 1}f(z,b,\theta,y) \\
= &  \prod_{i=1}^I\zeta^{b_i}(1-\zeta)^{1-b_i}\cdot \prod_{i=1}^I\prod_{j=1}^J\pi_j^{z_{ij}}\cdot \sum_{\sum_s\theta_{iks} = 1} \prod_{i=1}^I\prod_{k=1}^K\prod_{s=1}^S \left [f_{iks} \left (\prod_{j=1}^Jw_{jks}^{z_{ij}} \right )^{1-b_{i}}p_{is}^{b_i} \right ]^{\theta_{iks}}\\
= & \prod_{i=1}^I\zeta^{b_i}(1-\zeta)^{1-b_i}\cdot \prod_{i=1}^I\prod_{j=1}^J\pi_j^{z_{ij}}\cdot\prod_{i=1}^I\prod_{k=1}^K \left [\sum_{s=1}^Sf_{iks} \left (\prod_{j=1}^Jw_{jks}^{z_{ij}} \right )^{1-b_{i}}p_{is}^{b_i} \right ].\\
\end{aligned}
\end{equation}
Since 
\[
\sum_{s=1}^Sf_{iks}\left (\prod_{j=1}^Jw_{jks}^{z_{ij}} \right )^{1-b_{i}}p_{is}^{b_i}=\prod_{j=1}^J \left [\sum_{s=1}^Sf_{iks}w_{jks}^{1-b_{i}}p_{is}^{b_i} \right ]^{z_{ij}},
\]
Eqn. (\ref{eq:Zb}) can be rewritten as:
\begin{equation}\label{eq:Zb1}
f(z,b,y) = \prod_{i=1}^I\zeta^{b_i}(1-\zeta)^{1-b_i}\cdot \prod_{i=1}^I\prod_{j=1}^J\left [\pi_j\prod_{k=1}^K \left (\sum_{s=1}^Sf_{iks}w_{jks}^{1-b_{i}}p_{is}^{b_i} \right ) \right ]^{z_{ij}}.
\end{equation}
The joint distribution of $(b,Y)$ can be calculated from Eqn. (\ref{eq:Zb1}):
\begin{equation}\label{eq:b}
\begin{aligned}
f(b,y) = &\sum_{\sum_jz_{ij} = 1}f(z,b,y) \\
= &  \prod_{i=1}^I\zeta^{b_i}(1-\zeta)^{1-b_i}\cdot \sum_{\sum_jz_{ij} = 1}\prod_{i=1}^I\prod_{j=1}^J \left [\pi_j\prod_{k=1}^K \left (\sum_{s=1}^Sf_{iks}w_{jks}^{1-b_{i}}p_{is}^{b_i} \right ) \right ]^{z_{ij}}\\
= & \prod_{i=1}^I\zeta^{b_i}(1-\zeta)^{1-b_i}\cdot\prod_{i=1}^I \left [\sum_{j=1}^J\pi_j\prod_{k=1}^K \left (\sum_{s=1}^Sf_{iks}w_{jks}^{1-b_{i}}p_{is}^{b_i} \right ) \right ].
\end{aligned}
\end{equation}
We note that
\[
\sum_{j=1}^J\pi_j\prod_{k=1}^K \left (\sum_{s=1}^Sf_{iks}w_{jks}^{1-b_{i}}p_{is}^{b_i} \right )= \left [\sum_{j=1}^J\pi_j\prod_{k=1}^K \left(\sum_{s=1}^Sf_{iks}w_{jks} \right ) \right ]^{1-b_i} \left [\prod_{k=1}^K \left (\sum_{s=1}^Sf_{iks}p_{is} \right ) \right ]^{b_i}.
\]
Then, Eqn. (\ref{eq:b}) can be rewritten as:
\begin{equation}\label{eq:b1}
f(b,y) = \prod_{i=1}^I\left [(1-\zeta)\sum_{j=1}^J\pi_j\prod_{k=1}^K \left (\sum_{s=1}^Sf_{iks}w_{jks}\right ) \right ]^{1-b_i} \left [\zeta \prod_{k=1}^K \left (\sum_{s=1}^Sf_{iks}p_{is} \right ) \right ]^{b_i}.
\end{equation}
We can calculate the marginal density of $Y$, given in Eqn. (\ref{eq:marginlik}),  from Eqn. (\ref{eq:b1}) as:
\begin{equation}\label{eq:Y}
f(y) = \sum_{b_i\in\{0,1\}}f(b,y)=\prod_{i=1}^I \left [(1-\zeta)\sum_{j=1}^J\pi_j\prod_{k=1}^K \left (\sum_{s=1}^Sf_{iks}w_{jks} \right )+\zeta \prod_{k=1}^K \left (\sum_{s=1}^Sf_{iks}p_{is} \right ) \right ].
\end{equation}
Eqn.  (\ref{eq:thetalik}) can be obtained similarly. Moreover, we can  rewrite (\ref{eq:Zb}) as
\begin{equation}\label{eq:Zb2}
f(z,b,y)=\prod_{i=1}^I\left [\zeta\prod_{k=1}^K\sum_{s=1}^Sf_{iks}p_{is} \right ]^{b_i}\left [(1-\zeta)\prod_{j=1}^J \prod_{k=1}^K \left (\sum_{s=1}^Sf_{iks}w_{jks} \right )^{z_{ij}} \right]^{1-b_i}\cdot \prod_{i=1}^I\prod_{j=1}^J\pi_j^{z_{ij}}
\end{equation}
by using
\[
\sum_{s=1}^Sf_{iks}\left (\prod_{j=1}^Jw_{jks}^{z_{ij}}\right )^{1-b_{i}}p_{is}^{b_i}= \left[\prod_{j=1}^J\left (\sum_{s=1}^Sf_{iks}w_{jks} \right )^{z_{ij}}\right ]^{1-b_i}\left [\sum_{s=1}^Sf_{iks}p_{is} \right ]^{b_i}.
\]
Thus, the density of $(z,Y)$ can be calculated as:
\begin{equation}\label{eq:Z}
\begin{aligned}
f(z,y)=&\sum_{b_i\in\{0,1\}}\prod_{i=1}^I \left [\zeta\prod_{k=1}^K \left (\sum_{s=1}^Sf_{iks}p_{is} \right )\right ]^{b_i}\left[(1-\zeta)\prod_{j=1}^J\prod_{k=1}^K \left (\sum_{s=1}^Sf_{iks}w_{jks} \right )^{z_{ij}} \right ]^{1-b_i}\cdot \prod_{i=1}^I\prod_{j=1}^J\pi_j^{z_{ij}}\\
=&\prod_{i=1}^I \left [\zeta\prod_{k=1}^K \left (\sum_{s=1}^Sf_{iks}p_{is}\right )+(1-\zeta)\prod_{j=1}^J\prod_{k=1}^K \left (\sum_{s=1}^Sf_{iks}w_{jks} \right )^{z_{ij}} \right ]\cdot \prod_{i=1}^I\prod_{j=1}^J\pi_j^{z_{ij}}\\
=&\prod_{i=1}^I\prod_{j=1}^J\left [\pi_j\zeta\prod_{k=1}^K\left (\sum_{s=1}^Sf_{iks}p_{is} \right )+\pi_j(1-\zeta)\prod_{k=1}^K\left (\sum_{s=1}^Sf_{iks}w_{jks} \right ) \right ]^{z_{ij}}.
\end{aligned}
\end{equation}
Using Eqns. (\ref{eq:b1}) and (\ref{eq:Y}), we obtain Eqn. (\ref{eq:estep:b}) as
\begin{equation}\label{eq:b|y}
E[b_i|Y]=\frac{\zeta \prod_{k=1}^K \left (\sum_{s=1}^Sf_{iks}p_{is}\right )}{(1-\zeta)\sum_{j=1}^J\pi_j\prod_{k=1}^K\left (\sum_{s=1}^Sf_{iks}w_{jks} \right )+\zeta \prod_{k=1}^K\left (\sum_{s=1}^Sf_{iks}p_{is} \right)}.
\end{equation}
Similarly, using  Eqns. (\ref{eq:Z}) and (\ref{eq:Y}), we have
\begin{equation}\label{eq:z|y}
E[z_{ij}|Y]=\frac{\pi_j\zeta\prod_{k=1}^K \left (\sum_{s=1}^Sf_{iks}p_{is} \right )+(1-\zeta)\pi_j\prod_{k=1}^K\left (\sum_{s=1}^Sf_{iks}w_{jks} \right )}{\zeta\prod_{k=1}^K\sum_{s=1}^Sf_{iks}p_{is}+(1-\zeta)\sum_{j=1}^J\pi_j\prod_{k=1}^K\left (\sum_{s=1}^Sf_{iks}w_{jks} \right)}.
\end{equation}
Using Eqns.  (\ref{eq:Zb1}) and (\ref{eq:b}), we have
\begin{equation}\label{eq:z|b}
E[z_{ij}|b,Y]=\frac{\pi_j\prod_{k=1}^K \left (\sum_{s=1}^Sf_{iks}w_{jks}^{1-b_{i}}p_{is}^{b_i} \right )}{\sum_{j=1}^J\pi_j\prod_{k=1}^K\left (\sum_{s=1}^Sf_{iks}w_{jks}^{1-b_{i}}p_{is}^{b_i} \right )}.
\end{equation}
Eqns. (\ref{eq:z|b}) and (\ref{eq:b|y}) together results in Eqn. (\ref{eq:estep:zb}). 
Using Eqns. (\ref{eq:full}) and (\ref{eq:Zb}), we have
\begin{equation}\label{eq:theta|zb}
E[\theta_{iks}|z,b,Y]=\frac{f_{iks}\left (\prod_{j=1}^Jw_{jks}^{z_{ij}} \right )^{1-b_{i}}p_{is}^{b_i}}{\sum_{s=1}^Sf_{iks}\left (\prod_{j=1}^Jw_{jks}^{z_{ij}} \right )^{1-b_{i}}p_{is}^{b_i}}.
\end{equation}
Therefore, we obtain Eqn. (\ref{eq:estep:thetazb}) by using Eqns. (\ref{eq:b}), (\ref{eq:z|b}), and (\ref{eq:theta|zb}):
\begin{equation}\label{eq:thetazb}
\begin{aligned}
E[\theta_{iks}z_{ij}(1-b_i)|Y]=&[1-E(b_i|Y)]E(z_{ij}|b_i=0,Y)E(\theta_{iks}|b_i=0,z_{ij}=1,Y)\\
=&\frac{(1-\zeta)\sum_{j=1}^J\pi_j\prod_{k=1}^K\left (\sum_{s=1}^Sf_{iks}w_{jks}\right )}{(1-\zeta)\sum_{j=1}^J\pi_j\prod_{k=1}^K \left (\sum_{s=1}^Sf_{iks}w_{jks} \right )+\zeta \prod_{k=1}^K \left (\sum_{s=1}^Sf_{iks}p_{is} \right )} \cdot \\
& \frac{\pi_j\prod_{k=1}^K \left (\sum_{s=1}^Sf_{iks}w_{jks} \right )}{\sum_{j=1}^J\pi_j\prod_{k=1}^K \left (\sum_{s=1}^Sf_{iks}w_{jks} \right )} \cdot \frac{f_{iks}w_{jks}}{\sum_{s=1}^Sf_{iks}w_{jks}}.\\
\end{aligned}
\end{equation}
Finally, we obtain Eqn. (\ref{eq:estep:thetab}) using Eqns. (\ref{eq:theta|zb}) and (\ref{eq:b}):
\begin{equation}\label{eq:thetab}
\begin{aligned}
E(\theta_{iks}b_i|Y)=&E(b_i|Y)E(\theta_{iks}|b_i=1,Y)\\
=&\frac{\zeta \prod_{k=1}^K\left (\sum_{s=1}^Sf_{iks}p_{is} \right )}{(1-\zeta)\sum_{j=1}^J\pi_j\prod_{k=1}^K\left (\sum_{s=1}^Sf_{iks}w_{jks} \right )+\zeta \prod_{k=1}^K \left (\sum_{s=1}^Sf_{iks}p_{is} \right )}
\cdot \frac{f_{iks}p_{is}}{\sum_{s=1}^Sf_{iks}p_{is}}.\\
\end{aligned}
\end{equation}

\subsection{EM Algorithm with Mixture Data Distributions}\label{sec:multicomp}

An important extension of the MBASIC model is to allow multiple mixture components within each state. For example, our model in Section 4.2 models the data from state $s=2$ as a mixture of two negative binomial distributions following the 
well motivated model of \cite{Kuan11}: 

\[
Y_{ikl}-3 | \theta_{ik}=2 \sim \nu_{ikl}NB(\mu_{kl2},\sigma_{kl2})+(1-\nu_{ikl})NB(\mu_{kl3},\sigma_{kl3}),
\]
\[
\nu_{ikl}\sim Bernoulli( v_{kl} ),
\]
where  the constant $3$ denotes the minimum number of reads required to be in state $\theta=2$.
In this section, we describe the general algorithm for such extensions. We assume that data from state $s$ has a distribution of $m_s$ components:

\[
Y_{ikl}|\theta_{iks}=1 \sim \sum_{r=1}^{m_s}v_{klsr}f_{sr}(\cdot|\mu_{klsr}, \sigma_{klsr}, \gamma_{iklsr}).
\]

This can be written in a hierarchical form, using $\nu_{iklsr}$ as the hidden variable indicating the mixture component within the state:

\begin{equation}\label{eq:mixstate}
(\nu_{iklsr})_{1\leq r\leq m_s}\sim Multinom(1, (v_{klsr})_{1\leq r\leq m_s}),~Y_{ikl}|\theta_{iks}=1,\nu_{iklsr}=1~\sim~f_{sr}(\cdot|\mu_{klsr}, \sigma_{klsr}, \gamma_{iklsr}).
\end{equation}

Here, we allow the distribution parameters $\mu$ and $\sigma$ as well as the prior information derived $\gamma$ to depend on the component. Let $f_{iklsr}=f_{sr}(y_{ikl}|\mu_{klsr}, \sigma_{klsr}, \gamma_{iklsr})$. The joint density for this model is:

\begin{equation}\label{eq:jointmix}
\begin{aligned}
f(z,b,\theta,\nu,y) = & \prod_{i=1}^I\zeta^{b_i}(1-\zeta)^{1-b_i}\cdot \prod_{i=1}^I\prod_{j=1}^J\pi_j^{z_{ij}}\cdot \prod_{i=1}^I\prod_{k=1}^K\prod_{l=1}^{n_k}\prod_{s=1}^S\prod_{r=1}^{m_s}v_{iklsr}^{\nu_{klsr}}\\
\cdot &\prod_{i=1}^I\prod_{k=1}^K\prod_{s=1}^S\left [ \left ( \prod_{l=1}^{n_k}\prod_{r=1}^{m_s}f_{iklsr}^{\nu_{iklsr}} \right )\left (\prod_{j=1}^Jw_{jks}^{z_{ij}} \right )^{1-b_{i}}p_{is}^{b_i} \right ]^{\theta_{iks}}.
\end{aligned}
\end{equation}

Let $f_{iks}=\prod_{l=1}^{n_k}(\sum_{r=1}^{m_s}v_{klsr}f_{iklsr})$, then the joint density for $z, b, \theta, Y$ can be expressed exactly the same as Eqn. (\ref{eq:full}). Therefore, the M-step updates for $W$, $P$, $\zeta$ and $\pi$ are not changed, with the related E-step quantities computed as Eqns. (\ref{eq:estep:b}),  (\ref{eq:estep:zb}), (\ref{eq:estep:thetazb}), (\ref{eq:estep:thetab}). We only need to modify the algorithm to estimate variables that depend on the component index $r$: $\mu$, $\sigma$, and $v$.

The related quantities that need to be computed are $E[\nu_{iklsr}|Y]$ and  $E[\theta_{iks}\nu_{iklsr}|Y]$. By Eqn. (\ref{eq:jointmix}), we have

\[
P(\nu_{iklsr}=1|\theta_{iks}=0)=v_{klsr},~~P(\nu_{iklsr}=1|\theta_{iks}=1)=\frac{v_{klsr}f_{iklsr}}{\sum_{r=1}^{m_s}v_{klsr}f_{iklsr}}.
\]

Therefore, we have

\begin{equation}\label{eq:thetanu}
E[\theta_{iks}\nu_{iklsr}|Y]=E[\theta_{iks}|Y]\frac{v_{klsr}f_{iklsr}}{\sum_{r=1}^{m_s}v_{klsr}f_{iklsr}},
\end{equation}

where $E[\theta_{iks}|Y]=\sum_{j=1}^JE[\theta_{iks}(1-b_i)z_{ij}|Y]+E[\theta_{iks}b_i|Y]$ and can be computed by Eqns. (\ref{eq:thetazb}) and (\ref{eq:thetab}). As a result,

\begin{equation}\label{eq:nu}
E[\nu_{iklsr}|Y]=(1-E[\theta_{iks}|Y])v_{klsr}+E[\theta_{iks}\nu_{iklsr}|Y].
\end{equation}

Given Eqns. (\ref{eq:thetanu}) and (\ref{eq:nu}), the M-step update for $v_{klsr}$ is:

\[
v_{klsr}^{(t)}=\frac{\sum_{i=1}^IE[\nu_{iklsr}|\hat\phi^{(t-1)}]}{I}.
\]

The M-step updates for $\mu_{klsr}$, $\sigma_{klsr}$ can be derived using Eqn. (\ref{eq:thetanu}). For the negative binomial distribution, as in Section 4.2, we have

\begin{eqnarray*}
\hat \mu_{klsr}^{(t)}\sum_{i=1}^I\gamma_{ikls}E \left [\theta_{iks}v_{iklsr}|\hat\phi^{(t-1)} \right ]&=\sum_{i=1}^IE \left [\theta_{iks}\nu_{iklsr}|\hat\phi^{(t-1)} \right ](Y_{ikl}-3),\\
\sum_{i=1}^IE \left [\theta_{iks}v_{iklsr}|\hat\phi^{(t-1)} \right ]\left [ \hat \mu_{klsr}^{(t)2}\gamma_{ikls}^2\left(1+\frac{1}{\hat \sigma_{klsr}^{(t)}} \right )+ \hat \mu_{klsr}^{(t)}\gamma_{ikls} \right ]& = \sum_{i=1}^IE \left [\theta_{iks}v_{iklsr}|\hat\phi^{(t-1)} \right ](Y_{ikl}-3)^2.
\end{eqnarray*}

\section{Simulation Studies}\label{sec:appsimulation}

This section presents six broad simulation studies to evaluate the performance of MBASIC. Each simulation study had multiple settings as outlined in Table 1 of the main article. We introduced the following four families of distributions in our main article:

\begin{itemize}
\item \textit{Log-normal Distribution.}  $LN(\mu_{kls}\gamma_{ikls},\sigma_{kls})$ with a density function: 
\begin{equation}\label{eq:applognormal}
f_s(y|\mu_{kls},\sigma_{kls},\gamma_{ikls})=\frac{1}{\sqrt{2\pi}\sigma_{kls}}\exp \left \{-\frac{(\log(y+1)-\mu_{kls}\gamma_{ikls})^2}{2\sigma_{kls}^2} \right \}.
\end{equation}
\item \textit{Negative Binomial Distribution.}  $NB(\mu_{kls}\gamma_{ikls},\sigma_{kls})$ with a density function:
\begin{equation}\label{eq:appnegbin}
f_s(y|\mu_{kls},\sigma_{kls},\gamma_{ikls})=\frac{\Gamma(y+\sigma_{kls})}{\Gamma(\sigma_{kls})\Gamma(y+1)}\frac{(\mu_{kls}\gamma_{ikls})^y\sigma_{kls}^{\sigma_{kls}}}{(\mu_{kls}\gamma_{ikls}+\sigma_{kls})^{y+\sigma_{kls}}}.
\end{equation}
\item \textit{Binomial Distribution.} $Binom(\gamma_{ikls}, \mu_{kls})$ with a density function:
\begin{equation}\label{eq:appbinom}
f_s(y|\mu_{kls},\gamma_{ikls})= {\gamma_{ikls}\choose y} \mu_{kls}^y(1-\mu_{kls})^{\gamma_{ikls}-y}.
\end{equation}
\item \textit{Degenerate Distribution}:
\begin{equation}\label{eq:appdegen}
f_s(y|\mu_{kls},\sigma_{kls},\gamma_{ikls})=I(y=s).
\end{equation}
\end{itemize}

\subsection{Simulation Study 1}\label{sec:sim1}

The first simulation study investigated the performance of MBASIC when the true value of $J$ was known and there were no structured clusters. We set the number of observational units as $I=4000$ and the number of clusters as $J=20$. The number of conditions was set to $K=30$, and within each condition  the numbers of replicates varied as  $n_k=1,2, 3$, each with probability 0.3, 0.5, and 0.2. The size of the hidden state space was varied at three levels: $S=2,3,4$. We simulated data under three distributional families: log-normal (LN) (\ref{eq:lognormal}), negative binomial (NB) (\ref{eq:negbin}), and binomial (Bin) (\ref{eq:binom}). We also varied the proportion of singleton units $\zeta$ at 0, 0.1 and 0.4. For simplicity, we set $\gamma_{ikls}=1$ in all distributions.

\subsubsection{Parameter Settings}\label{sec:simsetting}

Parameters $w_{jks}$'s and $p_{is}$'s generate the hidden state variables $\theta_{ik}$'s. We set them as follows. For different values of $k$, $j$, and $i$, the vectors $w_{jk\cdot}=(w_{jks}:1\leq s \leq S)$ and $p_{i\cdot}=(p_{is}:1\leq s \leq S)$ were simulated independently, each following an S-dimensional Dirichlet distribution $Dir(\alpha,\cdots,\alpha)$. We chose a uniform concentration parameter of  $\alpha=0.2$ for all dimensions to ensure that for each vector $w_{jk\cdot}$ or $p_{i\cdot}$, the probability mass tended to concentrate on one component. This controlled the conditional variance of $(\theta_{ik}|b_i,z_{ij})$. An increased value of $\alpha$ would increase the conditional variance of $\theta_{ik}$, thus make it more difficult to recover $w_{jks}$'s and $p_{is}$'s.

The settings for parameters $\mu_{kls}$'s and $\sigma_{kls}$'s were important. These parameters connected hidden states $\theta_{ik}$'s to the observed values $Y_{ikl}$'s. In general, recovering hidden states from the observed data is more difficult if: (1) differences of the mean values $\mu_{kls}$'s between the states  are small; (2) variances of the distributions within each state are large. To control these two aspects at reasonable levels, we set these parameters as follows:

\begin{itemize}
\item For log-normal distributions (\ref{eq:lognormal}), we set $\xi_s=2+\log(4s-3)$, simulated $\mu_{kls}\sim N(\xi_s, 0.05^2)$, and set $\sigma_{kls}=0.5$;
\item For negative binomial distributions (\ref{eq:negbin}), we set $\xi_s=8s-6$, simulated $\mu_{kls}\sim N(\xi_s, 0.5^2)$, and set $\sigma_{kl1}=2.82$, $\sigma_{kls}=5$ for $s=2,~3,~4$;
\item For binomial distributions (\ref{eq:binom}), we simulated $\mu_{kls}\sim Beta(3s, 3(S+1-s))$, and $\gamma_{ikl1}=\gamma_{ikl2}=\cdots=\gamma_{iklS}\sim Pois(10)$.
\end{itemize}

\begin{figure}
\centering
\begin{tabular}{cc}
a) Real Data & b) Log-Normal\\
\includegraphics[width=0.43\textwidth,page=1]{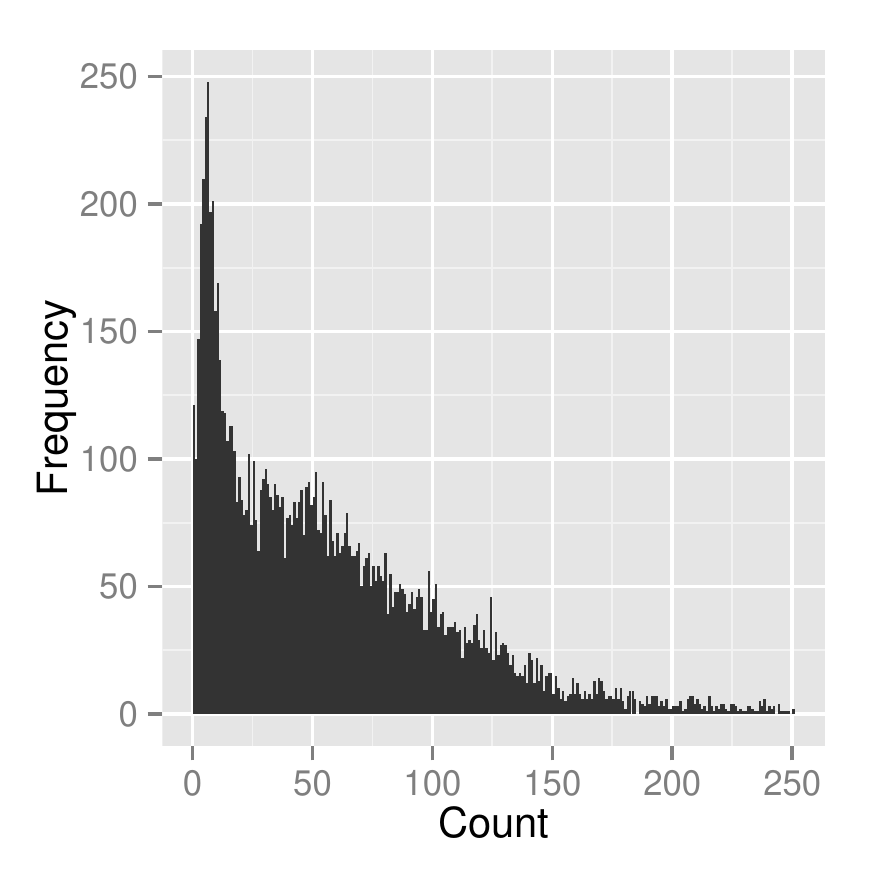} &
\includegraphics[width=0.43\textwidth,page=1]{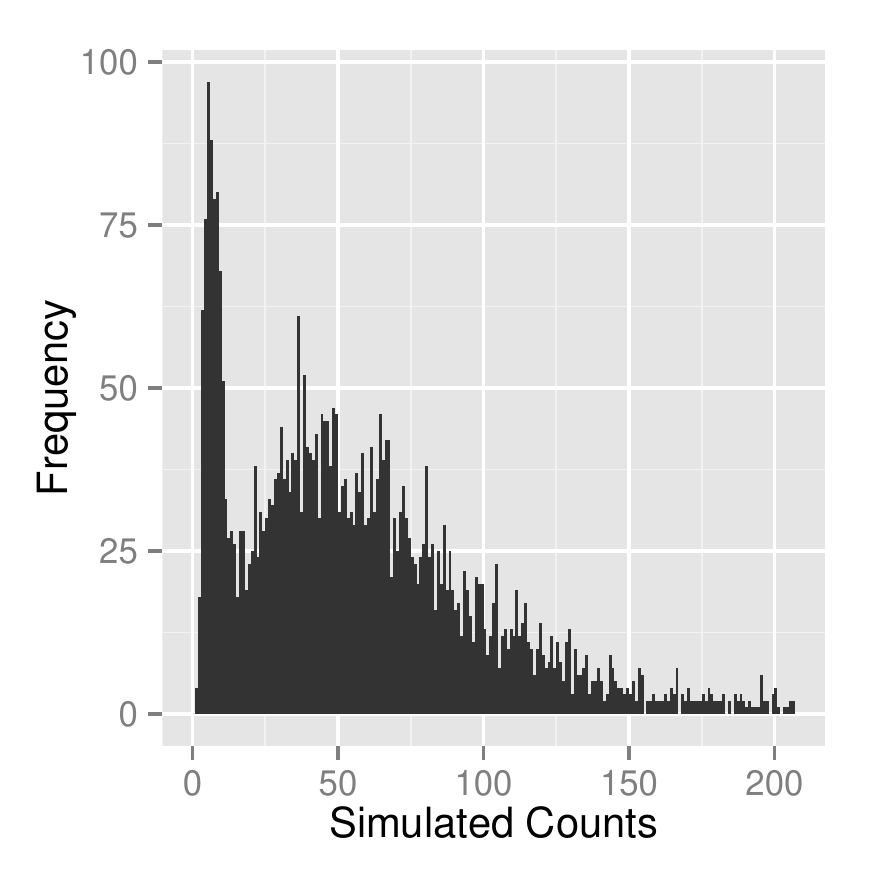} \\
c) Negative Binomial & d) Binomial \\
\includegraphics[width=0.43\textwidth,page=2]{simhist.pdf} &
\includegraphics[width=0.43\textwidth,page=3]{simhist.pdf} \\
\end{tabular}
\caption{Histograms for a) a real data set from a K562 Pol2 replicate in Section 4.1 of our main article; and simulated data from one condition based on one simulation for the b) Log-Normal, c) Negative Binomial and d) Binomial distribution with S=4 states.}\label{fig:simhist}
\end{figure}

 Figure \ref{fig:simhist} displays the histograms of $Y_{1,i,l}, \quad 1\leq i \leq I$ from one of the simulated data sets for all the three distributions with $S=4$ components. For comparison, we also present the histogram of an actual data set from the analysis in Section 4 of our main article. We observe that the mixture distribution of our simulated data with log-normal or negative-binomial distributions closely follow the real data.

\subsubsection{Alternative Approaches for Benchmarking MBASIC}

The MBASIC algorithm is summarized as Algorithm \ref{alg:mbasic}:
\begin{algorithm}                      
\caption{MBASIC}          
\label{alg:mbasic}                           
\begin{algorithmic}                    
  \FOR{$t=1,~2,\cdots$ until convergence}
  \STATE \textit{Expectation-Step}: Compute the conditional expectations $E[I(\theta_{ik}=s)|\hat \phi^{(t-1)}]$, $E[b_i|\hat \phi^{(t-1)}]$, $E[I(\theta_{ik}=s)b_i|\hat \phi^{(t-1)}]$, $E[z_{ij}(1-b_i)|\hat \phi^{(t-1)}]$, $E[I(\theta_{ik}=s)z_{ij}(1-b_i)|\hat \phi^{(t-1)}]$;
   \STATE \textit{Maximization-Step}: Update estimates for parameters $\mu_{kls}$, $\sigma_{kls}$, $\zeta$, $\pi_j$, $w_{jks}$, $p_{is}$.
   \ENDFOR
\end{algorithmic}
\end{algorithm}

To the best of our knowledge, there are currently no existing methods suited for the general setup of MBASIC. There are, however, algorithms tailored for analyzing specific data types with hierarchical state-space models similar to MBASIC. These algorithms largely fall into two categories. In the first category, estimation for the state-space variables are separated from state-space clustering. Some examples include \cite{Gernstein11}, \cite{Gernstein12}, \cite{histonePCA}, \cite{Neph12}. In the second category, distributional parameters for each experimental replicate are estimated first. These parameters are then fixed, and  the state-space variables and the clustering structure are estimated jointly conditional on these estimates. Examples of this approach include \cite{iaseq},  \cite{jMosaics},  and \cite{cormotif}.


\begin{table}
\centering
\caption{\textit{Simulation Study 1.} A summary of the  benchmark algorithms that are compared to MBASIC.
Neither SE-MC nor PE-MC  perform joint estimation of the model parameters. SE-* algorithms estimate the data-specific model parameters and state-space as a first step and then cluster the state variables. PE-* algorithms estimate data-specific model parameters and fixes these in joint estimation of the state-space and clustering. $^*$ Denotes distributional parameters for each experimental replicate.
}\label{tbl:s1summary}
\begin{tabular}{cccccc}
\hline
Algorithm & Is State-space & Is Parameter$^*$ &  Clustering  & Include \\
& Estimation Joint & Estimation Joint&  model & singletons\\
&   with Clustering? & with Clustering?  & &\\
\hline
\textbf{MBASIC} & Joint & Joint  & Mixture model & Yes\\
SE-HC & Separate & Separate & Hierarchical clustering & No\\
SE-MC & Separate & Separate   & Mixture model & Yes\\
PE-MC & Joint & Separate  & Mixture model & Yes\\
MBASIC0 & Joint & Joint  & Mixture model & No\\
SE-MC0 & Separate & Separate  & Mixture model & No\\
PE-MC0 & Joint & Separate   & Mixture model & No\\
\hline
\end{tabular}
\end{table}

To compare the general implementation of MBASIC as in Algorithm \ref{alg:mbasic}  with these existing model fitting ideas, we designed six benchmark algorithms.  Table \ref{tbl:s1summary} provides a summary of these algorithms. Two of these algorithms, SE-HC (State-space Estimation followed by Hierarchical Clustering) and SE-MC (State-space Estimation followed by Mixture model Clustering), treat the state-space mapping step and the state-space clustering separately. The third algorithm, PE-MC (Parameter Estimation followed by Mixture model Clustering), separates experiment-specific distributional parameter estimation from the joint estimation of other parameters.

For all the three algorithms, in the first step, observations from each experimental condition $\{Y_{ikl}:1\leq i \leq I, 1\leq l \leq n_k\}$ are fitted according to the following model:
\begin{equation}\label{eq:step1}
(Y_{ikl}|\theta_{ik}=s)\sim f_s(\cdot|\mu_{kls},\sigma_{kls}, \gamma_{ikls}), \quad P(\theta_{ik}=s)=q_{ks}.
\end{equation}
The standard E-M algorithm can be used for the first step and results in estimates of $q_{ks}$, $\mu_{kls}$, $\sigma_{kls}$ as well as the posterior estimates for the state space $P(\theta_{ik}=s|Y)$. In the second step, SE-MC and SE-HC cluster the observational units based on the estimated $P(\theta_{ik}=s|Y)$ from the first step. SE-HC (Algorithm \ref{alg:sehc}) uses hierarchical clustering, while SE-MC (Algorithm \ref{alg:semc}) uses MBASIC with degenerate distributions (\ref{eq:degen}) for clustering. The second step of PE-MC (Algorithm \ref{alg:pemc}) is similar to Algorithm \ref{alg:mbasic}, except that parameters $\mu_{kls}$, $\sigma_{kls}$'s are not updated.

In addition to  joint fitting of all model parameters, another important feature of MBASIC is its inclusion of the singleton cluster $\mathscr C_0$. To the best of our knowledge, this feature is not included in similar models such as   \cite{iaseq} and \cite{cormotif}. We conjecture that in practice, when some units can not be grouped together with other units due to their distinct state-space profiles, including this singleton cluster can enhance model estimation. To test this conjecture, we developed a version of each of the SE-MC, PE-MC, and MBASIC algorithms that ignore the singleton cluster, i.e., forces each unit into a cluster. This is achieved  simply by initializing $\zeta=0$ in the Algorithms \ref{alg:mbasic}, \ref{alg:semc}, and \ref{alg:pemc}. We refer to these algorithms by SE-MC0, PE-MC0, and MBASIC0.

\begin{algorithm}                      
\caption{State-space estimation followed by hierarchical clustering (SE-HC)}          
\label{alg:sehc}                           
\begin{algorithmic}                    
  \STATE Step 1:
  \FOR{$1\leq k \leq K$}
  \STATE Apply the standard E-M algorithm on data $\{Y_{ikl}:1\leq i \leq I, 1\leq l\leq n_k\}$ to estimate posterior probabilities $P(\theta_{ik}=s|Y)$.
\ENDFOR
\STATE Step 2:
\STATE Let $\tilde \theta_i= \{ P(\theta_{ik}=s|Y)\}_{1\leq k \leq K, 1\leq s \leq S}$. Cluster vectors $\tilde \theta_i$ into $J$ clusters using hierarchical clustering algorithm with the Euclidean distance. Estimate $w_{jks}$ as the means within each cluster.
\end{algorithmic}
\end{algorithm}

\begin{algorithm}                      
\caption{State-space estimation followed by mixture model clustering (SE-MC)}          
\label{alg:semc}                           
\begin{algorithmic}                    
\STATE Step 1:
  \FOR{$1\leq k \leq K$}
  \STATE Apply the standard E-M algorithm on data $\{Y_{ikl}:1\leq i \leq I, 1\leq l\leq n_k\}$ to estimate posterior probabilities $P(\theta_{ik}=s|Y)$.
\ENDFOR
\STATE Step 2:
\STATE Denote $\theta^*_{ik}=\arg_{s}\max P(\theta_{ik}=s|Y)$ for each $1\leq k \leq K, 1\leq i \leq I)$. Apply Algorithm \ref{alg:mbasic} with $\theta_{ik}\leftarrow \theta^*_{ik}$ and $f_{s}=I(y=s)$ to obtain estimates for $w_{jks}$, $p_{is}$, $\zeta$, and $\pi_j$. 
\end{algorithmic}
\end{algorithm}

\begin{algorithm}                      
\caption{Parameter estimation followed by mixture model clustering (PE-MC)}         
\label{alg:pemc}                           
\begin{algorithmic}                    
\STATE Step 1:
  \FOR{$1\leq k \leq K$}
  \STATE Apply the standard E-M algorithm on data $\{Y_{ikl}:1\leq i \leq I, 1\leq l\leq n_k\}$ to estimate $\mu_{kls}$, $\sigma_{kls}$ for each experiment.
\ENDFOR
\STATE Step 2:
\STATE Apply Algorithm \ref{alg:mbasic} without updating $\mu_{kls}$, $\sigma_{kls}$ in the Maximization step. 
\end{algorithmic}
\end{algorithm}

\subsubsection{Results}

\begin{figure}
\centering
\includegraphics[width=\textwidth,page=1]{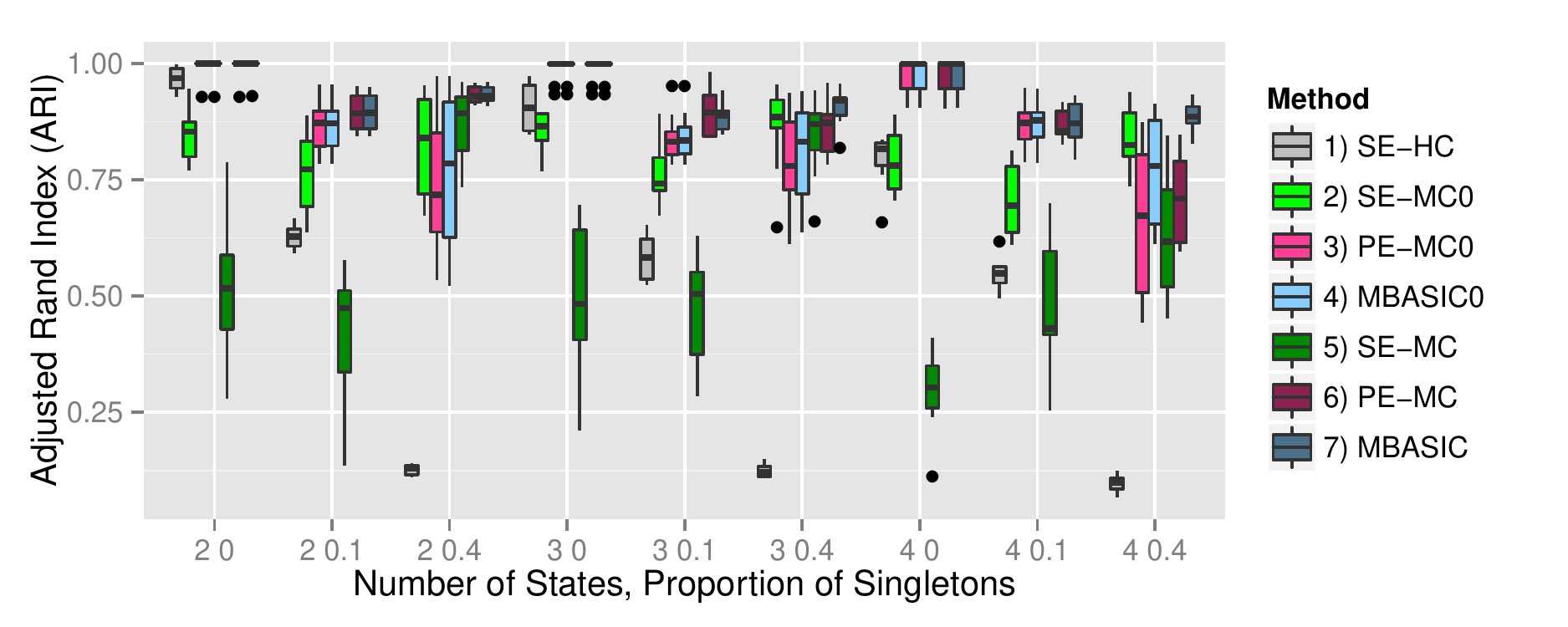}
\includegraphics[width=\textwidth,page=2]{simulation_S.pdf}
\includegraphics[width=\textwidth,page=5]{simulation_S.pdf}
\caption{\textit{Simulation Study 1, log-normal distribution.} Boxplots for ARI, MSE-W, and SPE across 10 simulated datasets. The number of states is varied at 2, 3, and 4,  and the proportion of singletons at 0, 0.1, 0.4.  Table \ref{tbl:s1summary} summarizes the methods compared.}\label{fig:sim_ln}
\end{figure}

\begin{figure}
\centering
\includegraphics[width=\textwidth,page=6]{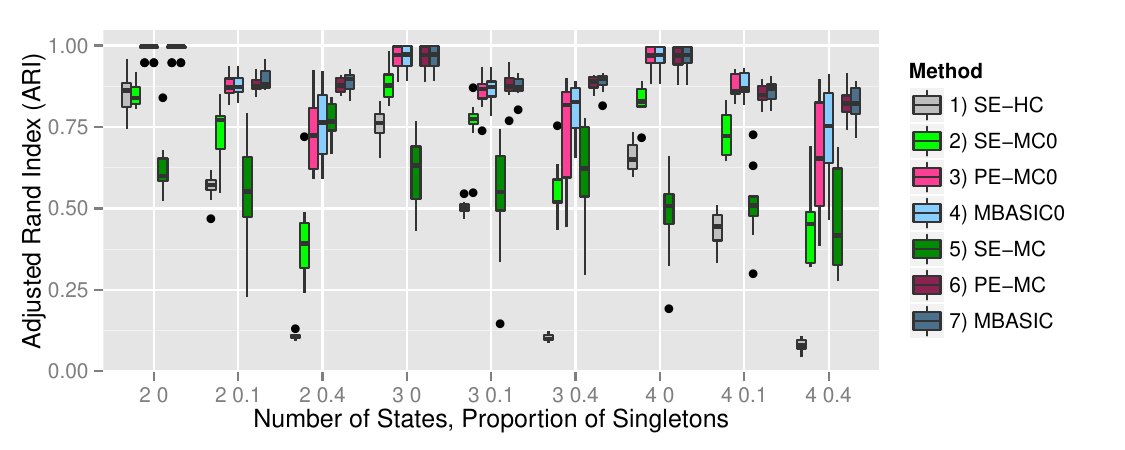}
\includegraphics[width=\textwidth,page=7]{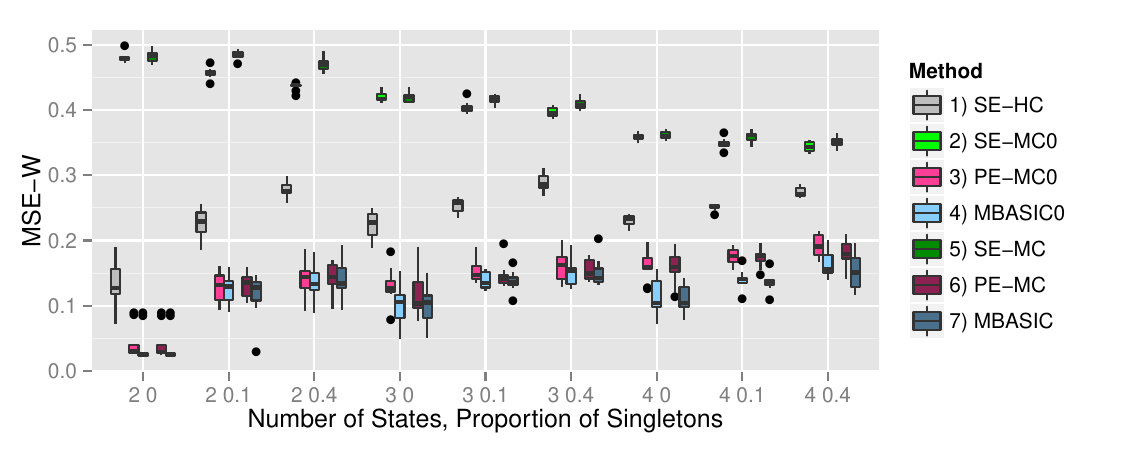}
\includegraphics[width=\textwidth,page=10]{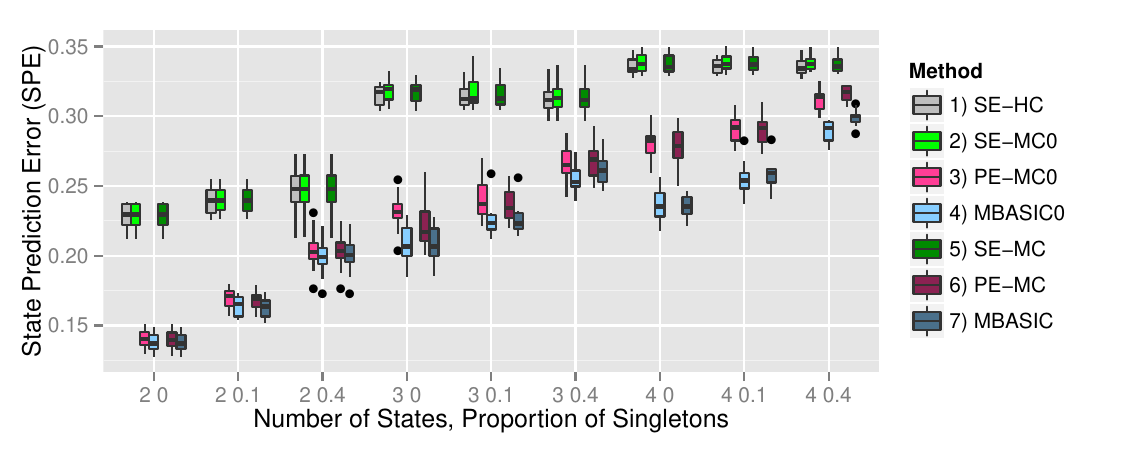}
\caption{\textit{Simulation Study 1, negative binomial distribution.} Boxplots for ARI, MSE-W, and SPE across 10 simulated datasets. The number of states is varied at 2, 3, and 4, and the proportion of singletons at 0, 0.1, 0.4.  Table \ref{tbl:s1summary} summarizes the methods compared.}\label{fig:sim_nb}
\end{figure}

\begin{figure}
\centering
\includegraphics[width=\textwidth,page=11]{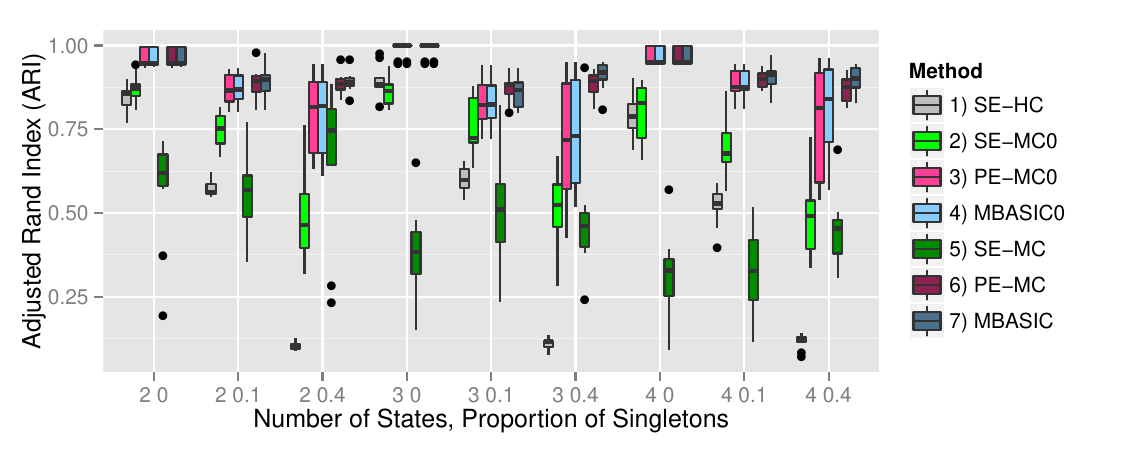}
\includegraphics[width=\textwidth,page=12]{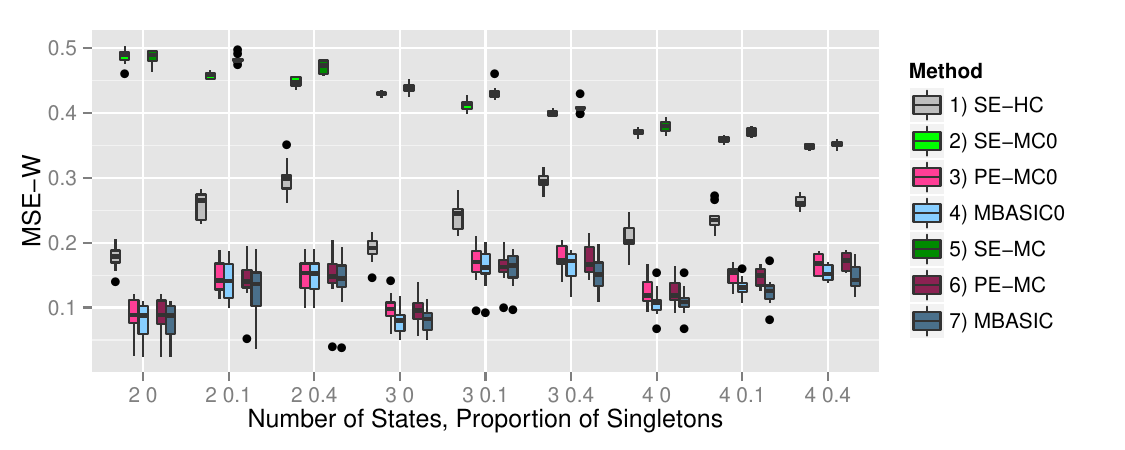}
\includegraphics[width=\textwidth,page=15]{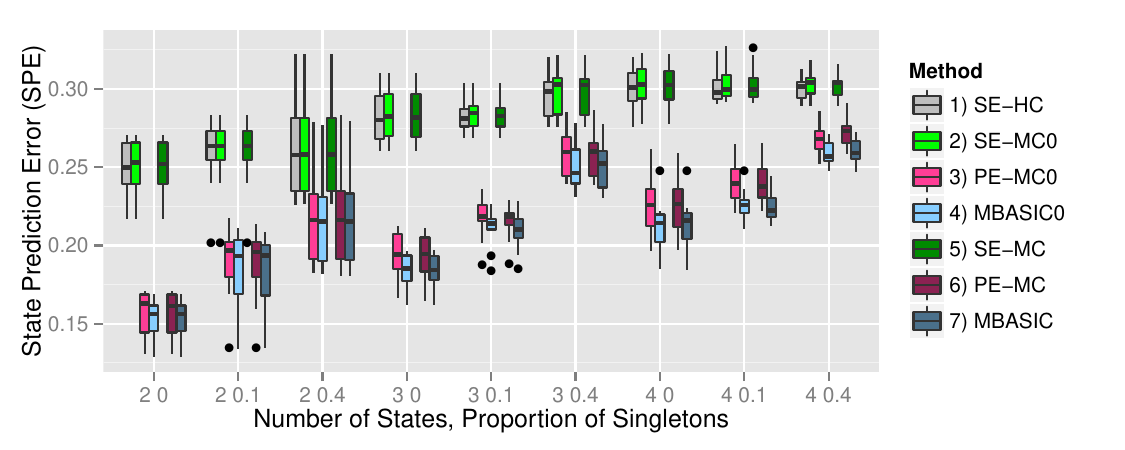}
\caption{\textit{Simulation study 1, binomial distribution.} Boxplots for ARI, MSE-W, and SPE across 10 simulated datasets. The number of states is varied at 2, 3, and 4,   and the proportion of singletons at 0, 0.1, 0.4.  Table \ref{tbl:s1summary} summarizes the methods compared.}\label{fig:sim_bin}
\end{figure}

We utilized several criteria to compare  the performance of MBASIC to the benchmark algorithms in Table \ref{tbl:s1summary}. To estimate how well the state space was characterized for each cluster, we computed the mean-squared error for $W$ (MSE-W) as MSE-W=$\sqrt{\sum_{j,k,s}(\hat w_{jks}-w_{jks})^2/(JKS)}$ . We also evaluated how well each method recovered the true state variables $\theta_{ik}$'s. This was reflected by the state prediction error (SPE) as the mean squared error between the simulated states $\theta_{ik}$'s and their posterior probabilities: SPE=$\sqrt{\sum_{i,k,s}[1\{\theta_{ik}=s\}-P(\theta_{ik}=s|Y)]^2/(IKS)}$. Finally, to compare the estimated clustering with the simulated true clustering, we computed the Adjusted Rand Index (ARI) \citep{randindex}. ARI is a measure for the similarity between two different clusterings of the data. Its value ranges between -1 and 1, with 1 indicating perfect match between the two clusterings.

ARI requires the true clusters denoted by $\mathscr C_j,~0\leq j \leq J$ and their estimates denoted by $\hat{\mathscr C}_j,~0\leq j \leq \hat J$. In our simulations, they were computed as:
\[
\mathscr C_0=\{1\leq i \leq I:b_i=1\};~\mathscr C_j=\{1\leq i \leq I: b_i=0,~z_{ij}=1\},~j\leq 1,
\]
where $\mathscr C_0$ denoted the set of singleton units. $\hat{\mathscr C}_j$ was computed from the posterior distributions as  $\hat{\mathscr C}_j=\{1\leq i\leq I: j=\arg\max_{0\leq j\leq J}P(i\in \mathscr C_j|Y)\}$, where $P(i\in \mathscr C_0|Y)=E(b_i|Y)$, and $P(i\in \mathscr C_j|Y)=E[(1-b_i)z_{ij}|Y]$.

The simulation results under various settings are summarized by the boxplots for each criterion in Figures \ref{fig:sim_ln}, \ref{fig:sim_nb}, and \ref{fig:sim_bin}. Across all different simulation settings, the performance of MBASIC was consistently among the best in all of the ARI, MSE-W, and SPE metrics. This shows that MBASIC could not only recover the clustering structure, but also achieve high accuracy in estimating individual states. SE-HC, SE-MC and SE-MC0 performed the worst in both detecting the clustering structure and estimating the individual states. This suggests that separating state-space inference from joint model fitting can significantly deteriorate model estimation. Different from the SE-* methods, performances of PE-MC and PE-MC0 were much closer to MBASIC. For the negative binomial and binomial distributions (Figures \ref{fig:sim_nb} and \ref{fig:sim_bin}), PE-MC achieved similar ARI levels to MBASIC and slightly larger SPE values. These observations show that by jointly estimating the clusters and the states, data under different conditions could borrow information from each other and thus substantially improve the state-space estimation. Overall, these observations are consistent with the results in \cite{cormotif} and \cite{iaseq}.

The simulation results also highlight the effect of modeling the singleton cluster in various settings. Comparing the performances of MBASIC with MBASIC0 and PE-MC with PE-MC0, we see that modeling the singleton cluster does not have a significant effect when the proportion of singletons is low, i.e., $\zeta=0$ or $0.1$; however, the improvement is highly significant when $\zeta=0.4$. When $\zeta=0.4$, including singletons significantly improved the performance with respect to ARI, but did not have an obvious effect on SPE. This has several implications in practice. First, the fact that MBASIC does not under-perform any other methods when $\zeta=0$ or $0.1$ indicates that increasing the model complexity by introducing singletons does not lead to unrobust inference. Because we are always agnostic on the existence of singletons for any real data, keeping them in our model would guard against their adverse influence in inferring the clustering structure. Second, although incorporating the singleton cluster does not improve estimating individual states, some epigenetic studies focus primarily on the association structure between units, as our example in 4.2. For such studies, the gain in estimating the clustering structure by including the singletons is essential. We note that in the comparison of  SE-MC0 with SE-MC for the negative binomial  and the binomial distributions (Figures \ref{fig:sim_nb} and \ref{fig:sim_bin}), modeling the singletons does not necessarily improve estimation for separate model fitting  even when the proportion of singletons is high, e.g., $\zeta=0.4$. This might suggest that the state-space estimation step is introducing additional noise to the clustering step, which in turn makes it less favorable to infer a complicated clustering structure with singletons.

\subsection{Simulation Study 2: Model Selection \label{sec:sim2}}

\begin{table*}
\centering
\caption{\textit{Simulation study 2, Scenario 1, unstructured clusters.} Simulation results for model selection without structural constraints. For each criterion, the mean is computed over 10 simulated data sets, with the standard deviation shown in the parentheses.}\label{tbl:unstruct}
\begin{tabular}{cc|cccc}
\hline
Dist. & $\zeta$ & $J$ & ARI & MSE-W & SPE \\
\hline
Bin & 0.1 &  20.8 ( 2.098 ) &  0.94 ( 0.036 ) &  0.096 ( 0.018 ) &  0.159 ( 0.014 ) \\  
Bin & 0.4 &  20.9 ( 1.101 ) &  0.914 ( 0.035 ) &  0.122 ( 0.034 ) &  0.204 ( 0.012 ) \\  
LN & 0.1 &  20.7 ( 0.823 ) &  0.989 ( 0.005 ) &  0.044 ( 0.03 ) &  0.086 ( 0.006 ) \\ 
LN & 0.4 &  21.3 ( 1.337 ) &  0.972 ( 0.007 ) &  0.095 ( 0.027 ) &  0.107 ( 0.008 ) \\  
NB & 0.1 &  21.6 ( 0.843 ) &  0.947 ( 0.021 ) &  0.089 ( 0.028 ) &  0.154 ( 0.007 ) \\  
NB & 0.4 &  20.6 ( 2.271 ) &  0.902 ( 0.026 ) &  0.112 ( 0.048 ) &  0.189 ( 0.007 ) \\  
\hline
\end{tabular}
\end{table*}

\begin{table*}
\centering
\caption{\textit{Simulation study 2, Scenario 2, structured clusters.} Simulation results for model selection with structural constraints. For each criterion, the mean is computed over 10 simulated data sets, with the standard deviation shown in the parentheses.}\label{tbl:struct}
\begin{tabular}{cc|ccccc}
\hline
Dist. & $\zeta$ & $J_1$ & $J$ & ARI & MSE-W & SPE \\ 
\hline
Bin & 0.1 & 10.3 ( 1.16 ) &  20.7 ( 1.494 ) &  0.934 ( 0.022 ) &  0.084 ( 0.035 ) &  0.162 ( 0.02 ) \\  
Bin & 0.4 & 10.3 ( 1.636 ) &  21 ( 2.625 ) &  0.897 ( 0.048 ) &  0.125 ( 0.03 ) &  0.196 ( 0.031 ) \\
LN & 0.1 & 10.4 ( 0.516 ) &  20.6 ( 0.516 ) &  0.984 ( 0.015 ) &  0.044 ( 0.032 ) &  0.086 ( 0.006 ) \\  
LN & 0.4 & 11.2 ( 1.619 ) &  22.5 ( 1.509 ) &  0.968 ( 0.01 ) &  0.108 ( 0.037 ) &  0.106 ( 0.006 ) \\  
NB & 0.1 & 10.9 ( 1.197 ) &  21 ( 1.054 ) &  0.955 ( 0.019 ) &  0.064 ( 0.035 ) &  0.155 ( 0.008 ) \\  
NB & 0.4 & 11.2 ( 1.814 ) &  22.2 ( 1.398 ) &  0.926 ( 0.014 ) &  0.108 ( 0.031 ) &  0.184 ( 0.013 ) \\  
\hline
\end{tabular}
\end{table*}

This second set of simulations aimed to evaluate the use of BIC to select the number of clusters as well as the structural constraints for each cluster. We simulated data sets under two scenarios. For the first scenario, each data set had $J=20$ clusters with $K=30$ experimental conditions, and none of the clusters had structural constraints. For the second scenario, each data set had $J=20$ clusters over $K=30$ conditions, but $J_1=10$ of the clusters were structurally constrained as follows:
\[
w_{j,k,s}=w_{j,k+K/2,s}, \forall j,  1 \leq j \leq J/2; \forall k,  1 \leq k \leq K/2.
\]
We refer the two scenarios as the \textit{unstructured scenario} and the \textit{structured scenario}, respectively. We considered log-normal distributions (\ref{eq:lognormal}), negative binomial distributions (\ref{eq:negbin}) and binomial distributions (\ref{eq:binom}) for both cases. We also varied the proportion of singleton units $\zeta$ at 0.1 and 0.4. The number of states was fixed at $S=2$. The remaining parameters were simulated following the same mechanism as in Section \ref{sec:simsetting}.

For each simulated data set, we fitted a number of candidate models. For the unstructured scenario, we varied the number of clusters $J$ from 10 to 30. For the structured scenario, we followed the two-phase procedure described in Section 3.4 of our main article. The best model was selected by the minimum BIC value. To assess the performances of these selected models, we computed the ARI, MSE-W and SPE metrics as described in Section \ref{sec:sim1}\footnote{When the actual $J$ and its estimate $\hat J$ are different, MSE-W is redefined as:

\[
MSE-W=\left [\frac{\sum_{1\leq j \leq J,k, s}(w_{jks}-\hat w_{c_1(j)ks})^2+ \sum_{1\leq j \leq \hat J,k,s}(\hat w_{jks}- w_{c_2(j)ks})^2}{KS(J+\hat J)} \right ]^{\frac{1}{2}},
\]

where 
\[
c_1(j)=\arg_{j'\leq \hat J}\min \sum_{k,s}(w_{jks}-\hat w_{jk's})^2~,~c_2(j)=\arg_{j'\leq J}\min \sum_{k,s}(\hat w_{jks}- w_{jk's})^2.
\]
}.

The simulation results are summarized in Tables \ref{tbl:unstruct} and \ref{tbl:struct}. 
Under each set of parameters, we computed the mean and the standard deviation for each of the criterion as well as the selected value of $J$ and $J_1$ under 10 simulated data sets. These tables show that the selected values for $J$ and $J_1$ were very close to the true values. Moreover, MBASIC performed uniformly well with respect to ARI, MSE-W, and SPE  under different settings. These results indicate that even if MBASIC may not identify the ``true'' structure that drives the actual data, the identified structures can still properly represent the state-space associations between units.

\subsection{Simulation Studies 3-5: Comparison with iASeq and CorMotif}

In this section, we compare MBASIC with two recently proposed models for integrative analysis of specific types of genomic data: CorMotif \citep{cormotif} and iASeq \citep{iaseq}. Both models have the similar state-space clustering structure as MBASIC. The main difference from MBASIC is that they each incorporate more complicated distributional assumptions targeting  specific genomic data types. The CorMotif model specifically addresses integrative differential expression analysis with  $n_{k1}$ case condition replicates and $n_{k0}$ control condition replicates for each experimental condition $k$.   It inherits the LIMMA \citep{limma} framework for differential analysis of gene-expression data and assumes mixture of Gaussian distributions with $S=2$ states: $s=1$ for the equally expressed state, and $s=2$ for the differentially expressed state. Specifically, the  CorMotif model  has the following state-space mapping structure:
\[
\begin{aligned}
\frac{n_ks_{k}^2}{\sigma_{ik}^2}\sim& \chi_{n_k}^2,\\
\mu_{ik}|\sigma_{ik}^2\sim & N(0, u_k\sigma_{ik}^2),\\
(Y_{ikl}|\theta_{ik}=1) \sim & N(\mu_{ik0}, \sigma_{ik}^2),~l=1,2,\cdots,n_{k1},\\
(Y_{ikl}|\theta_{ik}=2) \sim & N(\mu_{ik0}+\mu_{ik}, \sigma_{ik}^2),~l=1,2,\cdots,n_{k1},\\
X_{ikl} \sim & N(\mu_{ik0},\sigma_{ik}^2),~l=1,2,\cdots,n_{k0},\\
\end{aligned}
\]
where $X_{ikl}$'s are the observed data from control experiments, and $Y_{ikl}$ are the observed data from the case experiments. $n_k$ and $s_{k}^2$ are hyper parameters specific to each experiment to account for potential heterogeneity among units within the same state, and $u_{k}$ reflects the strength of differential expression. CorMotif assumes almost the same state-space clustering structure as MBASIC except that it does not include singletons. The iASeq model,  targeting at allele-specific binding problems has the following state-space mapping structure:
\[
\begin{aligned}
Y_{ikl}\sim & Binom(\gamma_{ikl}, p_{ik}),\\
p_{ik}|\theta_{ik}=2\sim&Beta(\alpha_k,\beta_k),\\
p_{ik}|\theta_{ik}=1\sim & Unif \left (0, \frac{\alpha_k}{\alpha_k+\beta_k}\right ),\\
p_{ik}|\theta_{ik}=3\sim & Unif\left (\frac{\alpha_k}{\alpha_k+\beta_k}, 1\right ),\\
\end{aligned}
\]
where the $\alpha_k$, $\beta_k$ are experiment-specific parameters, and $\gamma_{ikl}$ is the observed total number of reads between two alleles. The state-space mapping structure for iASeq is almost the same as MBASIC, except that it assumes no singletons, and that  one cluster is dedicated to equal binding/occupancy between the alleles (i.e., $w_{1k1}=1,~\forall 1\leq k\leq K$). 

There are two key differences between CorMotif/iASeq and MBASIC. First, both CorMotif and iASeq address the heterogeneity among the units within the same state, and they introduce additional hyper parameters to model the heterogeneous parameters associated with the distribution of individual units. Compared to MBASIC, where we assume the distributions within the same state are homogeneous, such heterogeneous distributional assumptions are much more realistic. Second, CorMotif and iASeq implement two-stage estimation procedures similar to PE-MC0, which separate parameter estimation from state-space clustering. \cite{cormotif} pointed out that once we have the heterogeneous distributional parameters within each state, joint model fitting for all parameters would require running a Markov Chain Monte-Carlo algorithm rather than the simple E-M algorithm we have developed for MBASIC. Therefore, the computational cost ensued might render its applicability for large real data sets.

In comparison of MBASIC to CorMotif and iASeq, we simulated data according to each of the assumed distributions of CorMotif/iASeq, but fitted MBASIC models using simplified distributions. For data simulated from the iASeq model, we used MBASIC to fit binomial distributions with $S=3$ states (\ref{eq:binom}). For data simulated from the CorMotif model, we first generated two versions of t-statistics as follows. For each unit and experiment, denote $\overline Y_{ik}=\sum_{l=1}^{n_{k1}}Y_{ikl}/n_{k1}$, $\overline X_{ik}=\sum_{l=1}^{n_{k0}}X_{ikl}/n_{k0}$, $\tilde s_{ik}^2=[\sum_{l=1}^{n_{k1}}(Y_{ikl}-\overline Y_{ik})^2+\sum_{l=1}^{n_{k0}}(X_{ikl}-\overline X_{ik})^2]/(n_{k1}+n_{k0}-2)$ and $v_k=1/n_{k1}+1/n_{k0}$. We computed the \textit{naive t-statistic} $T_{ik}$ as:

\begin{equation}\label{eq:naive-t}
T_{ik}=\frac{\overline Y_{ik} - \overline X_{ik}}{\sqrt{v_k}\tilde s_k}.
\end{equation}

We also computed the \textit{limma t-statistic} $\tilde T_{ik}$ by first fitting the data for each condition using LIMMA \citep{limma} to estimate $n_k$ and $s_k^2$, then computed:

\begin{equation}\label{eq:limma-t}
\tilde T_{ik}=\frac{\sqrt{n_k+n_{k1}+n_{k0}-2}(\overline Y_{ik} - \overline X_{ik})}{\sqrt{v_k[(n_{k1}+n_{k0}-2)\tilde s_k^2+n_ks_k^2]}}.
\end{equation}

For each set of $T_{ik}$'s and $\tilde T_{ik}$'s, we fitted the MBASIC model with $S=2$ components of scaled-t distributions:

\begin{equation}\label{eq:scaled-t}
T/\mu_{ks}|\theta_{iks}=1~\sim~t_{\sigma_{ks}},~~s=1,~2.
\end{equation}

Here, $\mu_{ks}$ is the scaling parameter, and $\sigma_{ks}$ is the degrees of freedom. Because we pooled the replicate level data to generate these t-statistics, the parameters $\mu$ and $\sigma$ no longer depended on $l$. We refer to the method using $\tilde T_{ik}$ as MBASIC-limma, and using $T_{ik}$ as MBASIC-t. Because there is no closed form maximum likelihood solution for t-distributions, we use the moment method to estimate $\mu_{ks}$'s and $\sigma_{ks}$'s in the M-step similar to the case of negative binomial distributions.

\begin{table}
\centering
\caption{Summary for the designs of the simulation settings in Simulation Study 4, originally designed by \cite{cormotif}.}\label{tbl:cormotif}
\begin{tabular}{cccc}
\hline
Simulation Setting & $I$ & $J$ & $K$ \\
\hline
1 & 10,000 & 4 & 4 \\
2 & 10,000 & 4 & 4 \\
3 & 10,000 & 5 & 8 \\
4 & 10,000 & 5 & 20\\
\hline
\end{tabular}
\end{table}

In Simulation Study 3, we simulated data following the iASeq model. We set $\alpha_k=\beta_k=2$, and simulated state-space variables the same as in Section \ref{sec:sim1} with $I=4000$. We set $J=10,~20$ and $\zeta=0,~0.1,~0.4$. Simulation Studies 4-5 compare MBASIC with CorMotif. In Simulation Study 4, we simulated data in four settings corresponding to Simulations 1-4 of \cite{cormotif} respectively. In these settings, we had $n_k=4$, $u_k=4$, $s_k^2=0.02$. Table \ref{tbl:cormotif} summarizes the settings for the number of clusters, experiment conditions, and units  for the state-space variables. We refer our readers to \cite{cormotif} for more details of the state-space design. We note that \cite{cormotif} simulations did not include singletons  (i.e., $\zeta = 0$) and furthermore, their settings assumed $w_{jks}\in\{0,~1\}$. This means that the state-space variables are completely determined by the clustering structure. In Simulation Study 5, we set $n_k$ and $s_k^2$ the same as in Simulation Study 4, but varied $u_k$ as $u_k=8$ for easier distinction between different states. However, we simulated $w_{jks}$ following S-dimensional Dirichlet distributions as in Simulation Study 1 to introduce noises in generating state-space variables. In addition, we simulated data with smaller number of units ($I=4000$), but more clusters ($J=10,~20$), and varied the proportion of singletons $\zeta=0,~0.1,~0.4$. The other details of generating state-space variables were the same as in Section \ref{sec:sim1}.

\begin{figure}
\centering
\includegraphics[width=0.8\textwidth, page = 1]{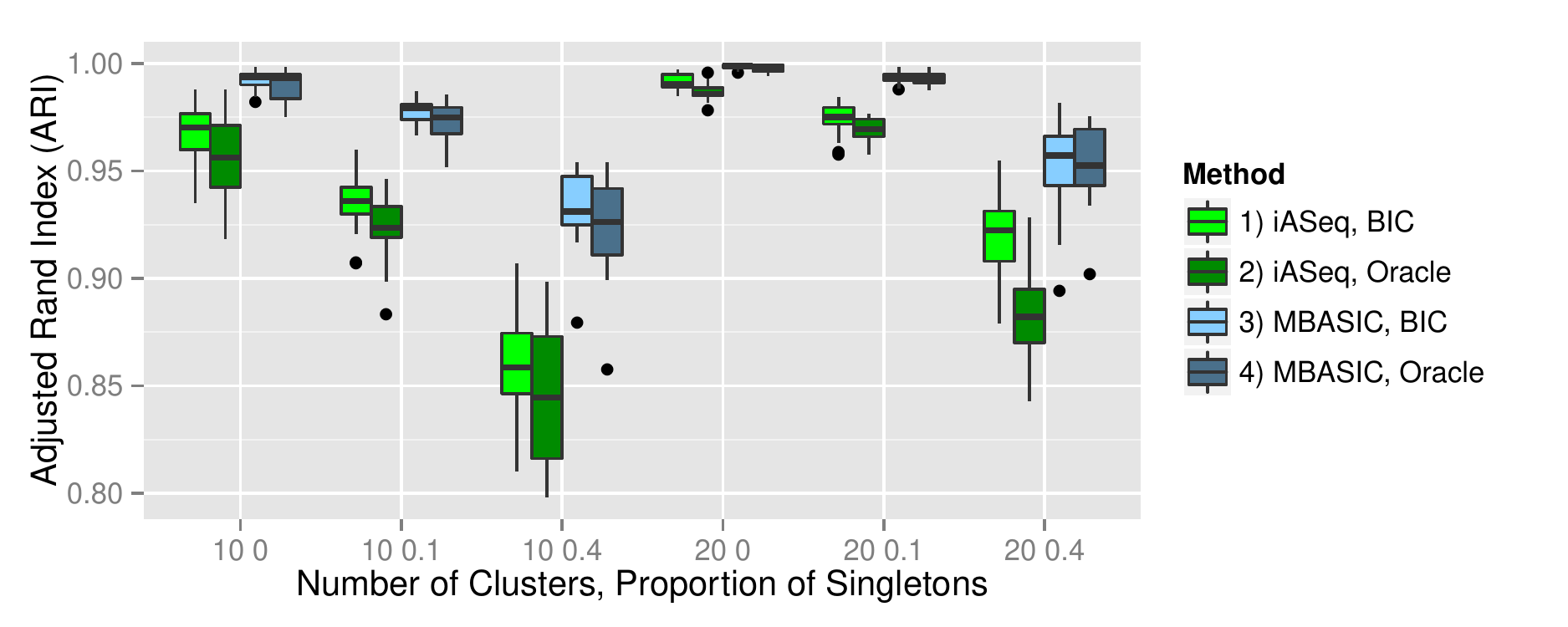}
\includegraphics[width=0.8\textwidth, page = 2]{simulation_iaseq.pdf}
\includegraphics[width=0.8\textwidth, page = 4]{simulation_iaseq.pdf}
\caption{\textit{Simulation Study 3, comparison between MBASIC and iASeq.} We varied the number of clusters at 10, 20 and the proportion of singletons at 0, 0.1 and 0.4. Results are summarized over 10 simulations under each setting.}\label{fig:iaseq}
\end{figure}

\begin{figure}
\centering
\includegraphics[width=0.8\textwidth, page = 1]{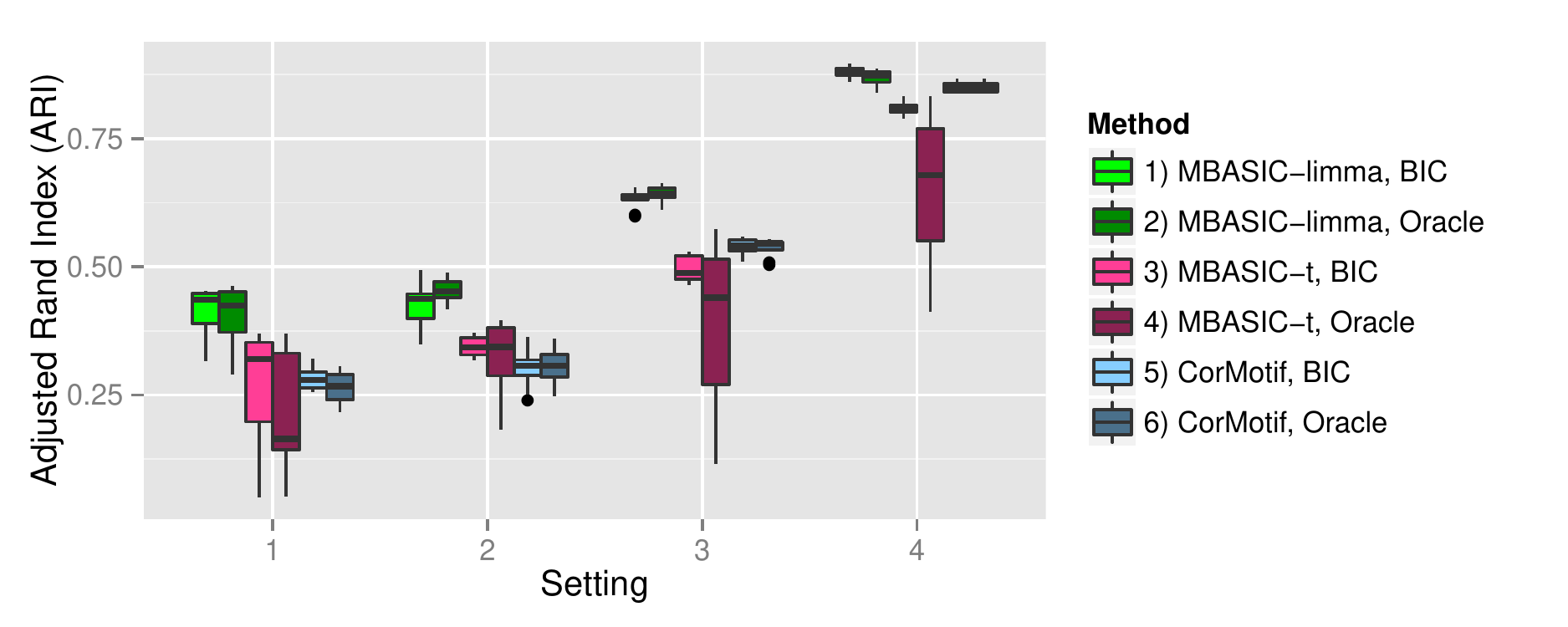}
\includegraphics[width=0.8\textwidth, page = 2]{simulation_cormotif_raw.pdf}
\includegraphics[width=0.8\textwidth, page = 4]{simulation_cormotif_raw.pdf}
\caption{\textit{Simulation Study 4, comparison between MBASIC and CorMotif.} We simulated data under four settings as in Table \ref{tbl:cormotif}. Results are summarized over 10 simulations under each setting.}\label{fig:cormotif1}
\end{figure}

\begin{figure}
\centering
\includegraphics[width=0.8\textwidth, page = 1]{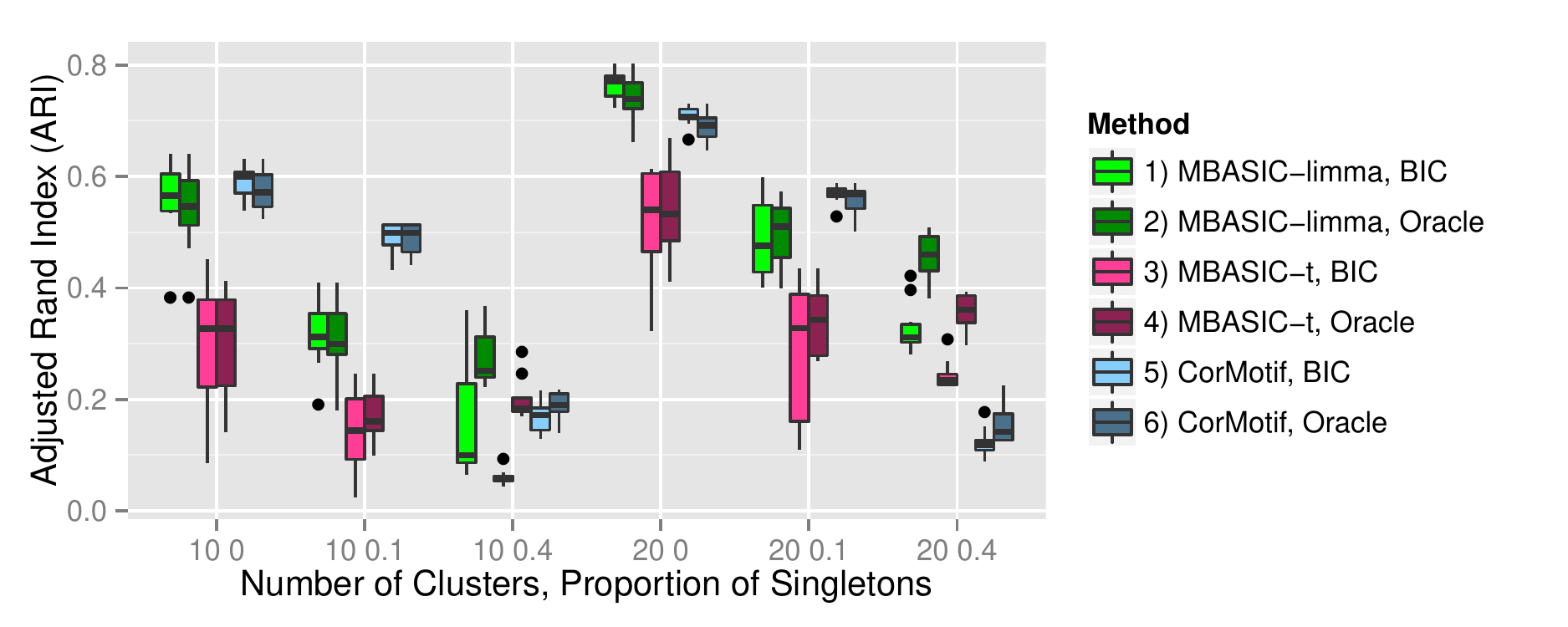}
\includegraphics[width=0.8\textwidth, page = 2]{simulation_cormotif.pdf}
\includegraphics[width=0.8\textwidth, page = 4]{simulation_cormotif.pdf}
\caption{\textit{Simulation Study 5, comparison between MBASIC and CorMotif.} We varied the number of clusters at 10, 20 and the proportion of singletons at 0, 0.1 and 0.4. Results are summarized over 10 simulations under each setting.}\label{fig:cormotif2}
\end{figure}

\begin{figure}
\centering
\begin{tabular}{cc}
Simulation Study 4, Setting 1 & Simulation Study 4, Setting 2\\
\includegraphics[width=0.45\textwidth]{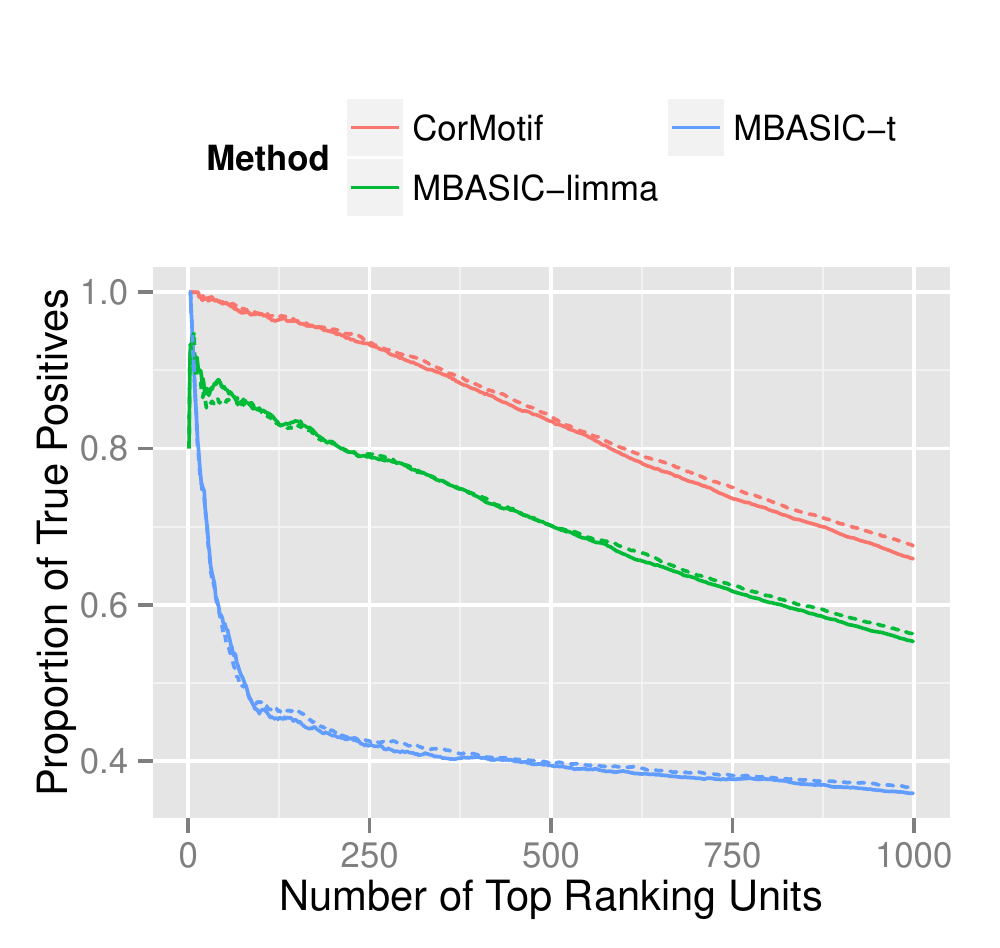}&
\includegraphics[width=0.45\textwidth]{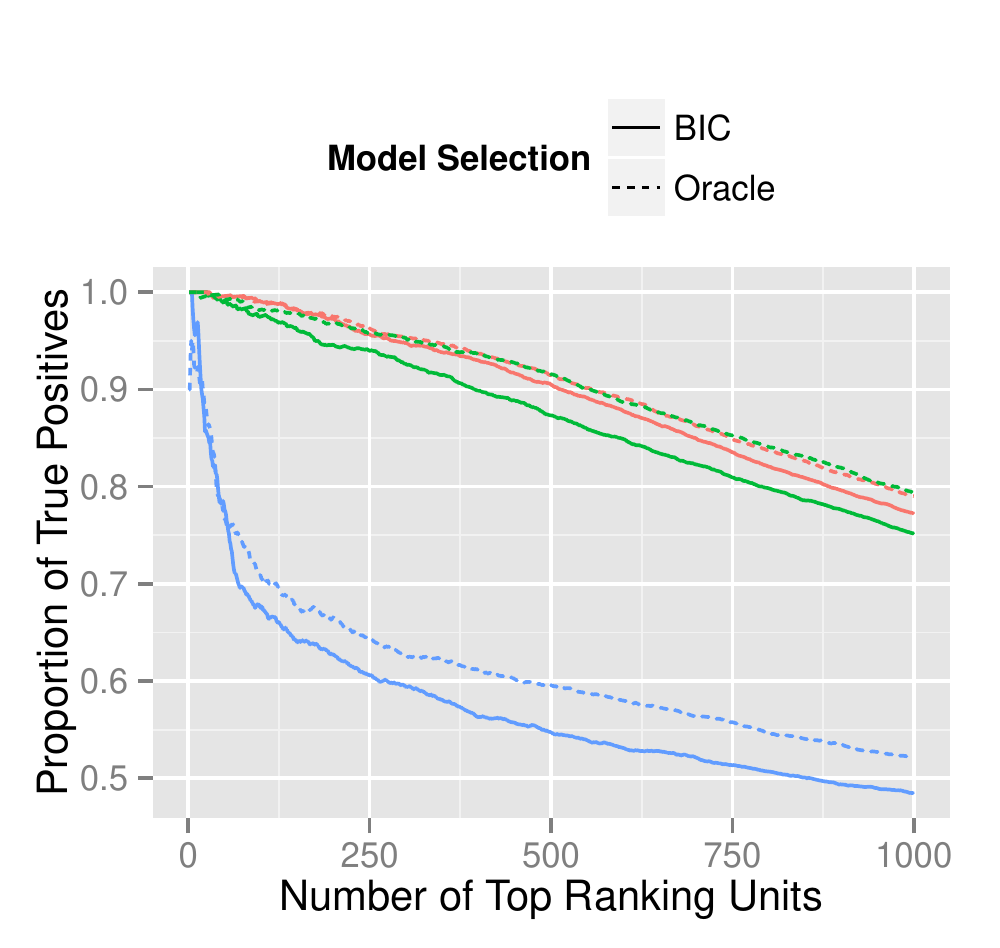}\\
Simulation Study 4, Setting 3 & Simulation Study 4, Setting 4\\
\includegraphics[width=0.45\textwidth]{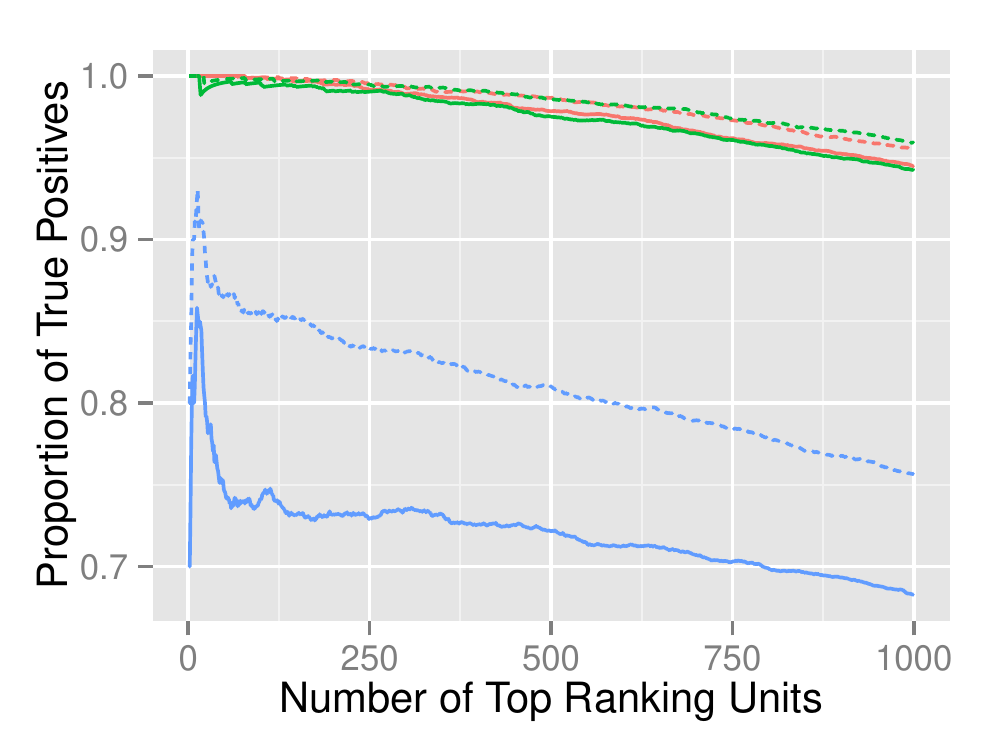}&
\includegraphics[width=0.45\textwidth]{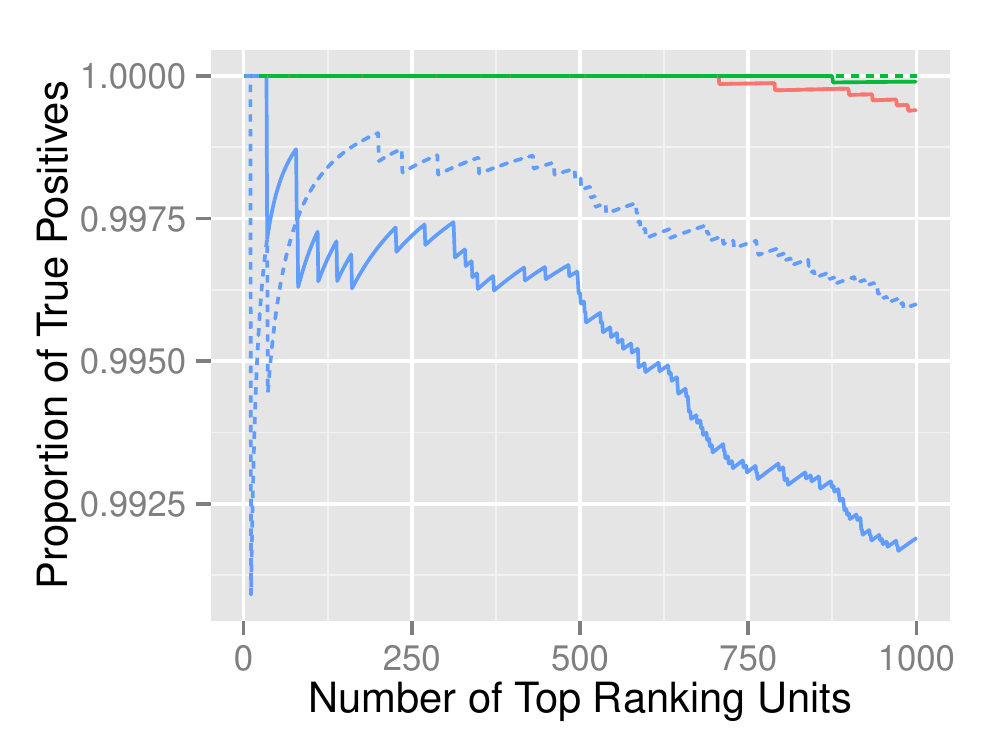}\\
\end{tabular}
\caption{\textit{Simulation Study 4, comparison between MBASIC and CorMotif.} The average true positive rate for the 10 simulations among the 1000 highest ranking   unit-experiment pairs for each of the four simulation settings in Table \ref{tbl:cormotif}. For each simulation, the ``true positive'' set consists of $(i,~k)$'s with $\theta_{ik}=2$, and the ranking is based on the posterior probability $P(\theta_{ik} =  2|Y)$. }\label{fig:tdr}
\end{figure}

\begin{table}
\centering
\caption{\textit{Simulation Studies 3-5.} A summary of the simulation designs, the fitting algorithms compared, and the figure numbers for the results.}\label{tbl:s345summary}
\begin{tabular}{llllll}
\hline
Study & $J$ & $\zeta$ & True model & Fitting algorithms & Related figures\\
\hline
3 & 10, 20 & 0, 0.1, 0.4 & iASeq & MBASIC, iASeq & Figure \ref{fig:iaseq}\\
4 & 4, 5 & 0 & CorMotif & MBASIC-limma, MBASIC-t, CorMotif & Figures \ref{fig:cormotif1}, \ref{fig:tdr}\\
5 & 10, 20 & 0, 0.1, 0.4 & CorMotif & MBASIC-limma, MBASIC-t, CorMotif & Figure \ref{fig:cormotif2}\\
\hline
\end{tabular}
\end{table}

Table \ref{tbl:s345summary} further summarizes the components of the Simulation Studies.  For each set of parameters, we simulated 10 data sets. We computed ARI, MSE-W, and SPE based on both the model with the number of clusters selected by BIC, and the oracle model where the number of clusters is set to its true value. The comparison between MBASIC and iASeq is shown in Figure \ref{fig:iaseq}. For all the different settings, MBASIC achieved better clustering performance, with higher ARI values.  However, iASeq performed better in SPE and MSE-W. When $\zeta=0$, iASeq performed overall better than MBASIC, with similar ARI values as MBASIC but much lower SPE. However, as $\zeta$ increased, iASeq's ARI value became significantly smaller than MBASIC, while its SPE value became closer to MBASIC's. In such cases, the benefits of modeling singletons seem to outweigh the loss of using simplified distributional assumptions.

The comparison between MBASIC and CorMotif is summarized in Figures \ref{fig:cormotif1} and \ref{fig:cormotif2}. In Simulation Study 4 (Figure \ref{fig:cormotif1}), because CorMotif models did not allow singletons,  we also excluded the singleton cluster in fitting MBASIC models. MBASIC-limma performed the best except in the first setting, where CorMotif achieved the best SPE. Figure \ref{fig:tdr}  depicts the average true positive rate in detecting states with $\theta_{ik}=2$ among the  1000 top ranking units for each of the four settings. In all but Setting 1, MBASIC-limma performed equally well as CorMotif. We note that Setting 1 has the fewest clusters $J=4$ and the fewest experimental conditions $K=4$, while the other settings have more complicated state-space structures. Performance of MBASIC-t was the worst in all the four settings. This suggests that neglecting the heterogeneity in these cases can significantly increase estimation error. Although MBASIC model alone does not address the heterogeneity issue, fitting MBASIC models after a data pre-processing step that incorporates the heterogeneity structure, such as computing $\tilde T_{ik}$ in MBASIC-limma, can significantly improve model inference. In Simulation Study 5 where we had stronger signals in separating distribution components but noisy state-space clusters, CorMotif resulted in the largest SPE values in all settings (Figure \ref{fig:cormotif2}). Although its performance in ARI was comparable with MBASIC-limma when $\zeta\leq 0.1$, it deteriorated with increasing proportion of singletons, i.e., $\zeta=0.4$.  Simulation Studies 4 and 5 collectively suggest that MBASIC's performance is competitive with CorMotif in  settings where we have less noise in clustering structure, small numbers of  clusters, and some level of singletons despite the fact that the distributional assumptions of MBASIC might be mis-specified. This indicates that for  real data sets where we are agnostic about the true data generating structure,  MBASIC might be a  more general and robust approach. 

\subsection{Simulation Study 6: Weak Clusters}

\cite{cormotif} pointed out that state-space clustering without accommodating singletons can lead to merging of small clusters with large clusters with distinct state-space profiles. 
Such a phenomenon may alter the interpretation of the $W$ matrix, because each column may represent the average state-space pattern of several small clusters that lack the data support. It is therefore important to investigate whether such a phenomenon still exists for MBASIC where we include a singleton cluster.

\begin{figure}
\centering
\begin{tabular}{cc}
(a) Simulations 1-3 & (b) Simulations 4-6\\
\includegraphics[width=0.48\textwidth]{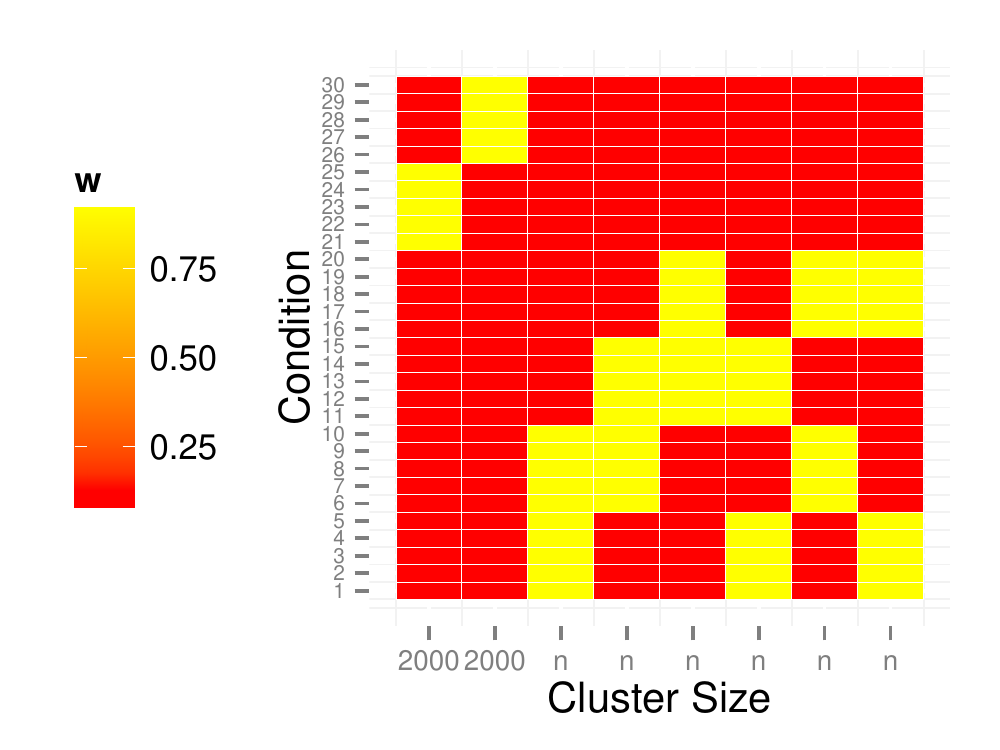} &
\includegraphics[width=0.48\textwidth]{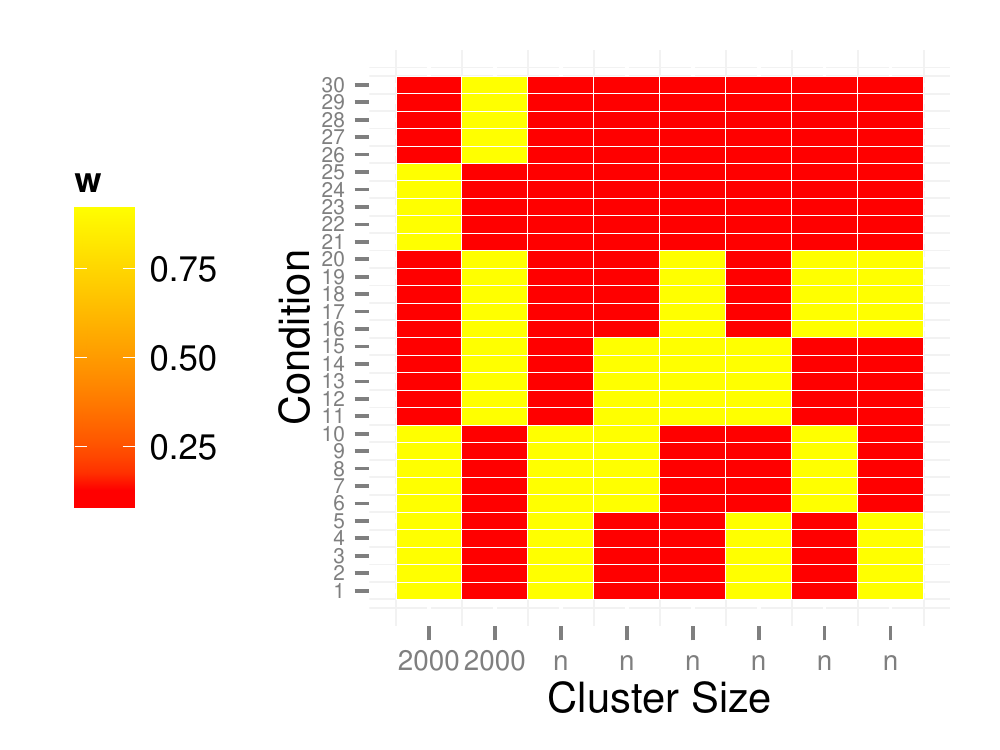} \\
\end{tabular}
\caption{\textit{Simulation Study 6.} Two settings of the true cluster patterns, represented by the matrix $(w_{jk2})_{1\leq j\leq 8,~1\leq k\leq 30}^T$.}\label{fig:s6true}
\end{figure}

We conducted six simulations in Simulation Study 6 to investigate this issue. We simulated data according to the log-normal distribution with $S=2$ states, $J=8$ clusters, and $K=30$ conditions. The number of replicates within each condition, as well as the distribution parameters within each state is the same as in Section \ref{sec:simsetting}. We set the sizes of the first two clusters as 2000, and varied the size of each of the other six clusters, that is $n_{\textnormal{small}}$, as 20 or 100. To vary the level of state-space similarity between the two big clusters and the small clusters, we had two settings for the state-space pattern as shown in Figure \ref{fig:s6true}. For Simulations 1-3, the conditions in which the small clusters have state $s=2$ are distinct from the two big clusters, while for Simulations 4-6, the patterns between the small and large clusters are more similar.  To control these cluster patterns, we set $w_{jks}\in \{0.1,~0.9\}$. Finally, we included $n_{\textnormal{singleton}}=$ 0 or 2000 singletons in each simulated data set. The states for the singleton units were generated the same as in Section \ref{sec:simsetting}.

\begin{figure}
\centering
\begin{tabularx}{\textwidth}{p{0.75in}>{\centering\arraybackslash}X>{\centering\arraybackslash}X}
&(a) MBASIC & (b) MBASIC0\\
\parbox[c]{\hsize}{$n_{\textnormal{small}}=20$\\ $n_{\textnormal{singleton}}=0$} &
\parbox[c]{\hsize}{\includegraphics[width=0.4\textwidth,page=1]{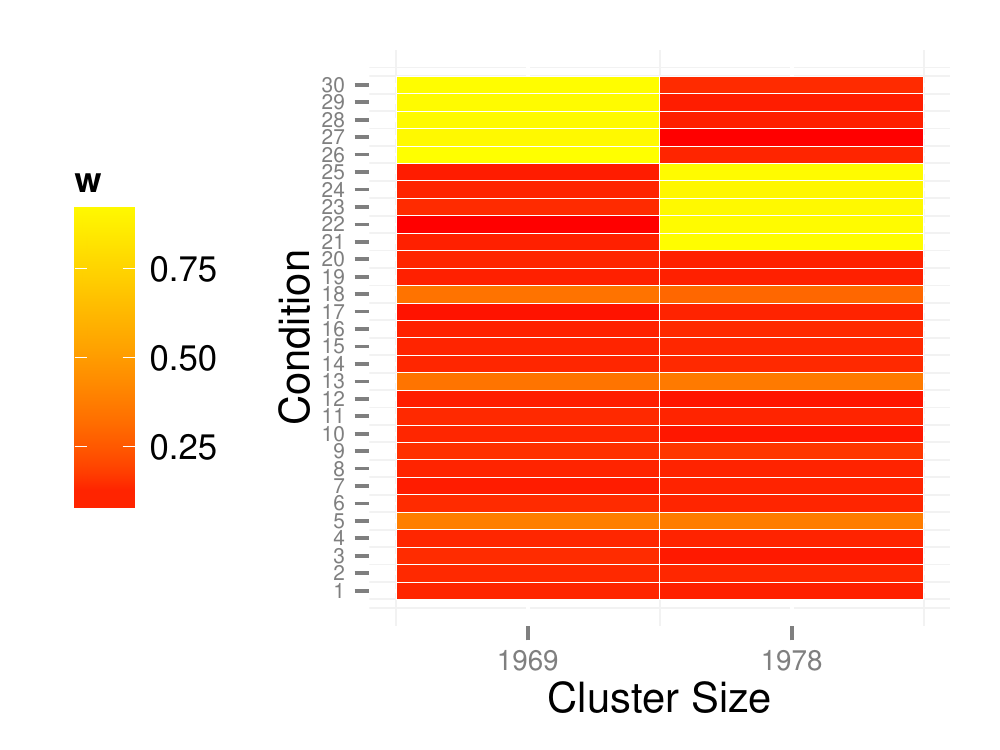}} &
\parbox[c]{\hsize}{\includegraphics[width=0.4\textwidth,page=2]{W_est_K_30_nc_20_ns_0_set1.pdf}} \\
&(c) MBASIC & (d) MBASIC0\\
\parbox[c]{\hsize}{$n_{\textnormal{small}}=20$\\ $n_{\textnormal{singleton}}=2000$} &
\parbox[c]{\hsize}{\includegraphics[width=0.4\textwidth,page=1]{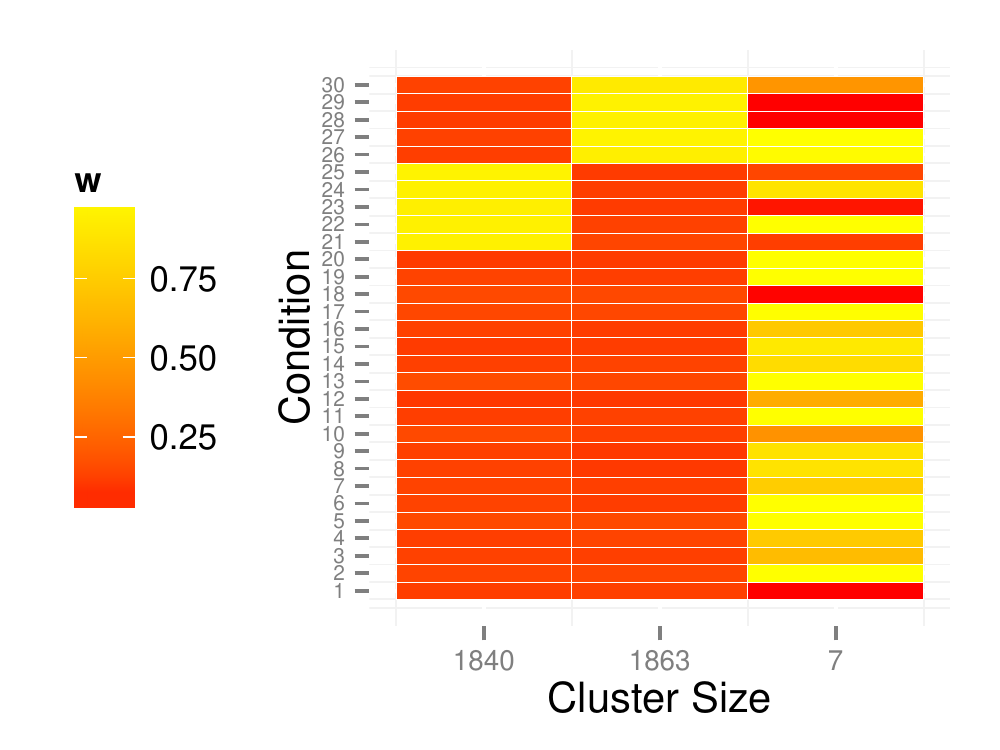}} &
\parbox[c]{\hsize}{\includegraphics[width=0.4\textwidth,page=2]{W_est_K_30_nc_20_ns_2000_set1.pdf}} \\
&(e) MBASIC & (f) MBASIC0\\
\parbox[c]{\hsize}{$n_{\textnormal{small}}=100$\\ $n_{\textnormal{singleton}}=0$} &
\parbox[c]{\hsize}{\includegraphics[width=0.4\textwidth,page=1]{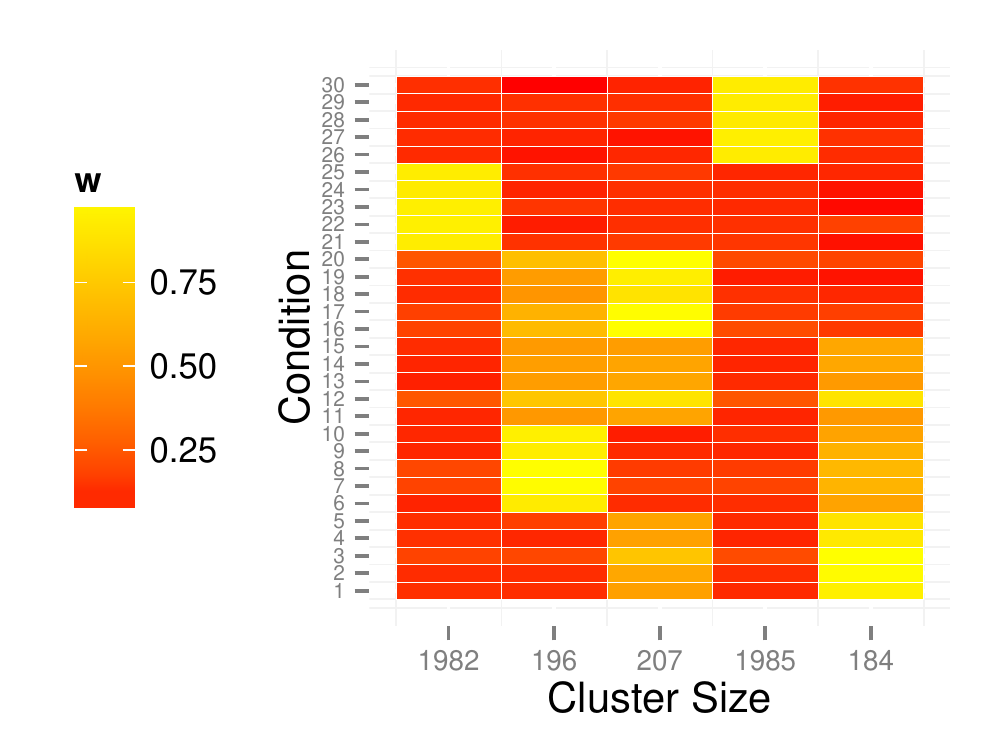}} &
\parbox[c]{\hsize}{\includegraphics[width=0.4\textwidth,page=2]{W_est_K_30_nc_100_ns_0_set1.pdf}} \\
\end{tabularx}
\caption{\textit{Simulation Study 6, Simulations 1-3.} Estimated cluster patterns by (a, c, e) MBASIC and (b, d, f) MBASIC0. The true clustering pattern is shown in Figure \ref{fig:s6true}(a).}\label{fig:s6sim1}
\end{figure}

\begin{figure}
\centering
\begin{tabularx}{\textwidth}{p{0.75in}>{\centering\arraybackslash}X>{\centering\arraybackslash}X}
&(a) MBASIC & (b) MBASIC0\\
\parbox[c]{\hsize}{$n_{\textnormal{small}}=20$\\ $n_{\textnormal{singleton}}=0$} &
\parbox[c]{\hsize}{\includegraphics[width=0.4\textwidth,page=1]{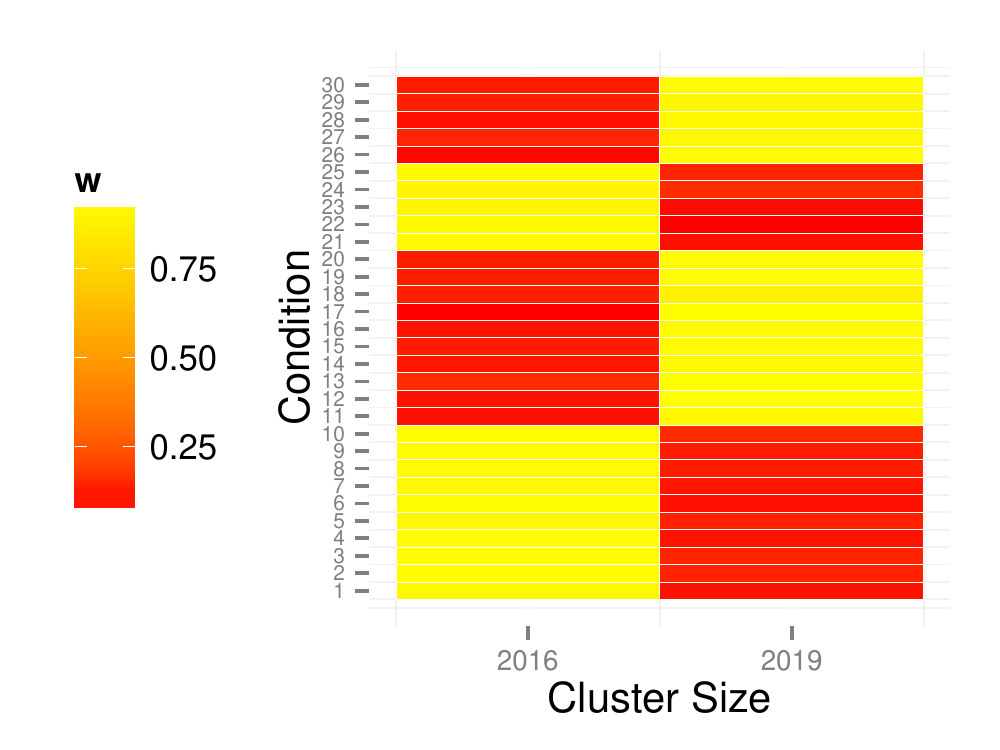}} &
\parbox[c]{\hsize}{\includegraphics[width=0.4\textwidth,page=2]{W_est_K_30_nc_20_ns_0_set2.pdf}} \\
&(c) MBASIC & (d) MBASIC0\\
\parbox[c]{\hsize}{$n_{\textnormal{small}}=20$\\ $n_{\textnormal{singleton}}=2000$} &
\parbox[c]{\hsize}{\includegraphics[width=0.4\textwidth,page=1]{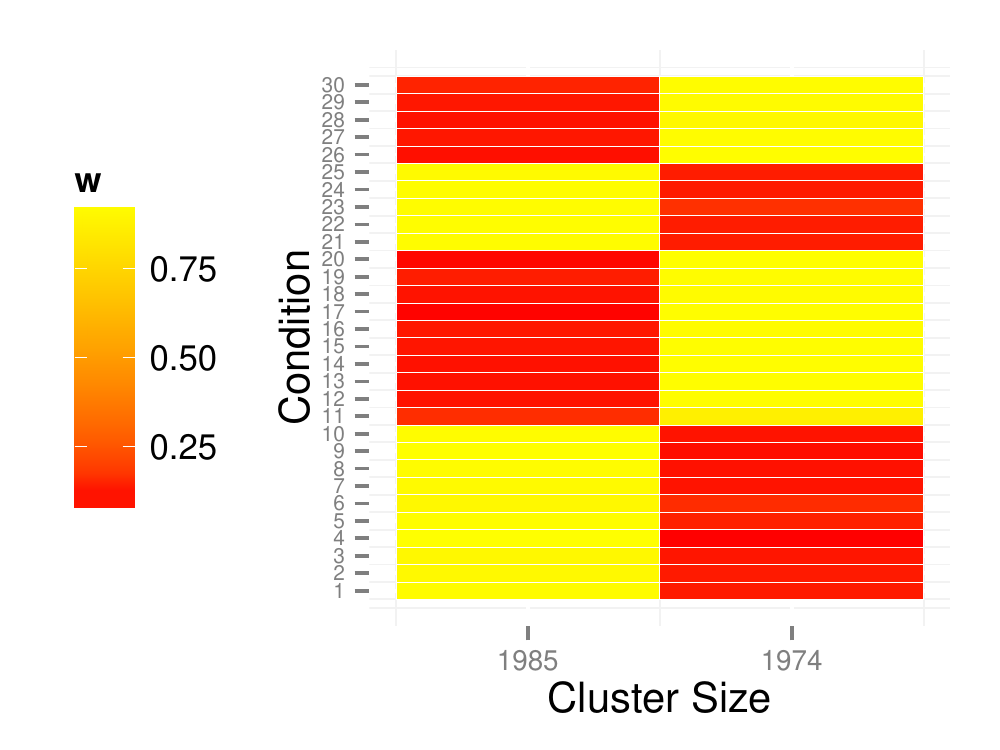}} &
\parbox[c]{\hsize}{\includegraphics[width=0.4\textwidth,page=2]{W_est_K_30_nc_20_ns_2000_set2.pdf}} \\
&(e) MBASIC & (f) MBASIC0\\
\parbox[c]{\hsize}{$n_{\textnormal{small}}=100$\\ $n_{\textnormal{singleton}}=0$} &
\parbox[c]{\hsize}{\includegraphics[width=0.4\textwidth,page=1]{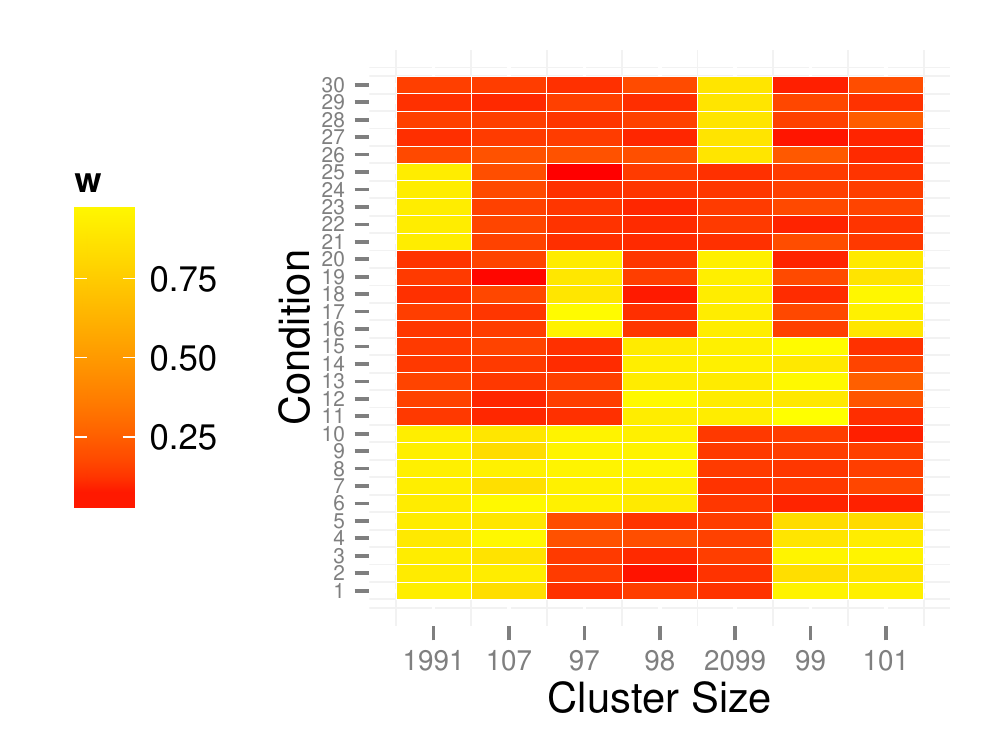}} &
\parbox[c]{\hsize}{\includegraphics[width=0.4\textwidth,page=2]{W_est_K_30_nc_100_ns_0_set2.pdf}} \\
\end{tabularx}
\caption{\textit{Simulation Study 6, Simulations 4-6.} Estimated cluster patterns by (a, c, e) MBASIC and (b, d, f) MBASIC0. The true clustering pattern is shown in Figure \ref{fig:s6true}(b).}\label{fig:s6sim2}
\end{figure}

\begin{table}
\centering
\caption{\textit{Simulation Study 6, Simulations 1-3.} Confusion matrix between the true clusters and the estimated clusters. The true cluster pattern is shown in Figure \ref{fig:s6true}(a).}\label{tbl:confusionmat1}
\begin{tabularx}{\textwidth}{>{\centering\arraybackslash}X>{\centering\arraybackslash}X}
\multicolumn{2}{c}{Simulation 1, $n_{\textnormal{small}}=20$, $n_{\textnormal{singleton}}=0$}\\
\parbox[c]{\hsize}{\centering\vspace{0.2in}
\begin{tabular}{c|ccc}
   & \multicolumn{3}{c}{MBASIC}\\
True & 0 & 1 & 2\\\hline
1 & 23 & 2 & 1975 \\
2 & 30 & 1967 & 3 \\
3 & 20 & 0 & 0 \\
4 & 20 & 0 & 0 \\
5 & 20 & 0 & 0 \\
6 & 20 & 0 & 0 \\
7 & 20 & 0 & 0 \\
8 & 20 & 0 & 0\\
\end{tabular}\vspace{0.2in}
}
&
\parbox[c]{\hsize}{\centering\vspace{0.2in}
\begin{tabular}{c|ccc}
 & \multicolumn{3}{c}{MBASIC0}\\
True & 1 & 2 & 3\\\hline
1 & 1998 & 2 & 0 \\
2 & 4 & 1995 & 1 \\
3 & 0 & 1 & 19 \\
4 & 0 & 1 & 19 \\
5 & 0 & 0 & 20 \\
6 & 0 & 1 & 19 \\
7 & 0 & 0 & 20 \\
8 & 0 & 1 & 19\\
\end{tabular}\vspace{0.2in}
}
\\
\multicolumn{2}{c}{Simulation 2, $n_{\textnormal{small}}=20$, $n_{\textnormal{singleton}}=2000$} \\
\parbox[c]{\hsize}{\centering\vspace{0.2in}
\begin{tabular}{c|cccc}
  & \multicolumn{4}{c}{MBASIC}\\
 True & 0 & 1 & 2 & 3\\\hline
0 & 1957 & 19 & 17 & 7 \\
1 & 180 & 1818 & 2 & 0 \\
2 & 153 & 3 & 1844 & 0 \\
3 & 20 & 0 & 0 & 0 \\
4 & 20 & 0 & 0 & 0 \\
5 & 20 & 0 & 0 & 0 \\
6 & 20 & 0 & 0 & 0 \\
7 & 20 & 0 & 0 & 0 \\
8 & 20 & 0 & 0 & 0\\
\end{tabular}\vspace{0.2in}}
&
\parbox[c]{\hsize}{\centering\vspace{0.2in}
\begin{tabular}{c|cccccc}
  & \multicolumn{6}{c}{MBASIC0}\\
True  & 1 & 2 & 3 & 4 & 5 & 6\\\hline
0 & 67 & 372 & 620 & 59 & 427 & 455 \\
1 & 1957 & 25 & 0 & 4 & 0 & 14 \\
2 & 8 & 23 & 3 & 1950 & 0 & 16 \\
3 & 0 & 0 & 1 & 0 & 0 & 19 \\
4 & 0 & 0 & 0 & 0 & 0 & 20 \\
5 & 0 & 0 & 1 & 0 & 0 & 19 \\
6 & 0 & 0 & 1 & 0 & 0 & 19 \\
7 & 0 & 0 & 1 & 0 & 0 & 19 \\
8 & 0 & 0 & 1 & 0 & 0 & 19\\
\end{tabular}\vspace{0.2in}}
\\
\multicolumn{2}{c}{Simulation 3, $n_{\textnormal{small}}=100$, $n_{\textnormal{singleton}}=0$} \\
\parbox[c]{\hsize}{\centering\vspace{0.2in}
\begin{tabular}{c|cccccc}
 & \multicolumn{6}{c}{MBASIC}\\
True  & 0 & 1 & 2 & 3 & 4 & 5\\\hline
1 & 21 & 1974 & 0 & 0 & 5 & 0 \\
2 & 12 & 7 & 0 & 1 & 1980 & 0 \\
3 & 1 & 1 & 9 & 0 & 0 & 89 \\
4 & 5 & 0 & 89 & 0 & 0 & 6 \\
5 & 6 & 0 & 2 & 92 & 0 & 0 \\
6 & 1 & 0 & 0 & 13 & 0 & 86 \\
7 & 0 & 0 & 96 & 4 & 0 & 0 \\
8 & 0 & 0 & 0 & 97 & 0 & 3\\
\end{tabular}\vspace{0.2in}}
&
\parbox[c]{\hsize}{\centering\vspace{0.2in}
\begin{tabular}{c|cccc}
  & \multicolumn{4}{c}{MBASIC0}\\
True & 1 & 2 & 3 & 4\\\hline
1 & 1993 & 0 & 7 & 0 \\
2 & 7 & 1 & 1991 & 1 \\
3 & 1 & 0 & 0 & 99 \\
4 & 0 & 2 & 0 & 98 \\
5 & 0 & 99 & 0 & 1 \\
6 & 0 & 6 & 0 & 94 \\
7 & 0 & 100 & 0 & 0 \\
8 & 0 & 99 & 0 & 1\\
\end{tabular}\vspace{0.2in}}\\
\end{tabularx}
\end{table}

\begin{table}
\centering
\caption{\textit{Simulation Study 6, Simulations 4-6.} Confusion matrix between the true clusters and the estimated clusters. The true cluster pattern is shown in Figure \ref{fig:s6true}(b).}\label{tbl:confusionmat2}
\begin{tabularx}{1.1\textwidth}{>{\centering\arraybackslash}X>{\centering\arraybackslash}X}
\multicolumn{2}{c}{Simulation 4, $n_{\textnormal{small}}=20$, $n_{\textnormal{singleton}}=0$}\\
\parbox[c]{\hsize}{\centering
\vspace{0.2in}
\begin{tabular}{c|ccc}
\hline 
 & \multicolumn{3}{c}{MBASIC}\\
True & 0 & 1 & 2\\\hline
1 & 5 & 1995 & 0 \\
2 & 1 & 0 & 1999 \\
3 & 1 & 19 & 0 \\
4 & 19 & 1 & 0 \\
5 & 1 & 0 & 19 \\
6 & 20 & 0 & 0 \\
7 & 18 & 1 & 1 \\
8 & 20 & 0 & 0\\
\hline
\end{tabular}\vspace{0.2in}}
&
\parbox[c]{\hsize}{\centering
\vspace{0.2in}\begin{tabular}{c|ccc}
\hline 
 & \multicolumn{3}{c}{MBASIC0}\\
True & 1 & 2 & 3\\\hline
1 & 1999 & 1 & 0 \\
2 & 0 & 0 & 2000 \\
3 & 17 & 3 & 0 \\
4 & 0 & 20 & 0 \\
5 & 0 & 5 & 15 \\
6 & 0 & 20 & 0 \\
7 & 0 & 19 & 1 \\
8 & 1 & 19 & 0\\
\hline
\end{tabular}\vspace{0.2in}}\\
\multicolumn{2}{c}{Simulation 5, $n_{\textnormal{small}}=20$, $n_{\textnormal{singleton}}=2000$}\\
\parbox[c]{\hsize}{\centering
\vspace{0.2in}\begin{tabular}{c|ccc}
\hline 
 & \multicolumn{3}{c}{MBASIC}\\
True & 0 & 1 & 2\\\hline
0 & 1986 & 7 & 7 \\
1 & 31 & 1969 & 0 \\
2 & 45 & 0 & 1955 \\
3 & 11 & 9 & 0 \\
4 & 20 & 0 & 0 \\
5 & 8 & 0 & 12 \\
6 & 20 & 0 & 0 \\
7 & 20 & 0 & 0 \\
8 & 20 & 0 & 0\\
\hline
\end{tabular}\vspace{0.2in}}
&
\parbox[c]{\hsize}{\centering
\vspace{0.2in}\begin{tabular}{c|cccccc}
\hline 
 & \multicolumn{6}{c}{MBASIC0}\\
True & 1 & 2 & 3 & 4 & 5 & 6\\\hline
0 & 16 & 393 & 457 & 561 & 560 & 13 \\
1 & 1990 & 0 & 0 & 5 & 5 & 0 \\
2 & 0 & 0 & 0 & 5 & 12 & 1983 \\
3 & 12 & 0 & 0 & 1 & 7 & 0 \\
4 & 0 & 0 & 0 & 1 & 19 & 0 \\
5 & 0 & 0 & 0 & 0 & 6 & 14 \\
6 & 0 & 0 & 0 & 0 & 20 & 0 \\
7 & 0 & 0 & 0 & 0 & 20 & 0 \\
8 & 0 & 0 & 0 & 1 & 19 & 0\\
\hline
\end{tabular}\vspace{0.2in}}\\
\multicolumn{2}{c}{Simulation 6, $n_{\textnormal{small}}=100$, $n_{\textnormal{singleton}}=0$}\\
\parbox[c]{\hsize}{\centering
\vspace{0.2in}\begin{tabular}{c|cccccccc}
\hline 
 & \multicolumn{8}{c}{MBASIC}\\
True & 0 & 1 & 2 & 3 & 4 & 5 & 6 & 7\\\hline
1 & 3 & 1983 & 14 & 0 & 0 & 0 & 0 & 0 \\
2 & 1 & 0 & 0 & 0 & 0 & 1999 & 0 & 0 \\
3 & 0 & 8 & 92 & 0 & 0 & 0 & 0 & 0 \\
4 & 1 & 0 & 0 & 1 & 98 & 0 & 0 & 0 \\
5 & 1 & 0 & 0 & 0 & 0 & 99 & 0 & 0 \\
6 & 1 & 0 & 0 & 0 & 0 & 0 & 99 & 0 \\
7 & 1 & 0 & 1 & 96 & 0 & 1 & 0 & 1 \\
8 & 0 & 0 & 0 & 0 & 0 & 0 & 0 & 100\\
\hline
\end{tabular}\vspace{0.2in}}
&
\parbox[c]{\hsize}{\centering
\vspace{0.2in}\begin{tabular}{c|ccccccc}
\hline 
 & \multicolumn{7}{c}{MBASIC0}\\
True & 1 & 2 & 3 & 4 & 5 & 6 & 7\\\hline
1 & 1976 & 24 & 0 & 0 & 0 & 0 & 0 \\
2 & 0 & 0 & 0 & 0 & 2000 & 0 & 0 \\
3 & 7 & 93 & 0 & 0 & 0 & 0 & 0 \\
4 & 0 & 0 & 99 & 0 & 0 & 1 & 0 \\
5 & 0 & 0 & 0 & 0 & 99 & 0 & 1 \\
6 & 0 & 0 & 1 & 0 & 0 & 0 & 99 \\
7 & 0 & 2 & 1 & 1 & 1 & 95 & 0 \\
8 & 0 & 0 & 0 & 100 & 0 & 0 & 0\\
\hline
\end{tabular}\vspace{0.2in}}\\
\end{tabularx}
\end{table}

\begin{table}
\centering
\caption{\textit{Simulation Study 6.} ARI, MSE-W, and SPE in all simulations.}\label{tbl:s6metric}
\begin{tabular}{ccc|ccc|ccc}
\hline
&&& \multicolumn{3}{c|}{MBASIC} & \multicolumn{3}{c}{MBASIC0}\\
Simulation & $n_{\textnormal{small}}$ & $n_{\textnormal{singleton}}$ & ARI & MSE-W & SPE & ARI & MSE-W & SPE\\
\hline
1 & 20 & 0 & 0.967 & 0.433 & 0.185 & 0.991 & 0.29 & 0.189 \\
2 & 20 & 2000 & 0.79 & 0.442 & 0.192 & 0.727 & 0.34 & 0.193\\
3 & 100 & 0 & 0.969 & 0.203 & 0.187 & 0.975 & 0.233 & 0.193\\
4 & 20 & 0 & 0.977 & 0.384 & 0.169 & 0.983 & 0.260 & 0.167\\
5 & 20 & 2000 & 0.926 &	0.384 &	0.185 & 0.773 &	0.333 &	0.184\\
6 & 100	& 0 & 0.949 & 0.087 & 0.172 & 0.947 & 0.087 & 0.171\\
\hline
\end{tabular}
\end{table}

In each simulation, we fitted the data using MBASIC and MBASIC0 with BIC to select the number of clusters. MBASIC0 differs from MBASIC only by the exclusion of the singleton feature. Therefore, comparing these two methods allow us to assess how fitting a singleton cluster may affect the small cluster merging problem. Tables \ref{tbl:confusionmat1} and \ref{tbl:confusionmat2} compare the confusion matrices between the fitted and the true clusters. We also display the state-space patterns of the fitted models in Figures \ref{fig:s6sim1} and \ref{fig:s6sim2}. In Simulations 1 and 4, with $n_{\textnormal{small}}=20$ units in each of the small clusters, MBASIC classified the units of small clusters as singletons, while MBASIC0 merged them to form a spurious cluster. The state-space pattern estimated by MBASIC represented the two real big clusters. When the data included singletons, as in Simulations 2 and 5, MBASIC0 formed more spurious clusters, while MBASIC continued to allocate the small clusters as singletons. When we had $n_{\textnormal{small}}=100$ units in each of the small clusters, both methods identified these small cluster patterns in Simulation 6 (Figure \ref{fig:s6sim2}), but formed spurious clusters in Simulation 3 (Figure \ref{fig:s6sim1}). We compare the resulting ARI, MSE-W, and SPE between MBASIC and MBASIC0 in Table \ref{tbl:s6metric}. Performances of these two methods are close when we have no singletons, but differentiate otherwise.  Based on these simulations, we conclude that fitting a singleton cluster can substantially avoid merging weak clusters. The state-space patterns estimated by the $W$ matrix are more likely to reflect true underlying clusters rather than the average of several small clusters. We acknowledge that how well modeling the singletons can avoid merging weak clusters requires further investigation in more dynamic settings as we vary the similarity among clusters, the difference among the states, as well as other variables that influence cluster structures such as $J$, $K$, $S$. We leave such potential investigations as future research.

\clearpage
\section{Additional Tables and Figures}\label{sec:add}

\begin{figure}[hbtp]
\centering
\begin{tabular}{ccc}
\includegraphics[width=0.8\textwidth, page = 1]{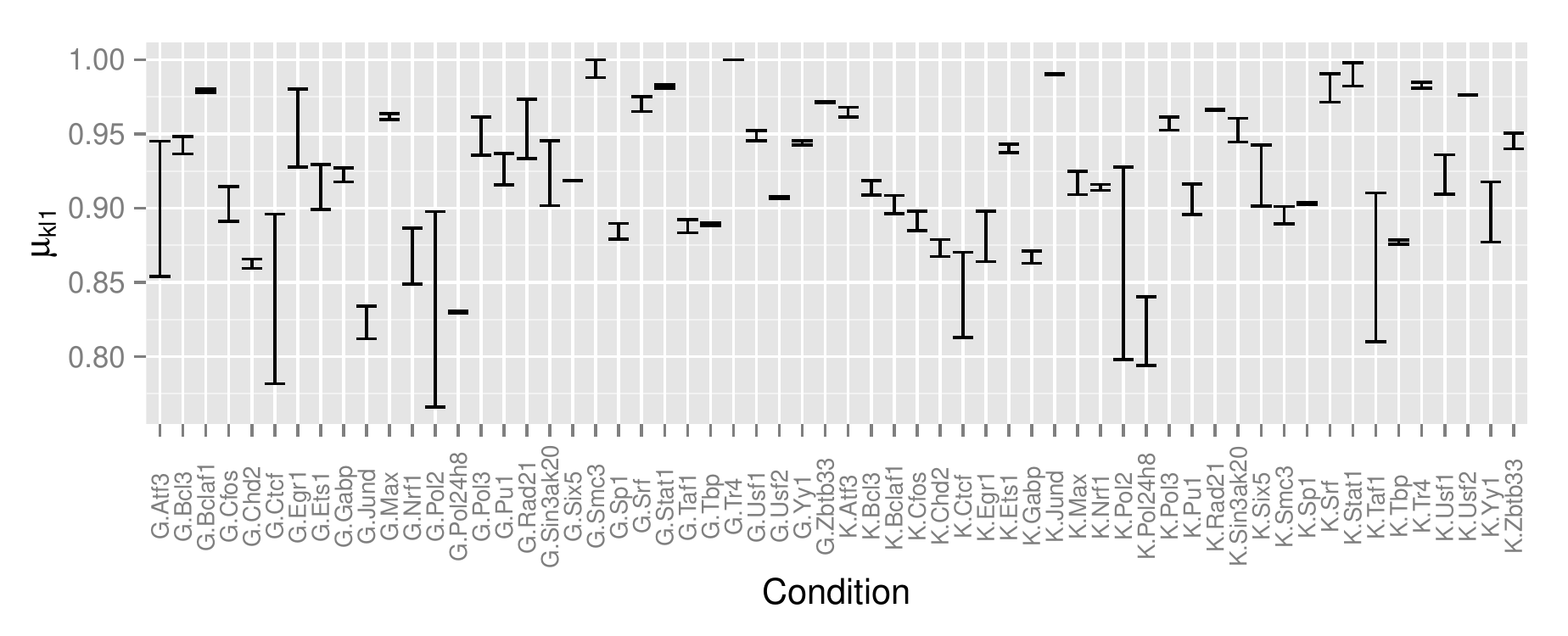}\\
\includegraphics[width=0.8\textwidth, page = 2]{par_gene.pdf}\\
\includegraphics[width=0.8\textwidth, page = 3]{par_gene.pdf}\\
\includegraphics[width=0.8\textwidth, page = 4]{par_gene.pdf}\\
\end{tabular}
\caption{The range of the estimated parameters $\mu_{kls}$ and $\sigma_{kls}$ among the different replicates under the same experimental condition for the transcription factor enrichment network data in Section 4.1.}\label{fig:par_gene}
\end{figure}

\begin{figure}[hbtp]
\centering
\begin{tabular}{ll}
(a) & (b) \\
\includegraphics[width=0.45\textwidth,page=1]{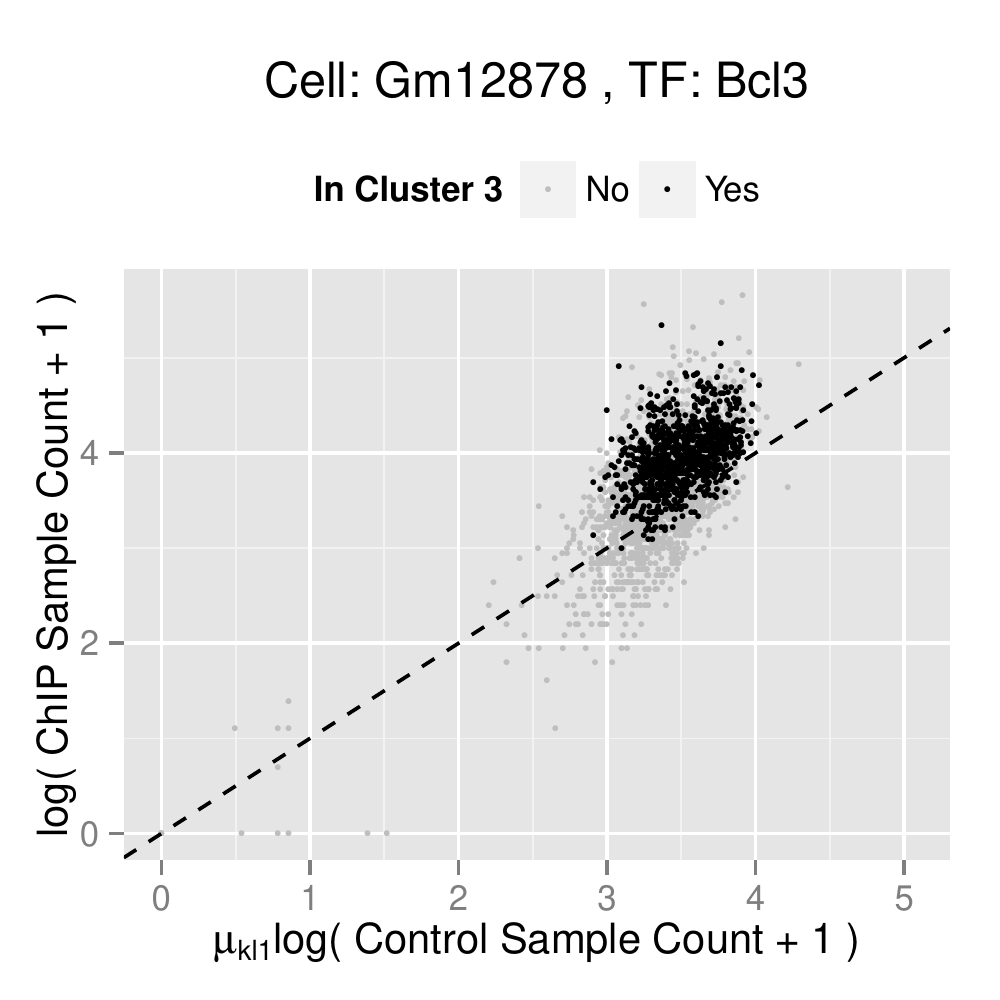} &
\includegraphics[width=0.45\textwidth,page=2]{data_compare_Gm12878_Bcl3_Bclaf1m33_c3.pdf}\\
\end{tabular}
\caption{Plots of the transformed ChIP sample read counts against the transformed control sample read counts for all units in the Gm12878 cell for (a) Bcl3 and (b) Bclaf1. Data from unenriched units are expected to locate around the 45 degree dashed line.}\label{fig:bcl}
\end{figure}

\begin{figure}[hbtp]
\centering
\begin{tabular}{ll}
(a) & (b)\\
\includegraphics[width=0.45\textwidth,page=1]{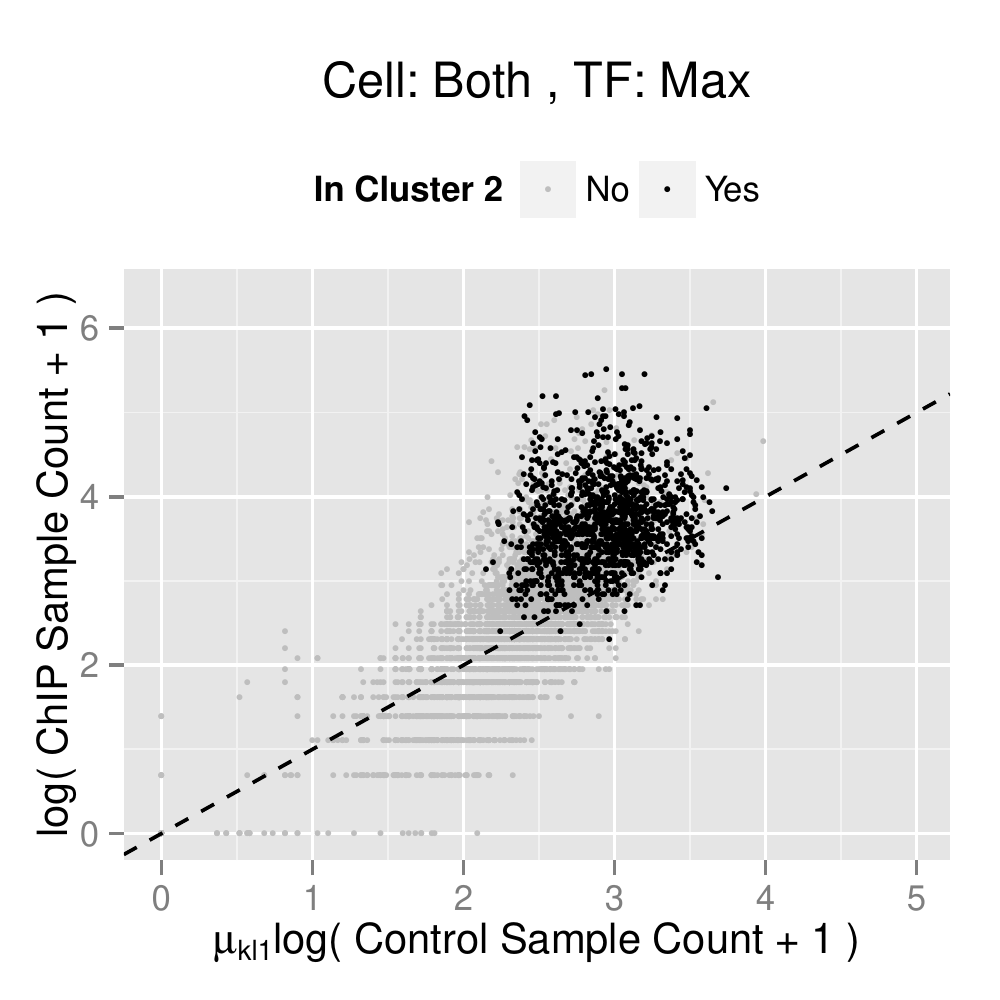}&
\includegraphics[width=0.45\textwidth,page=2]{data_compare_Both_Max_Usf1_c2.pdf}\\
\end{tabular}
\caption{Plots of the transformed ChIP sample read counts against the transformed control sample read counts for all units in both Gm12878 and K562 cells for (a) Max and (b) Usf1. Data from unenriched units are expected to locate around the 45 degree dashed line.}\label{fig:maxusf1}
\end{figure}

\begin{table}[hb]
\centering
\caption{Enriched cell type-TF combination for each cluster  in the TF enrichment network analysis of Section 4.1 of the main text. TFs with estimated enrichment probability $>95\%$ are listed for each cluster.}\label{tbl:profile}
\begin{tabular}{p{0.07\textwidth}|p{0.07\textwidth}|p{0.27\textwidth}|p{0.27\textwidth}|p{0.27\textwidth} }
\hline
Cluster & \# of Loci & Common TF & Gm12878 Specific & K562 Specific \\
\hline 
 1  &  34  &  Bcl3, Max, Sp1, Taf1, Zbtb33 & Atf3, Ets1, Jund, Nrf1, Pol24h8, Sin3ak20, Tr4 & Egr1, Pu1, Rad21, Usf2 \\ \hline 2  &  317  &  Bcl3, Chd2, Max, Pol24h8, Sp1, Taf1 & Atf3, Ets1, Nrf1, Sin3ak20, Usf2 & Bclaf1, Egr1, Pu1, Smc3, Tbp, Zbtb33 \\ \hline 3  &  490  &  Ets1, Pol2, Pol24h8, Sin3ak20, Taf1, Tbp, Usf1 & Atf3, Chd2, Nrf1, Usf2 & Bcl3, Bclaf1, Max, Sp1 \\ \hline 4  &  555  &  Bcl3, Bclaf1, Chd2, Pol2, Pol24h8, Sp1, Taf1, Tbp & Atf3, Ets1, Nrf1, Sin3ak20, Six5, Smc3 & Pu1 \\ \hline 5  &  428  &  Bcl3, Chd2, Ets1, Pol24h8, Sp1, Taf1 & Atf3, Ctcf, Nrf1, Rad21, Sin3ak20 & Bclaf1, Pol2, Smc3, Tbp \\ \hline 6  &  729  &  Bcl3, Chd2, Ets1, Pol2, Pol24h8, Sin3ak20, Smc3, Sp1, Taf1 & Atf3, Nrf1 & Bclaf1, Egr1, Tbp \\ \hline 7  &  391  &   &  & Bcl3, Bclaf1, Chd2, Egr1, Ets1, Smc3, Sp1 \\ \hline 8  &  133  &  Ctcf & Atf3, Chd2, Rad21 & Smc3, Srf \\ \hline 9  &  146  &  Ctcf &  & Bclaf1, Pol24h8, Smc3, Tbp \\ \hline 10  &  469  &  Pol2, Pol24h8, Taf1 & Smc3 & Bclaf1, Ets1, Sin3ak20, Tbp \\ \hline 11  &  440  &  Gabp, Pol24h8, Taf1 & Ets1, Smc3 & Pol2 \\ \hline 12  &  184  &   & Bcl3, Chd2, Ets1, Nrf1, Pol24h8, Taf1 &  \\ \hline 13  &  277  &  Chd2, Pol2, Pol24h8, Sp1, Taf1, Tbp & Smc3 & Cfos \\ \hline 14  &  156  &  Chd2, Pol2, Pol24h8, Tbp, Zbtb33 & Ets1, Smc3, Taf1 &  \\ \hline 15  &  412  &  Pol2, Pol24h8 &  &  \\ \hline 16  &  327  &   & Pol24h8 &  \\ \hline 17  &  213  &  Usf1 & Atf3, Usf2 & Max \\ \hline 18  &  241  &  Six5 & Ets1, Smc3 &  \\ \hline 19  &  187  &  Chd2, Sp1 &  & Cfos \\ \hline 20  &  222  &   &  & Ets1 \\ \hline 21  &  385  &   &  & Pol24h8 \\ \hline 22  &  343  &  Ctcf &  & Smc3 \\ \hline 23  &  449  &   & Nrf1, Smc3 &  \\ \hline 24  &  1674  &   &  &  \\ \hline
\end{tabular}
\end{table}

\begin{figure}[hbtp]
\centering
\includegraphics[ width = \textwidth, page = 3 ]{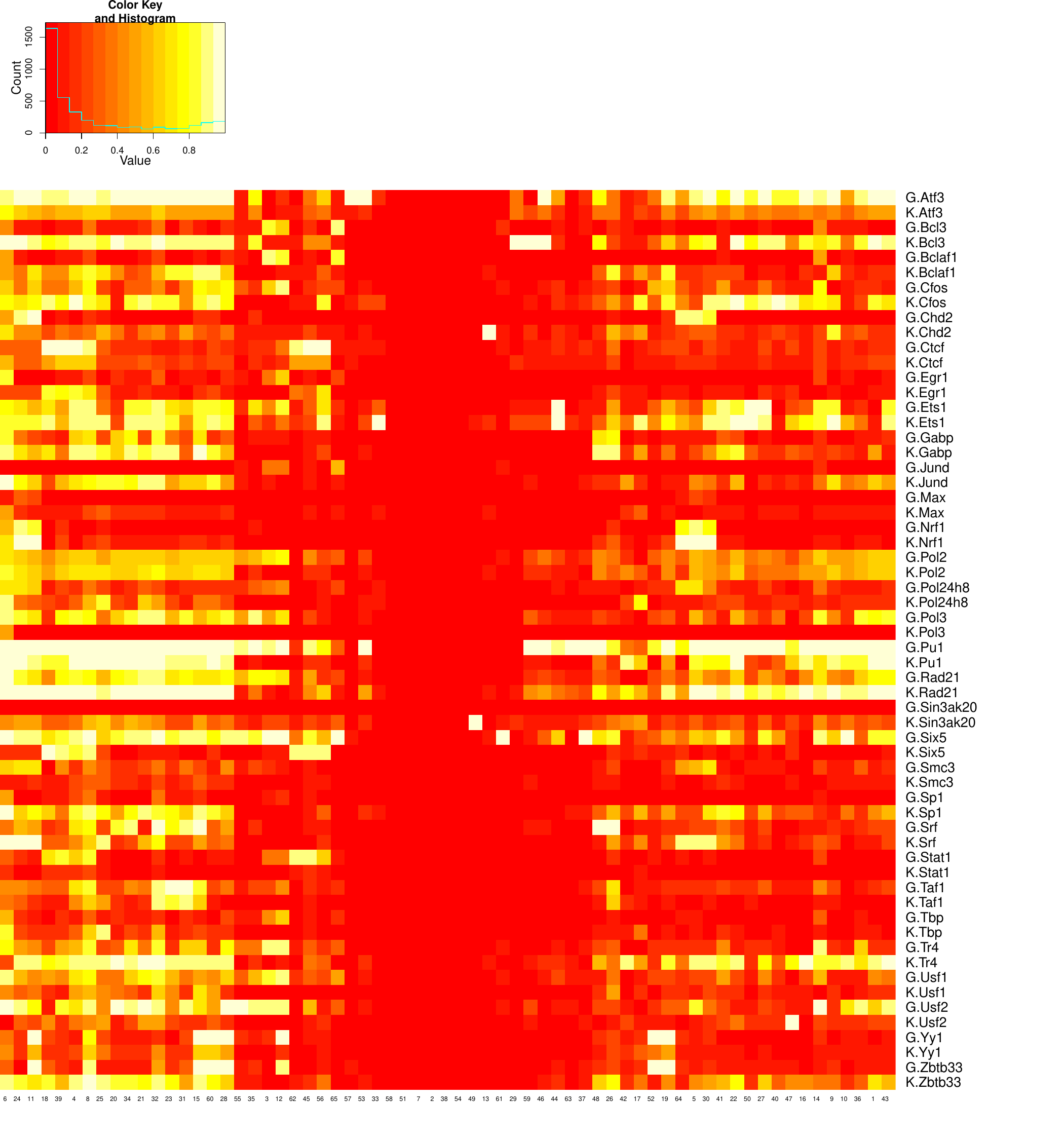}
\caption{ Estimated enrichment probability  for each of the 90 clusters identified by MClust. }
\label{fig:W_mclust}
\end{figure}

\begin{figure}[hbtp]
\centering
\begin{tabular}{ccc}
\includegraphics[width=0.8\textwidth, page = 1]{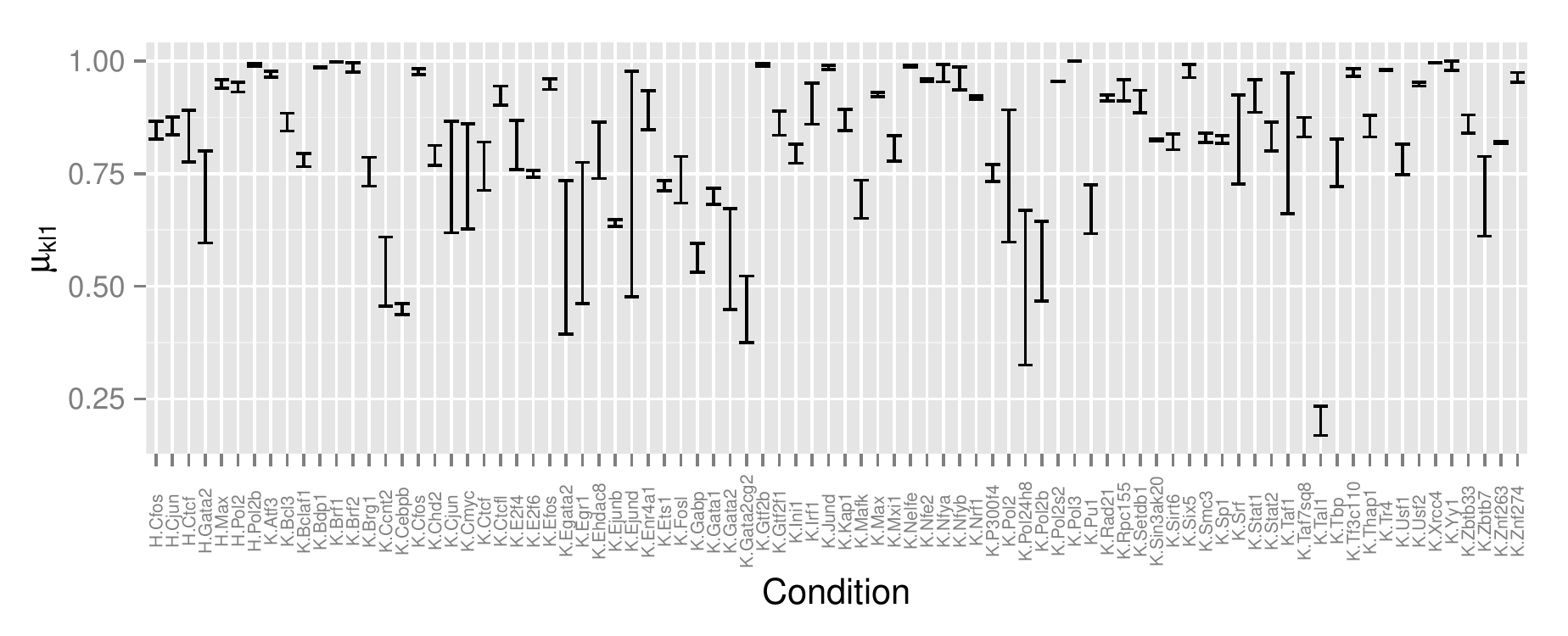}\\
\includegraphics[width=0.8\textwidth, page = 2]{par_ebox.pdf}\\
\includegraphics[width=0.8\textwidth, page = 3]{par_ebox.pdf}\\
\end{tabular}
\caption{The range of the estimated parameters $\mu_{kls}$ among the different replicates under the same experimental condition for the +9.5 composite element data in Section 4.2.}\label{fig:par_ebox_mu}
\end{figure}

\begin{figure}[hbtp]
\centering
\begin{tabular}{ccc}
\includegraphics[width=0.8\textwidth, page = 4]{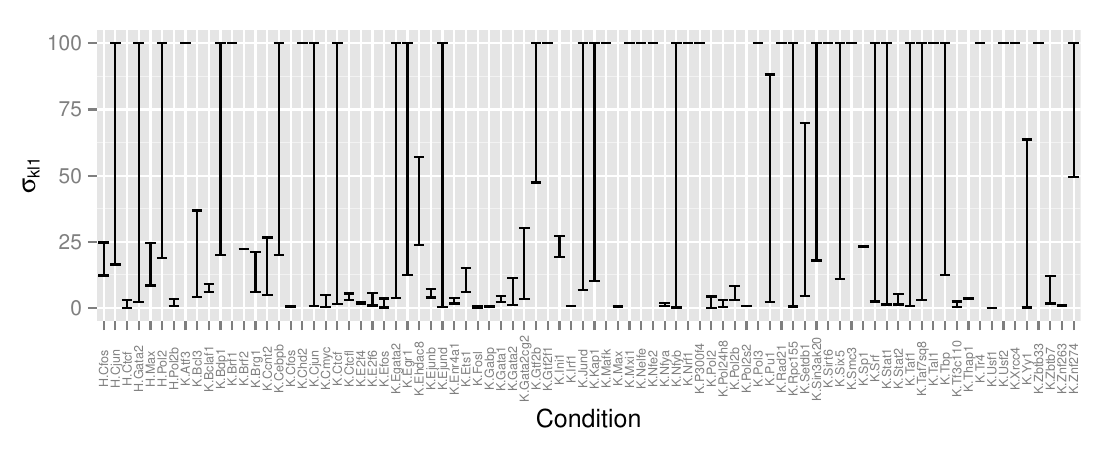}\\
\includegraphics[width=0.8\textwidth, page = 5]{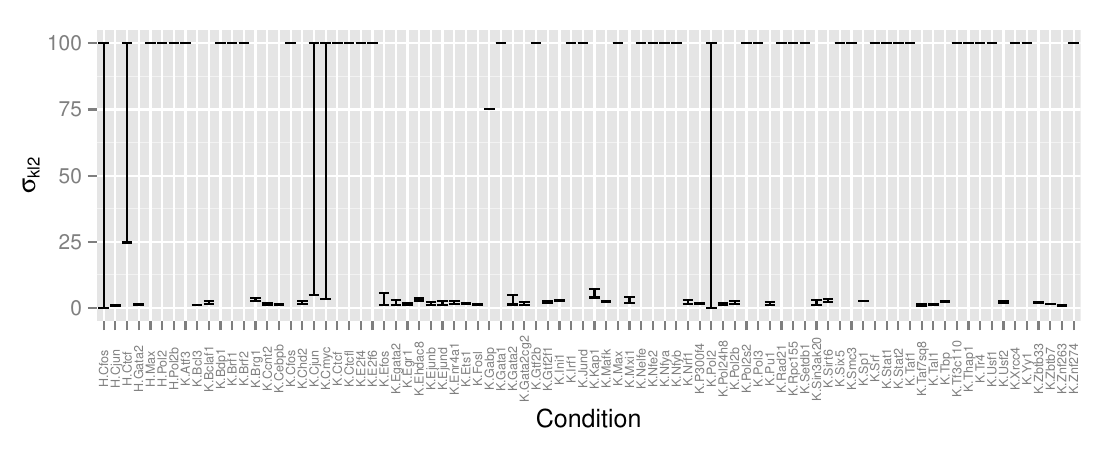}\\
\includegraphics[width=0.8\textwidth, page = 6]{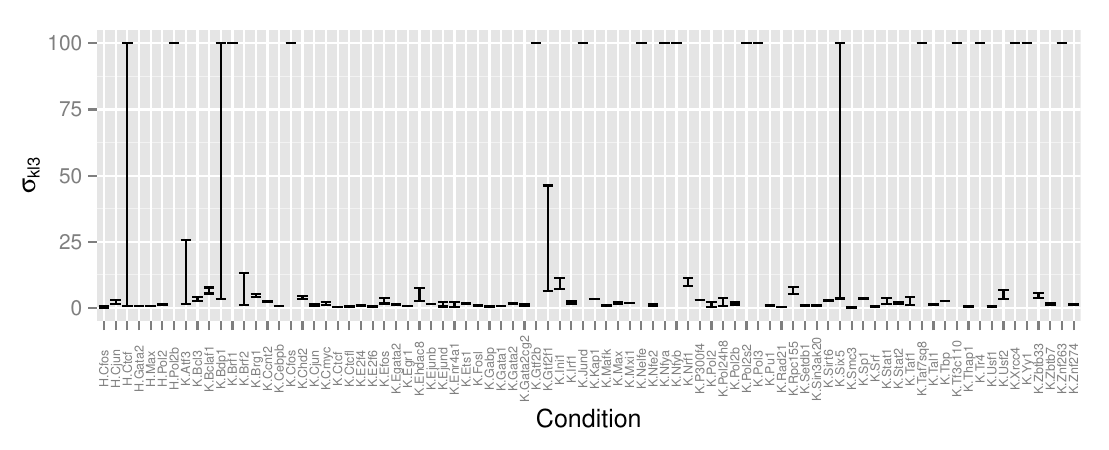}\\
\end{tabular}
\caption{The range of the estimated parameters $\sigma_{kls}$ among the different replicates under the same experimental condition for the +9.5 composite element data in Section 4.2.  For a subset of the replicates, the data for an individual state can be under-dispersed, resulting in a negative value for the estimated size parameter in the negative binomial distribution. In that case, we set $\sigma_{kls}=100$. Although our model do not fully capture the under-dispersion patterns, the large variations in $\sigma_{kls}$ for a subset of the experimental conditions suggest that assuming replicate-specific distributions is quite necessary.}\label{fig:par_ebox_sigma}
\end{figure}

\begin{figure}[hbtp]
\centering
\includegraphics[width=0.8\textwidth,height=\textheight,page=2]{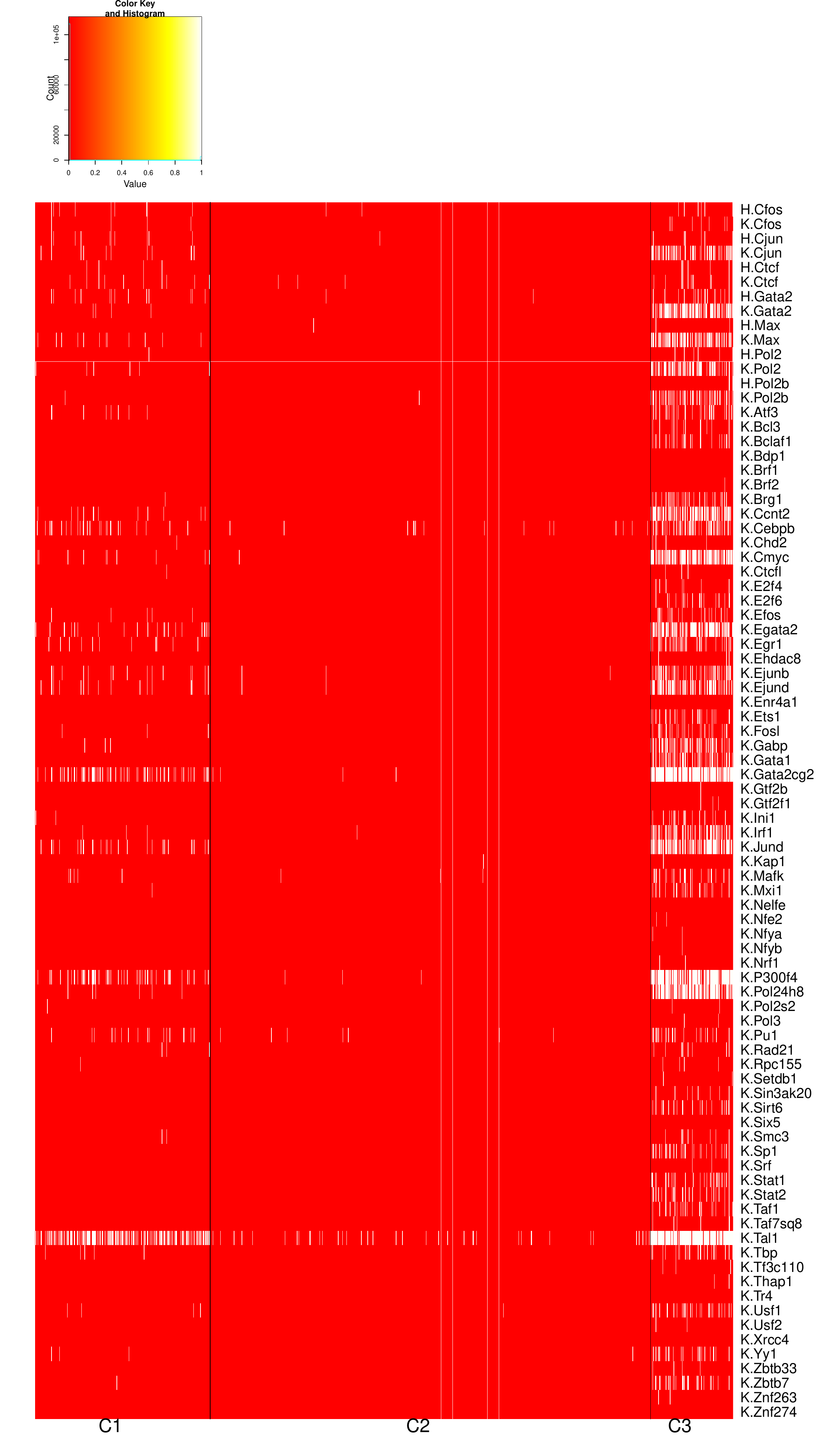}
\caption{Enrichment states provided by the ENCODE peak profiles. Seven empty rows corresponding to the TFs that lack ENCODE peak profiles. Note that only a small percentage of the composite elements which harbor the canonical GATA binding site are identified as enriched for GATA family transcription factors. This suggests that the ENCODE peak profiles can be conservative.}\label{fig:encode_ebox}
\end{figure}

\begin{figure}[hbtp]
\centering
\begin{tabular}{cc}
(a) & (b) \\
\multicolumn{2}{c}{\includegraphics[width=\textwidth,page=2]{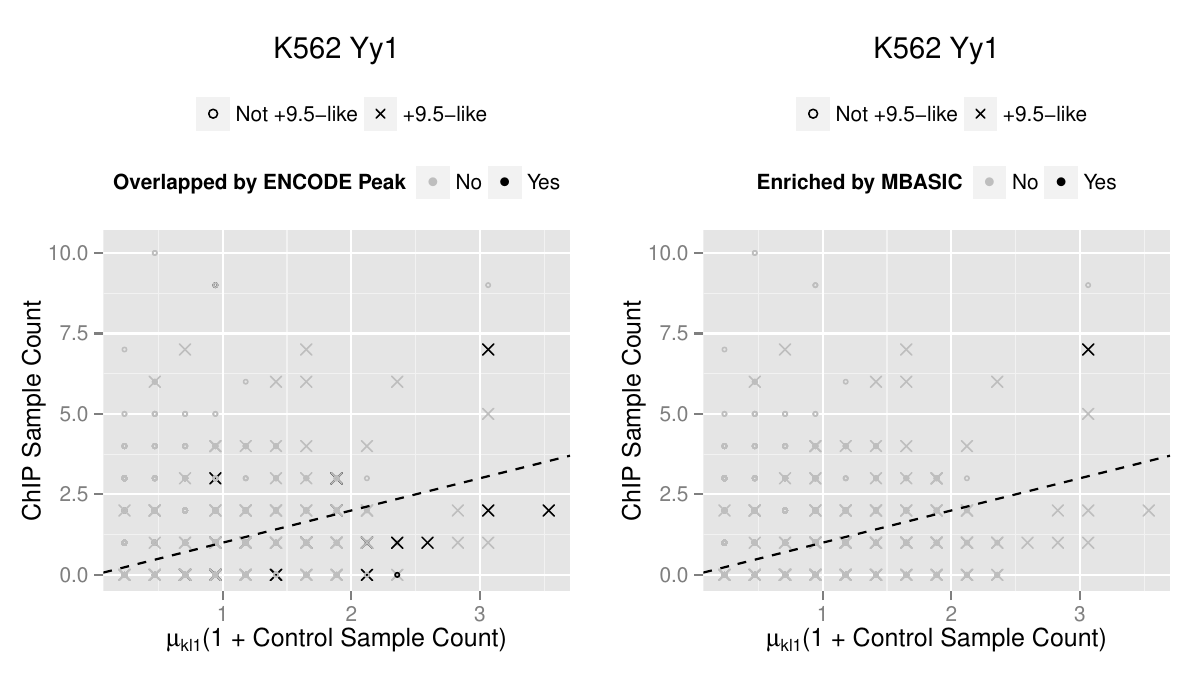}}
\end{tabular}
\caption{(a, b) ChIP sample read counts against control sample read counts for one replicate with K562-Yy1. Enrichment status are annotated by (a) the ENCODE peak profiles and (b) MBASIC prediction.}\label{fig:yy1}
\end{figure}

\begin{landscape}
\scriptsize
\raggedleft
\begin{longtable}{ccccccccccc}
\caption{Annotations for +9.5 Element-like loci in 5p1 (2Kb upstream of transcription start site (TSS)), 5p2 (2Kb to 10Kb upstream of TSS) and intronic regions.}\label{tbl:pa_result}\\
\toprule
Ref ID & Gene & Chr & Strand & Gene Start & Gene End & Region & Distance & Peak Start & Peak End & +9.5 Similarity\\
\endfirsthead
\toprule
Ref ID & Gene & Chr & Strand & Gene Start & Gene End & Region & Distance & Peak Start & Peak End & +9.5 Similarity\\
\midrule
\endhead
\midrule 
\multicolumn{11}{c}{Continued on the next page.}
\endfoot
\bottomrule
\endlastfoot
\midrule
NM\_001145662 & GATA2 & chr3 & - & 128198264 & 128206764 & intron & 4601 & 128202079 & 128202248 & 0.964\\
NM\_001145661 & GATA2 & chr3 & - & 128198264 & 128207373 & intron & 5210 & 128202079 & 128202248 & 0.964\\
NM\_032638 & GATA2 & chr3 & - & 128198264 & 128212030 & intron & 9867 & 128202079 & 128202248 & 0.964\\
NM\_005225 & E2F1 & chr20 & - & 32263292 & 32274210 & 5p2 & -3886 & 32278012 & 32278182 & 0.774\\
NM\_001166 & BIRC2 & chr11 & + & 102217965 & 102249394 & 5p2 & -5004 & 102212877 & 102213046 & 0.753\\
NM\_203343 & EPB41 & chr1 & + & 29213602 & 29446558 & intron & 39968 & 29253487 & 29253655 & 0.74\\
NM\_203342 & EPB41 & chr1 & + & 29213602 & 29446558 & intron & 39968 & 29253487 & 29253655 & 0.74\\
NM\_001166007 & EPB41 & chr1 & + & 29213602 & 29446558 & intron & 39968 & 29253487 & 29253655 & 0.74\\
NM\_004437 & EPB41 & chr1 & + & 29213602 & 29446558 & intron & 39968 & 29253487 & 29253655 & 0.74\\
NM\_001166005 & EPB41 & chr1 & + & 29213602 & 29446558 & intron & 39968 & 29253487 & 29253655 & 0.74\\
NM\_001166006 & EPB41 & chr1 & + & 29241087 & 29391731 & intron & 12483 & 29253487 & 29253655 & 0.74\\
NM\_173485 & TSHZ2 & chr20 & + & 51588876 & 52103965 & intron & 203292 & 51792084 & 51792253 & 0.735\\
NM\_007077 & AP4S1 & chr14 & + & 31494682 & 31555007 & intron & 13234 & 31507832 & 31508001 & 0.733\\
NM\_001128126 & AP4S1 & chr14 & + & 31494682 & 31562634 & intron & 13234 & 31507832 & 31508001 & 0.733\\
NM\_001430 & EPAS1 & chr2 & + & 46524540 & 46613842 & intron & 42757 & 46567214 & 46567382 & 0.728\\
NM\_018119 & POLR3E & chr16 & + & 22308740 & 22345341 & intron & 833 & 22309489 & 22309658 & 0.719\\
NM\_181442 & ADNP & chr20 & - & 49506882 & 49547527 & intron & 27423 & 49520020 & 49520189 & 0.718\\
NM\_015339 & ADNP & chr20 & - & 49506882 & 49547527 & intron & 27423 & 49520020 & 49520189 & 0.718\\
NM\_020359 & PLSCR2 & chr3 & - & 146151081 & 146213722 & 5p1 & -921 & 146214559 & 146214728 & 0.718\\
NM\_006257 & PRKCQ & chr10 & - & 6469104 & 6622238 & intron & 106706 & 6515449 & 6515617 & 0.713\\
NM\_018309 & TBC1D23 & chr3 & + & 99979685 & 100044078 & intron & 28727 & 100008329 & 100008497 & 0.711\\
NM\_020382 & SETD8 & chr12 & + & 123868703 & 123893898 & intron & 4170 & 123872789 & 123872958 & 0.71\\
NM\_015385 & SORBS1 & chr10 & - & 97071530 & 97321171 & intron & 29191 & 97291896 & 97292065 & 0.709\\
NM\_024991 & SORBS1 & chr10 & - & 97071530 & 97321171 & intron & 29191 & 97291896 & 97292065 & 0.709\\
NM\_000440 & PDE6A & chr5 & - & 149237519 & 149324356 & intron & 4584 & 149319688 & 149319857 & 0.699\\
NM\_012091 & ADAT1 & chr16 & - & 75632997 & 75657154 & intron & 2033 & 75655038 & 75655206 & 0.693\\
NM\_005033 & EXOSC9 & chr4 & + & 122722471 & 122738175 & 5p2 & -6644 & 122715743 & 122715912 & 0.691\\
NM\_001034194 & EXOSC9 & chr4 & + & 122722471 & 122738175 & 5p2 & -6644 & 122715743 & 122715912 & 0.691\\
NM\_004099 & STOM & chr9 & - & 124101353 & 124132545 & intron & 388 & 124132073 & 124132243 & 0.689\\
NM\_198194 & STOM & chr9 & - & 124101356 & 124132545 & intron & 388 & 124132073 & 124132243 & 0.689\\
NM\_014395 & DAPP1 & chr4 & + & 100737980 & 100791344 & intron & 25687 & 100763583 & 100763752 & 0.682\\
NM\_181078 & IL21R & chr16 & + & 27413722 & 27462115 & intron & 28911 & 27442549 & 27442718 & 0.681\\
NM\_181079 & IL21R & chr16 & + & 27414422 & 27462115 & intron & 28211 & 27442549 & 27442718 & 0.681\\
NM\_021798 & IL21R & chr16 & + & 27438578 & 27462115 & intron & 4055 & 27442549 & 27442718 & 0.681\\
NM\_002492 & NDUFB5 & chr3 & + & 179322574 & 179342287 & intron & 10827 & 179333318 & 179333486 & 0.68\\
NM\_021831 & AGBL5 & chr2 & + & 27274490 & 27293489 & intron & 11452 & 27285858 & 27286027 & 0.678\\
NM\_020132 & AGPAT3 & chr21 & + & 45285115 & 45407474 & 5p2 & -4281 & 45280751 & 45280919 & 0.677\\
NM\_007356 & LAMB4 & chr7 & - & 107663995 & 107770801 & intron & 50003 & 107720714 & 107720883 & 0.677\\
NM\_001010985 & MYBPHL & chr1 & - & 109834986 & 109849663 & 5p1 & -613 & 109850192 & 109850361 & 0.67\\
NM\_006253 & PRKAB1 & chr12 & + & 120105760 & 120119428 & 5p2 & -2184 & 120103492 & 120103661 & 0.668\\
NM\_015226 & CLEC16A & chr16 & + & 11038344 & 11276044 & intron & 27935 & 11066196 & 11066364 & 0.667\\
NM\_020448 & NIPAL3 & chr1 & + & 24742244 & 24799472 & intron & 22892 & 24765053 & 24765221 & 0.663\\
NM\_015560 & OPA1 & chr3 & + & 193310932 & 193415599 & intron & 67680 & 193378528 & 193378697 & 0.659\\
NM\_130832 & OPA1 & chr3 & + & 193310932 & 193415599 & intron & 67680 & 193378528 & 193378697 & 0.659\\
NM\_130831 & OPA1 & chr3 & + & 193310932 & 193415599 & intron & 67680 & 193378528 & 193378697 & 0.659\\
NM\_130834 & OPA1 & chr3 & + & 193310932 & 193415599 & intron & 67680 & 193378528 & 193378697 & 0.659\\
NM\_130837 & OPA1 & chr3 & + & 193310932 & 193415599 & intron & 67680 & 193378528 & 193378697 & 0.659\\
NM\_130836 & OPA1 & chr3 & + & 193310932 & 193415599 & intron & 67680 & 193378528 & 193378697 & 0.659\\
NM\_130835 & OPA1 & chr3 & + & 193310932 & 193415599 & intron & 67680 & 193378528 & 193378697 & 0.659\\
NM\_130833 & OPA1 & chr3 & + & 193310932 & 193415599 & intron & 67680 & 193378528 & 193378697 & 0.659\\
NM\_001004342 & TRIM67 & chr1 & + & 231298673 & 231357314 & 5p2 & -2591 & 231295998 & 231296167 & 0.659\\
NM\_020201 & NT5M & chr17 & + & 17206679 & 17250975 & intron & 1325 & 17207920 & 17208089 & 0.658\\
NM\_173054 & RELN & chr7 & - & 103112232 & 103629963 & intron & 331913 & 103297966 & 103298135 & 0.657\\
NM\_005045 & RELN & chr7 & - & 103112232 & 103629963 & intron & 331913 & 103297966 & 103298135 & 0.657\\
NM\_014206 & C11orf10 & chr11 & - & 61556602 & 61560085 & 5p2 & -8290 & 61568291 & 61568461 & 0.657\\
NR\_030342 & MIR611 & chr11 & - & 61559967 & 61560033 & 5p2 & -8342 & 61568291 & 61568461 & 0.657\\
NM\_173685 & NSMCE2 & chr8 & + & 126104082 & 126379367 & intron & 242235 & 126346233 & 126346402 & 0.655\\
NM\_001127511 & APC & chr5 & + & 112043217 & 112181935 & 5p2 & -3243 & 112039890 & 112040060 & 0.653\\
NM\_021926 & ALX4 & chr11 & - & 44282277 & 44331716 & intron & 39937 & 44291695 & 44291864 & 0.652\\
NM\_016213 & TRIP4 & chr15 & + & 64680019 & 64747500 & intron & 42584 & 64722519 & 64722688 & 0.649\\
NM\_007217 & PDCD10 & chr3 & - & 167401696 & 167452594 & intron & 10656 & 167441855 & 167442023 & 0.649\\
NM\_145860 & PDCD10 & chr3 & - & 167401696 & 167452630 & intron & 10692 & 167441855 & 167442023 & 0.649\\
NM\_145859 & PDCD10 & chr3 & - & 167401696 & 167452651 & intron & 10713 & 167441855 & 167442023 & 0.649\\
NM\_203318 & MYO18A & chr17 & - & 27400527 & 27507407 & intron & 17382 & 27489941 & 27490111 & 0.645\\
NM\_078471 & MYO18A & chr17 & - & 27400527 & 27507407 & intron & 17382 & 27489941 & 27490111 & 0.645\\
NM\_001626 & AKT2 & chr19 & - & 40736224 & 40791265 & intron & 12426 & 40778755 & 40778924 & 0.643\\
NM\_004767 & GPR37L1 & chr1 & + & 202092028 & 202098633 & 5p2 & -4769 & 202087175 & 202087344 & 0.642\\
NM\_015531 & C2CD3 & chr11 & - & 73745479 & 73882064 & intron & 84354 & 73797626 & 73797796 & 0.637\\
NM\_002738 & PRKCB & chr16 & + & 23847299 & 24231930 & intron & 166844 & 24014060 & 24014228 & 0.634\\
NM\_212535 & PRKCB & chr16 & + & 23847299 & 24231930 & intron & 166844 & 24014060 & 24014228 & 0.634\\
NM\_004571 & PKNOX1 & chr21 & + & 44394642 & 44453688 & intron & 15907 & 44410465 & 44410634 & 0.632\\
NR\_026749 & SKINTL & chr1 & - & 48567386 & 48648100 & intron & 4923 & 48643093 & 48643262 & 0.629\\
NM\_005560 & LAMA5 & chr20 & - & 60884122 & 60942368 & intron & 9836 & 60932449 & 60932617 & 0.626\\
NM\_001080826 & SGK223 & chr8 & - & 8175258 & 8239257 & intron & 9672 & 8229501 & 8229670 & 0.622\\
NM\_130465 & TSPAN17 & chr5 & + & 176074387 & 176086058 & intron & 1462 & 176075765 & 176075934 & 0.621\\
NM\_012171 & TSPAN17 & chr5 & + & 176074387 & 176086058 & intron & 1462 & 176075765 & 176075934 & 0.621\\
NM\_001006616 & TSPAN17 & chr5 & + & 176074387 & 176086058 & intron & 1462 & 176075765 & 176075934 & 0.621\\
NM\_013326 & C18orf8 & chr18 & + & 21083461 & 21111742 & 5p2 & -4363 & 21079015 & 21079183 & 0.616\\
NM\_138371 & FAM113B & chr12 & + & 47610051 & 47630441 & 5p2 & -8921 & 47601047 & 47601215 & 0.616\\
NM\_182498 & ZNF428 & chr19 & - & 44111376 & 44124014 & intron & 3381 & 44120549 & 44120718 & 0.615\\
NM\_025179 & PLXNA2 & chr1 & - & 208195589 & 208417665 & intron & 207170 & 208210411 & 208210580 & 0.614\\
NM\_020133 & AGPAT4 & chr6 & - & 161551056 & 161695107 & intron & 12103 & 161682920 & 161683089 & 0.611\\
NM\_013427 & ARHGAP6 & chrX & - & 11155662 & 11683821 & intron & 252809 & 11430929 & 11431097 & 0.609\\
NM\_006125 & ARHGAP6 & chrX & - & 11161516 & 11683821 & intron & 252809 & 11430929 & 11431097 & 0.609\\
NM\_001669 & ARSD & chrX & - & 2822011 & 2847392 & intron & 6868 & 2840441 & 2840609 & 0.607\\
NM\_009589 & ARSD & chrX & - & 2831654 & 2847392 & intron & 6868 & 2840441 & 2840609 & 0.607\\
NM\_032359 & C3orf26 & chr3 & + & 99536677 & 99897476 & intron & 243457 & 99780050 & 99780220 & 0.605\\
NM\_182909 & FILIP1L & chr3 & - & 99551988 & 99833349 & intron & 53215 & 99780050 & 99780220 & 0.605\\
NM\_001042459 & FILIP1L & chr3 & - & 99566772 & 99833349 & intron & 53215 & 99780050 & 99780220 & 0.605\\
NM\_194298 & SLC16A9 & chr10 & - & 61410521 & 61469649 & 5p2 & -3726 & 61473291 & 61473460 & 0.603\\
NM\_007356 & LAMB4 & chr7 & - & 107663995 & 107770801 & intron & 39404 & 107731314 & 107731482 & 0.594\\
NM\_203456 & PPIE & chr1 & + & 40204529 & 40229585 & intron & 18825 & 40223270 & 40223439 & 0.594\\
NM\_152726 & EFHA1 & chr13 & - & 22066839 & 22178307 & intron & 81061 & 22097162 & 22097331 & 0.592\\
NM\_001025107 & ADAR & chr1 & - & 154554535 & 154600437 & 5p1 & -1420 & 154601773 & 154601942 & 0.592\\
NM\_001130966 & TBXAS1 & chr7 & + & 139478046 & 139720123 & intron & 75241 & 139553203 & 139553372 & 0.591\\
NM\_001166254 & TBXAS1 & chr7 & + & 139478046 & 139720123 & intron & 75241 & 139553203 & 139553372 & 0.591\\
NM\_030984 & TBXAS1 & chr7 & + & 139528951 & 139720123 & intron & 24336 & 139553203 & 139553372 & 0.591\\
NM\_001166253 & TBXAS1 & chr7 & + & 139528951 & 139720123 & intron & 24336 & 139553203 & 139553372 & 0.591\\
NM\_001061 & TBXAS1 & chr7 & + & 139528951 & 139720123 & intron & 24336 & 139553203 & 139553372 & 0.591\\
NR\_029394 & TBXAS1 & chr7 & + & 139528951 & 139720123 & intron & 24336 & 139553203 & 139553372 & 0.591\\
NM\_173542 & PLBD2 & chr12 & + & 113796370 & 113827458 & intron & 20911 & 113817197 & 113817366 & 0.591\\
NM\_001159727 & PLBD2 & chr12 & + & 113796370 & 113827458 & intron & 20911 & 113817197 & 113817366 & 0.591\\
NM\_138356 & SHF & chr15 & - & 45459413 & 45493373 & intron & 31722 & 45461567 & 45461736 & 0.589\\
NM\_021908 & ST7 & chr7 & + & 116593380 & 116863955 & intron & 92727 & 116686023 & 116686192 & 0.588\\
NM\_018412 & ST7 & chr7 & + & 116593380 & 116870073 & intron & 92727 & 116686023 & 116686192 & 0.588\\
NM\_017681 & NUP62CL & chrX & - & 106366657 & 106449670 & intron & 53243 & 106396343 & 106396512 & 0.587\\
NM\_020845 & PITPNM2 & chr12 & - & 123468026 & 123594975 & intron & 73786 & 123521105 & 123521274 & 0.587\\
NM\_001135054 & SIGIRR & chr11 & - & 405715 & 414999 & 5p2 & -7383 & 422299 & 422467 & 0.586\\
NM\_021805 & SIGIRR & chr11 & - & 405715 & 417397 & 5p2 & -4985 & 422299 & 422467 & 0.586\\
NM\_001135053 & SIGIRR & chr11 & - & 405715 & 417397 & 5p2 & -4985 & 422299 & 422467 & 0.586\\
NM\_001012302 & ANO9 & chr11 & - & 417929 & 442011 & intron & 19629 & 422299 & 422467 & 0.586\\
NM\_001098816 & ODZ4 & chr11 & - & 78364328 & 79151695 & intron & 773881 & 78377730 & 78377899 & 0.582\\
NM\_178865 & SERINC2 & chr1 & + & 31885962 & 31907524 & 5p2 & -2738 & 31883140 & 31883309 & 0.581\\
NM\_004481 & GALNT2 & chr1 & + & 230202955 & 230417875 & 5p1 & -672 & 230202200 & 230202368 & 0.579\\
NM\_032427 & MAML2 & chr11 & - & 95711439 & 96076344 & intron & 19672 & 96056588 & 96056758 & 0.577\\
NM\_021961 & TEAD1 & chr11 & + & 12695968 & 12966298 & intron & 202724 & 12898608 & 12898778 & 0.577\\
NM\_016436 & PHF20 & chr20 & + & 34359922 & 34538288 & intron & 130764 & 34490602 & 34490771 & 0.573\\
NM\_003128 & SPTBN1 & chr2 & + & 54683453 & 54898582 & intron & 117920 & 54801290 & 54801458 & 0.573\\
NM\_178313 & SPTBN1 & chr2 & + & 54785530 & 54889444 & intron & 15843 & 54801290 & 54801458 & 0.573\\
NM\_001037165 & FOXK1 & chr7 & + & 4721929 & 4811074 & intron & 30769 & 4752614 & 4752783 & 0.573\\
NM\_005802 & TOPORS & chr9 & - & 32540542 & 32552601 & 5p1 & -1681 & 32554199 & 32554367 & 0.573\\
NM\_182739 & NDUFB6 & chr9 & - & 32553522 & 32573182 & intron & 18900 & 32554199 & 32554367 & 0.573\\
NM\_002493 & NDUFB6 & chr9 & - & 32553522 & 32573182 & intron & 18900 & 32554199 & 32554367 & 0.573\\
NM\_004466 & GPC5 & chr13 & + & 92050934 & 93519485 & intron & 7890 & 92058740 & 92058909 & 0.569\\
NM\_001145169 & GPR113 & chr2 & - & 26531040 & 26541970 & intron & 2083 & 26539803 & 26539972 & 0.563\\
NM\_153835 & GPR113 & chr2 & - & 26531040 & 26569685 & intron & 29798 & 26539803 & 26539972 & 0.563\\
NM\_001145168 & GPR113 & chr2 & - & 26532812 & 26541917 & intron & 2030 & 26539803 & 26539972 & 0.563\\
NM\_000593 & TAP1 & chr6 & - & 32812986 & 32821748 & 5p2 & -3584 & 32825248 & 32825417 & 0.549\\
NM\_002800 & PSMB9 & chr6 & + & 32821937 & 32827626 & intron & 3395 & 32825248 & 32825417 & 0.549\\
NM\_148954 & PSMB9 & chr6 & + & 32821937 & 32827626 & intron & 3395 & 32825248 & 32825417 & 0.549\\
NM\_033104 & STON2 & chr14 & - & 81736910 & 81864927 & intron & 94218 & 81770625 & 81770794 & 0.546\\
NM\_001001894 & TTC3 & chr21 & + & 38445570 & 38575406 & intron & 12717 & 38458203 & 38458372 & 0.544\\
NM\_003316 & TTC3 & chr21 & + & 38455246 & 38575406 & intron & 3041 & 38458203 & 38458372 & 0.544\\
NM\_000147 & FUCA1 & chr1 & - & 24171573 & 24194859 & 5p2 & -4297 & 24199072 & 24199241 & 0.544\\
NM\_016063 & HDDC2 & chr6 & - & 125596495 & 125623282 & 5p1 & -844 & 125624043 & 125624211 & 0.544\\
NM\_000404 & GLB1 & chr3 & - & 33038099 & 33138694 & intron & 89213 & 33049397 & 33049566 & 0.541\\
NM\_001135602 & GLB1 & chr3 & - & 33038099 & 33138694 & intron & 89213 & 33049397 & 33049566 & 0.541\\
NM\_001079811 & GLB1 & chr3 & - & 33038099 & 33138314 & intron & 88833 & 33049397 & 33049566 & 0.541\\
NM\_017803 & DUS2L & chr16 & + & 68057203 & 68113183 & intron & 17562 & 68074682 & 68074850 & 0.54\\
NM\_001101417 & ISPD & chr7 & - & 16127151 & 16460947 & intron & 160970 & 16299893 & 16300063 & 0.532\\
NM\_001101426 & ISPD & chr7 & - & 16127151 & 16460947 & intron & 160970 & 16299893 & 16300063 & 0.532\\
NM\_002736 & PRKAR2B & chr7 & + & 106685177 & 106802255 & intron & 31362 & 106716455 & 106716624 & 0.532\\
NR\_024448 & LOC91316 & chr22 & - & 23980676 & 24059610 & intron & 32642 & 24026884 & 24027054 & 0.529\\
NM\_153615 & RGL4 & chr22 & + & 24033047 & 24041358 & 5p2 & -6079 & 24026884 & 24027054 & 0.529\\
NM\_033631 & LUZP1 & chr1 & - & 23410515 & 23495351 & intron & 53595 & 23441673 & 23441841 & 0.526\\
NM\_001142546 & LUZP1 & chr1 & - & 23410515 & 23495351 & intron & 53595 & 23441673 & 23441841 & 0.526\\
NM\_001134492 & HS2ST1 & chr1 & + & 87380334 & 87564124 & intron & 77077 & 87457327 & 87457496 & 0.52\\
NM\_012262 & HS2ST1 & chr1 & + & 87380334 & 87575680 & intron & 77077 & 87457327 & 87457496 & 0.52\\
NM\_001085481 & MAP1LC3B2 & chr12 & + & 116997185 & 117014425 & intron & 2387 & 116999488 & 116999657 & 0.499\\
NM\_079834 & SCAMP4 & chr19 & + & 1905372 & 1926011 & intron & 2224 & 1907513 & 1907681 & 0.487\\
NM\_138422 & ADAT3 & chr19 & + & 1905416 & 1913443 & intron & 2180 & 1907513 & 1907681 & 0.487\\
NM\_032932 & RAB11FIP4 & chr17 & + & 29718641 & 29865232 & intron & 33750 & 29752307 & 29752476 & 0.48\\
NM\_033129 & SCRT2 & chr20 & - & 642240 & 656823 & 5p2 & -8983 & 665723 & 665891 & 0.47\\
NM\_012079 & DGAT1 & chr8 & - & 145538246 & 145550567 & intron & 1560 & 145548924 & 145549092 & 0.443\\\end{longtable}
\end{landscape}
\clearpage


\bibliographystyle{natbib}
\bibliography{mbasic}

\end{document}